\documentclass{article}
\usepackage{graphicx}
\usepackage{caption,caption3}
\usepackage{amsmath,latexsym,amssymb,nccmath}
\usepackage{color,appendix}
\usepackage[caption = false]{subfig}
\usepackage[sort&compress,square,comma,authoryear]{natbib}
  \usepackage{movie15}
\usepackage{hyperref}
\usepackage{authblk}
\usepackage{blindtext}
\begin{document}

\providecommand{\keywords}[1]
{
  \small	
  \textbf{\textit{Keywords---}} #1
}
\title{On the linear evolution of disturbances in plane Poiseuille flow}
\author[]{Usha Srinivasan \thanks{usha.s@nal.res.in}}
 \author[]{Rangachari Kidambi \thanks{kidambi@nal.res.in}}
\affil{Computational and Theoretical Fluid Dynamics Division, National Aerospace Laboratories, Bengaluru 560017, India}
\newcommand{\norm}[1]{\left\lVert#1\right\rVert}
\maketitle
\begin{abstract}
The linear evolution of disturbances due to a ribbon vibrating at frequency $\omega_0$ in plane Poiseuille flow is computed by solving the associated initial boundary value problem in the Fourier-Laplace plane, followed by inversion. A novel algorithm for identifying the temporal modes of the Orr-Sommerfeld equation (OSE) in the complex wavenumber plane, which are required in the inversion, is presented. Unlike in many prior studies, the performance of the Laplace integral first, not only avoids complicated causality arguments and confusion, in locating upstream and downstream modes, that is prevalent in literature but also yields a spatio-temporally uniform solution.
It also reveals that the solution consists of a time-periodic part at $\omega_0$, associated with the relevant spatial mode (the Tollmein-Schlichting wave) and a transient wavepacket, associated mainly with the saddle points of the OSE and is computed by the method of steepest descents, which can also include contributions from the spatial pole.  
Which of these parts dominates depends on the Reynolds number and $\omega_0$.  A secondary stability analysis of this dominant part is seen to explain the disturbance growth observed in the seminal experiments of Nishioka, Iida \& Ichikawa ({\em {J. Fluid Mech.}}, vol.72 , 1975, p.731)  and  Nishioka, Iida \& Kanbayashi (NASA TM-75885,  1981). Threshold amplitudes for instability at a subcritical Reynolds number $Re=5000$ are obtained from the time-averaged three dimensional disturbances, by combining the secondary base states and the growing Floquet modes. 
The observed minima of the
threshold amplitude curves in the experiments are explained in terms of the instabilities of these two base states.
Computations, for another subcritical (4000) and a supercritical (6000) Reynolds number, are also validated with the experimental data.
\end{abstract}
\section{Introduction}\label{intro}
It is well-known that shear flows become turbulent much below the critical Reynolds number   predicted by linear stability theory (LST), a phenomenon commonly referred to as `subcritical transition'. Despite many studies, it remains not completely understood even in simple shear flows like plane Poiseuille flow (pPf), even when subject to controlled conditions like a vibrating ribbon. \\

Of the handful of studies involving vibrating ribbons in pPf, the pioneering \cite{nish1} (N75, hereafter) is the most exhaustive, when it comes to studying in great detail, the beginning stages of the transition process. \cite{nish1a} (N81, hereafter) is a subsequent study aimed at clarifying the role of three-dimensionality in the transition process.  Apart from being the first study to observe two-dimensional Tollmein-Schlichting (2D TS) waves in pPf, N75 also sketched the downstream disturbance evolution in the linear regime as in figure 15 (N75F15; the reference to figure x of 
N75 will appear as N75Fx and that of N81 as N81Fx) and provided, for the first time, a plot of the threshold disturbance amplitude as a function of the ribbon frequency $\omega_0$ for three $Re$ (N75F16). The results presented in N75 and N81 are what concern us predominantly in this work.\\

Ever since it became evident that LST could not explain real transition scenarios, alternative mechanisms that could produce disturbance growth in subcritical conditions, have been sought. A popular one was the transient growth framework which was used to develop theories based on, for example, optimal disturbances \citep{schmid}. However, none of these optimals have been seen, for example, in N75 and there is a gap between how, if at all, the observed features evolve into these optimals. The second one, which actually appeared earlier than the first, is the concept of secondary instability, whose proponents, for example \cite{herb2}, have argued that the linearisation about a base state different from the parabolic profile could give rise to growing secondary modes even in subcritical conditions. The discovery of nonlinear equilibria and quasi-equilibria in pPf \citep{ors}, apart from the aforementioned optimals, suggested alternative base states for linearisation. \cite{red} sketched the supposed route to transition in a controlled disturbance environment-\\

$ 2D \, TS \, wave \rightarrow 2D \, state \rightarrow 2D \, state \, breakdown \rightarrow transition.$\\

The first component in the above pathway is the least stable spatial Orr-Sommerfeld (OS) mode of finite amplitude, at frequency $\omega_0$. This  is supposed to evolve nonlinearly into a 2D equilibrium state for $Re > 2900$ , which being unstable to 3D disturbances, breaks down and leads to transition. It is commonly held (for example, p.270 of \citealp{red}), that `this transition scenario agrees qualitatively and quantitatively with experiments where a two-dimensional TS wave is introduced in the flow via a vibrating ribbon'. However, this description a) assumes that the starting point is always the 2D TS wave and b) begs the question of how the subsequent 2D (nonlinear) state comes about in a vibrating ribbon experiment. We will show later that the starting point may not always be the 2D TS wave. As regards (b), neither have there been any 
experimental observations of such a state nor has a numerical simulation shown such a state emerge from the 2D TS wave.\\

How the disturbance input from a vibrating ribbon gets transformed into a disturbance field in the flow, also known as receptivity, is a key step in the evolution of the disturbance.  Receptivity study involves solving the linearised NS equations for given initial-boundary conditions, the so-called initial-boundary value problem (IBVP). A variety of analytic solution procedures for the IBVP in parallel shear flows have been advanced over the last 50 years. Most of these procedures have the same two key steps - i) solution of an ODE in the transformed plane and ii) inversion of the transformed solution back to the physical plane; the issues that crop up repeatedly are a) should the problem be treated temporally or spatially or as a combination, b) how to account for contributions of different modes, c) which modes contribute upstream and which downstream and d) how to fix the lines of integration for the inversion. \\

 \cite{gaso}(G65 hereafter) was probably the first to try to simulate the vibrating ribbon problem for the flat plate boundary layer; he was looking to reproduce the spatially growing waves that \cite{SS1948} had observed downstream of a disturbance source in a boundary layer. Till G65, theoreticians mostly used only temporally growing modes and in
fact, had reservations about the usage of spatial modes (See Section 47.2 of \citealp{draz} ). Thus, the experimental observations were sought to be explained by calculation of temporal growths which were converted to spatial growths by using the group velocity. The ribbon was approximated by a delta function as has been done here and the possibility of an exponentially growing mode allowed.  The inhomogeneous BC was incorporated into the integrand of the inversion integral. G65 fixed the line of integration (LOI) for the $\omega$ inversion in the UHP above all the poles, as per standard Laplace transform theory but then moved it below the real line and claimed that the contribution from the straight line segment vanished in the infinite time limit. 
This in turn led to  the fixing of the LOI in the $\alpha$ plane in one of
two ways - a) contour A passing above the pole and b) contour B passing below the pole,
dictated by which one would fit the initial unperturbed condition; this was in turn shown
to be linked to the sign of $\partial \alpha_r(\omega_0)
/ \partial \omega_r$  with the LOI being chosen as A if the latter is
negative and B if it is positive. For BL flows, the sign of this quantity is shown to be
the same sign as that of the group velocity and approximately equal to its reciprocal.
After performing the Fourier inversion, Laplace inversion was done by assuming that
the positive frequency axis could be closed by a 'suitable contour' on which the integral
vanished. The major shortcoming in G65 is due to non-appreciation of the possibility
of non-negligible contributions from this part of the contour. In fact, even in subcritical
flows, there is at least one saddle in the $\omega$ plane which makes the regular Jordan arcs
untenable due to intervening hills of the saddle. Thus, parts of the `suitable contour'
have to pass through the saddle, whose contribution then has to be included.\\

 \cite{ash1} (AR90 hereafter) revisit the vibrating ribbon problem for the same flow; their formulation, again in terms of Fourier transforms, is identical to that of G65, though following \cite{gus}, the continuous spectrum and branch cuts, that G65 had missed, were included.
 AR90 differs from G65 in including branch cuts associated with the continuous modes which are shown to have an upstream influence. It is similar to G65 in obtaining the time asymptotic solution by lowering the integration contour below the real axis in the $\omega$ plane. 
 After Fourier inversion, AR90 obtain the same solution as G65  for the discrete part, apart from the fact that the possibility of upstream propagating modes are allowed, if there exist poles in the left half $\alpha$ plane.
 They claim that the location of the pole determines the direction of wave propagation
with the left half poles (negative phase velocity) contributing to upstream propagating waves and the right half ones (positive phase velocity) to downstream ones. However, the choice of the integration contours, in figure 5, for example, is same as that of Gaster and consequently should have been dictated by considering the sign of the group velocity.  While this makes no difference in cases where the phase and group velocities are of the same sign, (ai, bii, ci and dii of figure 5), the sign of the group velocity decides which mode contributes where in the other cases (aii, bi, cii and di). Similar to G65, their simplistic treatment of the Jordan arc, ignoring the saddle points,
 leads to a non-accounting of possible downstream-growing wavepackets for supercritical Reynolds numbers. \\

  \cite{hill} considers the receptivity of a boundary layer flow to a variety of disturbances including freestream and boundary disturbances. The Lagrange identity and adjoint velocity, pressure and stress fields are used to compute the response for a given OS eigenmode. The response for the vibrating ribbon problem is shown to match that obtained by AR90. The emphasis of \cite{hill} is on computing the responses solely from the adjoint field; the physical fields do not seem to be computed and hence issues related to inversion are not 
discussed.\\
 
 \cite{tum} considers disturbance evolution due to blowing / suction in a wall slot in pipe Poiseuille flow. Assuming that the response is at the forced frequency $\omega_0$ and disturbances decay both upstream and downstream, the solution is written in two alternate ways - as an inverse Fourier transform and as a (countably infinite) sum over all the `downstream moving' spatial OS modes corresponding to $\omega_0$.  The computed solution is valid only downstream of the disturbance source. the receptivity coefficients are calculated by using bi-orthogonality between 
the original disturbance eigenfunctions and a set of adjoint eigenfunctions. However, the  procedure has several deficiencies.  Apart from the use of the spatial modes, mentioned earlier, it is not clear how the `downstream moving' modes have been identified. Even though spatial eigenvalues are found in the first, second and fourth quadrants, only the contributions from the first quadrant poles seem to have been considered. Though this simplifies the fixing of the inversion contour 
(which can be taken as the real line), it is questionable. Another deficiency of
 this method is to  represent the solution as a sum of spatial eigenmodes, whose completeness has not been established till date. Hence, the vibrating ribbon problem necessarily has to be formulated as an IBVP unlike \cite{tum}. Therefore, receptivity cannot be established by a set of coefficients called receptivity coefficients; it is in fact a spatio-temporal evolution  as 
will be shown in the next section. \\

\cite{ma} assume Fourier series in the $\theta$ and $t$ 
variables, an eigenmode expansion in  $r$ and derive an ODE system in the streamwise variable $x$. Adjoint eigenmodes are used in deriving this system. The issue of which modes to include in which part of the domain ( $x > 0, < 0$) is settled by appealing to the `well-known' linear stability of  Hagen-Poseuille flow and also by comparison with the DNS results; thus first and second quadrant modes are taken as contributing to the downstream field and the third and fourth quadrant modes contributing to the upstream disturbance field. We will formulate and solve the problem in a way that will obviate the need for making such extraneous assumptions.\\

  \cite{man} studied time dependent disturbances of pPf, with the disturbance generator being a triangular vibrator of finite length $l,$ mounted on the upper wall, oscillating sinusoidally at frequency $\omega_0$. 
Symmetric and antisymmetric modes are considered separately and subcritical and supercritical cases are distinguished. For the former case, it is shown that, for  $x > l (< -l$ resp.), the LOI has to be closed in the UHP 
(LHP resp.) and the solution is written as a sum of discrete spatial modes corresponding to $\omega_0$, with UHP poles
selected for $x > 0$ and LHP poles for $x < 0.$ As we will see later, this is incorrect.\\

In summary, all the works cited above (except \citealt{hill}) solve the ODE in the transformed plane and then invert the solution back to the physical plane. The method of solution varies and is different from the lifting procedure employed here. Both temporal and spatial inversions are performed in general and a variety of strategies for choice of inversion contours are used; all of these seem to invoke a radiation condition without stating explicitly. Also, all the analyses (apart from G65 who briefly mentions the transients but in a different context) deal with large time asymptotics; these methods are applicable only for $x/t\approx 0$. Hence, it is not possible to use these methods at all $x$ even at large $t$, in particular for $x/t \rightarrow \infty$. The solutions from these methods are spatially unbounded for supercritical flows. 
 This is a crucial deficiency as Fourier transforms require spatial boundedness at any given time. This not only makes these procedures mathematically suspect but also render them incapable of producing solutions that can be compared with experimental results.  \\
 
 The use of radiation condition is unavoidable if the disturbance response is assumed to be at the
  forcing frequency; this assumption results in a boundary value problem. A properly posed problem 
  has to be an IBVP, requiring only boundedness conditions at infinity; the resulting solution is unique and hence there is no need for an additional radiation condition (\citealp{schot}).
 In the present approach to evaluating the integrals (described elsewhere in detail), the global topography of the $\alpha$ and $\omega$ modes is computed and the contributions of the various saddles, branch points and poles are assessed. This leads, in a natural way, without having to make ad hoc assumptions, to the correct spatial decay at infinity.\\

 Such IBVPs have been solved routinely, and in a clear manner, with none of the confusion described earlier, in wave propagation problems in other branches of physics like optics, geophysics and atomic and molecular physics (for e.g. \citealp{felsen}).  A recent exemplary application in fluid mechanics is \cite{gor} which describes a procedure for inverting similar integrals, and give numerical results for the forced Ginzburg-Landau equation and the forced Kelvin-Helmholtz problems. By performing the Laplace inversion first, the structure of the solution is laid bare - the solution is seen to consist of a term that gives a response at the signal frequency and a second term which gives the transient and whose evaluation is based on the well-known method of steepest descent.  This method
has been used in stability studies, but mostly to distinguish between convective and absolute instabilities (\citealp{gas1}, \citealp{juniper},  \citealp{ling1}). With this procedure, conceptual issues surrounding fixing of inversion contours, choice of upstream / downstream modes and the role of the transient are clarified. We formulate and sketch the solution in \S \ref{sec:methods1}.\\
 
The Fourier inversion involves complex integration in the wavenumber plane and hence, the map of each temporal eigenvalue, as a function of the complex wavenumber, has to be obtained. An understanding of the topography specified by these maps as signified by the knowledge of critical points of the map, like saddles and branch points, is crucial to a correct solution and the absence of a reliable method to sort the eigenmodes correctly has hampered a proper investigation of these problems. \cite{koch} was one of the first to attempt a sort of mode-tracing for pPF in the $\omega$ and $\alpha$ planes as a function of real $\omega$ and $\alpha$ respectively. His procedure was sensitive to mode jumping and his 
mode indexing (for example in figure 4 of that paper) can only be considered tentative. It can be seen from his figure that what is designated the principal instability mode, in fact is clearly not the least damped mode at all frequencies considered in that figure. The problems with mode tracing and partial solutions are described extensively in \cite{sus1}, wherein the author details successively the problems of sorting eigenvalues based on real parts, imaginary parts or
even switching between the two procedures. While the first two fail whenever there are `collisions' of temporal branches, the last one fails for `true collisions' apart from being difficult to implement numerically. He suggested tracking the eigenvalues based on a quantity $\gamma_r$; the eigenvalue with the largest value of $\gamma_r$ corresponds to the dominant mode in a frame
with speed $\rightarrow 0.$ However, this method can also fail at true collisions. We have used the analytic properties to develop an algorithm to sort eigenmodes corresponding to eigenvalues that have been produced from an eigenvalue solver. It works even at a true collision (for e.g. a double root in the $\omega$ plane) and correctly produces branch points and branch cuts. The algorithm is sketched in Appendix \ref{sec:appA}. Some of the relevant modal maps will be presented in \S \ref{sec:maps}. Key aspects of the solution can be deduced by an asymptotic analysis using the method of steepest descent; details are presented in this section.\\
 
  The spatio-temporal evolution is in general complicated but two idealized secondary states can be identified  - a) the TS wave (related to the dominant spatial OS pole at $\omega_0$) and b) a wavepacket (related to the saddle of the phase function of the dominant OS mode). Depending on $\omega_0$ and $Re$, either of these may be dominant; there are also mixed regions where both may be important. Some of these solutions will be presented in \S \ref{sec:results1a}. The two states have very different decay rates. While the characteristics of TS wave are obtained directly from the  OS dispersion equation, those of the wavepacket have to be deduced from the IBVP solution; the decay rate and the propagation velocity of the wavepacket are discussed in detail 
  in this section. The IBVP solutions are used to explain the features of the linear developments presented in N75. \\

With the solution of the IBVP as a guide, we choose an appropriate secondary base state (either of (a) or (b) above) and perform a Floquet analysis to get the secondary growth rate. 
Ideally a spatial secondary analysis should be performed at the given drive frequency. However, this is much more complicated than a temporal analysis and also the spatial growth rate can be obtained (approximately) from the temporal one, following Herbert et al(1987). Following most studies in this area, we also do a temporal secondary analysis. The results of the secondary instability analysis are presented in \S 5. \\

In the subsequent \S \ref{sec:comparison}, we present comparisons with experimental results of N75 and N81. We concentrate, in particular, on two figures of these papers N75F16 and N81F15, that have been reproduced in figure 12 and 13, for convenience.  Since N75F16 is based on N75F15, we discuss that first.\\

N75F15 shows growth / decay of a subcritical disturbance at 72 Hz at an $Re = 5000$. It is clear from the figure that the disturbances, below a threshold level, show an initial growth and eventual decay in the streamwise direction; above the threshold, they grow continuously. \cite{klei} claims to see similar behavior in the DNS of an initial value problem in pPf; however, instead of a vibrating ribbon, initial conditions were prescribed. \cite{tref} show similar behavior for a $2 \times 2$ nonlinear model. Below some nonlinearity threshold, the curves rise and then decay; above these, they grow and saturate at some amplitude. The former is attributed to non-normal transient growth, the latter to the added nonlinearity, which though not directly contributing to the growth, redistributes the energy such that explosive growth can occur. However, no comparisons with experiments are attempted and the discussion on the relevance of this model to actual flows (p.582)is only at the level of conjecture. To our knowledge, it is yet to be substantiated, say by comparison with an experiment like N75. More importantly, both these studies, and the pathway sketched in \cite{red}, involve nonlinearities whereas the lower curves in N75 are in the linear regime, as we will see in detail later. We seek to throw some light on this behavior by a study of the IBVP solution.\\

N75F16 shows threshold amplitudes $A_t$ vs. $\omega_0$ for three Re (two subcritical Reynolds numbers 4000, 5000 and one mildly supercritical, 6000). We assume that the threshold was deduced by observing an unchanging $u_{max}/ U$ with $x$, much like the (mostly) flat curve (iv) in figure 15. \\

The threshold curves have the following features -\\

For fixed $Re,$\\

a) Two minima $Mi_1, Mi_2$ separated by a maximum $Ma$. ($Mi_2$  is not a true minimum but an endpoint of the interval at which $A_t$ attains a global minimum.)\\

b) $Mi_1$ is roughly that at which the spatial decay rate is the minimum.\\

c) $A_t(Mi_2) < A_t(Mi_1).$ \\

For fixed $\omega_0,$\\

d) $A_t$ is a decreasing function of $Re.$\\

Also shown in the figure are the nonlinear calculations of $A_t$ (\citealt{ito}). Though N75 claims good agreement with Itoh's nonlinear calculations, the fact is that only (d), which  is qualitative in nature, is reproduced. Not only does \cite{ito} not produce clear minima at all $Re$, it does not produce $Mi_2$ at all. N75 speculates (p.750) that transition at higher drive frequencies is triggered `directly by spot-like fluctuations appearing before the fundamental has grown sufficiently.' and also express their belief that `this may be due to the highly three-dimensional nature of a disturbance with a large $\beta.$' 
Not only are we not aware of any subsequent published study, including N81, that 
 clarifies the issue, but also an exhaustive search of the 
 literature did not reveal an explanation for $Ma$ or $Mi_2.$\\
 
 \cite{zhou}, \cite{sen} and \cite{sus2} are other nonlinear calculations that tried to reproduce this figure. The first of these does produce threshold amplitude curves that are remarkably similar to the experimental curves, though higher; the difference is attributed to neglect of three-dimensional effects. However, the procedure involves artificially splitting the flow field into a base flow, with real fundamental eigenvalue, and a perturbation and is done to produce solvability conditions, which similar earlier studies could not, with the TS mode. The second of these uses the earlier formulation of \cite{rey} but claims to sum the resulting Stuart-Landau series more accurately by using Shanks method. They also critically discuss Zhou's results and conclude that there could be convergence issues, without a clear resolution of which, `it is perhaps too early to reach any conclusions regarding Zhou's results.'  The final work solves a cubic Landau equation, based on the Watson model, thus retaining more physical features of the problem, as compared to \cite{zhou}. The threshold amplitude curves in the last two studies do not show the features of the N75 curve; the second one does not show $Mi_2$ whereas the third shows scattered values with no clear trend. \cite{dhan} is a nonlinear analysis based on higher 
order amplitude expansions; the important difference from the other nonlinear studies  is that the basic state here is three-dimensional, by considering channel walls in the form of small amplitude spanwise waves. However, no higher order terms are included and the analysis does not capture $Mi_2.$  \\

We seek to throw light on the local extrema in the threshold amplitude curves  by secondary instability analysis. In particular, we provide evidence that the two minima are linked to the secondary instabilities of the TS wave and wavepacket states.

Concluding remarks are presented in \S \ref{sec:conclude}.

\section{The IBVP}\label{sec:methods1}
\subsection{Formulation}
The setting for the problem is a plane channel between the walls $y = \pm 1$ with the base flow being the unidirectional, plane Poiseuille flow $U(y) = 1 - y^2.$ We consider the problem of creation and evolution of disturbances in such a flow that is subjected to a local unsteady forcing, typically on $y=-1.$
The forcing is supposed to simulate the effect of a vibrating ribbon, or a blowing / suction device in a controlled transition experiment.\\

 The linearized equations governing the normal disturbance velocity $v$ and vorticity $\eta$ are the well-known (\citealt{schmid}) OS and Squire equations respectively -

\begin{fleqn}
\begin{align*}
L_{OS}(v)=\left( \frac{\partial}{\partial t}+U\frac{\partial}{\partial x} \right) 
\left( \nabla_h^2+{D^2} \right)v-D^2U 
\frac{\partial v}{\partial x}-\frac{1}{Re}
\left( \nabla_h^2 + D^2 \right)^2 v = 0, \nonumber\tag{2.1a}
\label{eq1}
 \end{align*}
\end{fleqn}
\begin{fleqn}
\begin{align*}
L_{SQ}(\eta) = \left( \frac{\partial}{\partial t}+U\frac{\partial}{\partial x} -\frac{1}{Re} \left[\nabla_h^2 +{D^2}\right]\right)  \eta
=-DU \frac{\partial v}{\partial z}.\tag{2.1b}
\end{align*}
\label{eq1b}
\end{fleqn}
\noindent where $\nabla_h^2 = \partial_{xx} + \partial_{zz}$ and $D$ denotes the differential w.r.t $y$. \\

The system (\ref{eq1}) has to be solved with appropriate initial / boundary conditions. Many facets of the eigenvalue problem, where (\ref{eq1}) is solved with homogeneous BCs have been extensively studied, with the literature running to hundreds of papers. In the present case, we are interested in formulating and solving an initial-boundary value problem (IBVP) for {\eqref{eq1}}. In particular, we prescribe the external forcing as a time-dependent boundary condition for the normal disturbance velocity $v$ on the lower wall $y = -1$. 
\begin{fleqn} 
\begin{align*}
 v(x,-1,z,t)= \delta(x) g(z) h(t) \ \ \ \hbox{and}\ \ \ Dv(x,-1,z,t)=0.
\tag{2.2}
 \label{eq2}
\end{align*}
\end{fleqn}
\noindent $h(t)$ is usually assumed to be periodic-in-time, starting from $t=0$. For $t<0, \ \ h(t)=0$. Here, the disturbance source is at $x=0$. The positive and negative values of $x$ denote streamwise positions downstream and upstream of the disturbance source respectively. Spanwise conditions, 
expressed through $g(z)$, usually take a periodic form with $g(z) = e^{i \beta_0 z}$ with $\beta_0$ real, with the understanding that the appropriate part (real or imaginary) of the final solution will be taken.\\

On the top wall $y=1$, $v$ and $Dv$ satisfy homogeneous boundary conditions:

\begin{fleqn}
\begin{equation*}
 v(x,1,z,t)= 0 \ \ \ \hbox{and}\ \ \ Dv(x,1,z,t)=0.\tag{2.3} \label{eq3}
\end{equation*}
\end{fleqn}

At  $x,z= \pm\infty$,  the disturbances and all its derivatives are assumed to decay as $t\rightarrow \infty$.  
We assume zero initial conditions i.e. $v(x,y,z,0) = 0.$ This means that the disturbance generator starts from rest. 
It is
well known that for subcritical and slowly growing supercritical pPf in the linear regime,  the least stable disturbances are two-dimensional. We hence restrict the present study to  two-dimensional,  $z$-independent wall forcing; $g(z)=1$. 
\subsection{Solution}\label{sec:solution}
The disturbance evolution is governed by equations \eqref{eq1}
together with the initial and boundary conditions \eqref{eq2} and \eqref{eq3}. 
\eqref{eq1} is a homogeneous ODE system for $v$ with non-homogeneous BC. It 
turns out to be convenient to transform this to a 
inhomogeneous ODE system for the auxiliary variable $v_1$
satisfying homogeneous BC.
This can be achieved by a suitable lifting procedure (p.436, \citealt{lanc})which involves expressing $v(x,y,t)$ as
\begin{fleqn}
 \begin{equation*}
 v(x,y,t)=v_1(x,y,t)+ \delta(x) h(t) f(y) 
\quad \hbox{where} \quad f(y)=\frac{2-3y+y^3}{4}. \tag{2.4}\label{eq4}
\end{equation*}
\end{fleqn}

Substituting (\ref{eq4}) into (\ref{eq1}), the inhomogeneous ODE for $v_1$ 
is obtained as
\begin{fleqn}
\begin{equation*}
 L_{OS}(v_1(x,y,t)) = - L_{OS}\left[\delta(x) h(t) f(y) \right],  
\tag{2.5}\label{eq5}
\end{equation*}
\end{fleqn}
Assuming $h(0)=0$, $v_1$ satisfies zero initial condition and homogeneous boundary conditions.\\ 

Fourier and Laplace transforming (2.5) and its homogeneous boundary and initial conditions,
 w.r.t $x$ and  $t$ respectively,
the well-known Orr-Sommerfeld equation with an inhomogeneous term is obtained -  
\begin{fleqn}
\begin{equation*}
[\mathcal{L} - i \omega \mathcal{M} ] \hat{v}_1 = -\tilde{h}(\omega)  [\mathcal{L} - i \omega \mathcal{M}] f(y), 
\tag{2.6}\label{eq6}
\end{equation*}
\end{fleqn}
\noindent where 
\begin{fleqn}
\begin{align*}
\mathcal{L} = i \alpha U (D^2 - \alpha^2) - i \alpha D^2 U - \frac{1}{Re}
(D^2 - \alpha^2)^2, \ \hbox{and} \ \mathcal{M} = D^2 - \alpha^2. 
\end{align*}
\end{fleqn}
\noindent The hat symbol $ \hat{}  $ denotes the Fourier-Laplace 
transform of a given 
function and the overbar  denotes Fourier transform w.r.t $x$.\\

We are interested in sinusoidal forcing, starting from rest i.e. we take
$h(t) = \sin \omega_0 t$. 
For a given $\alpha$ (real) completeness of temporal OS eigenfunctions 
(\citealt{prima})
allows eigenfunction expansion for $\hat{v_1}$ as
\begin{fleqn}
\begin{equation*}
 \hat{v_1}(\alpha,y,\omega)=\sum_{n=1}^{\infty}
C_n(\alpha,\omega)
\phi_n^{(\alpha)}(y) \tag{2.7}\label{eq7}
\end{equation*}
\end{fleqn}
\noindent Using the bi-orthogonality of $\phi_n^{\alpha}(y)$ and the adjoint eigenfunctions $\xi_n^{\alpha}(y)$\\
\noindent (\citealt{schmid}),
\begin{fleqn}
\begin{eqnarray}
\tilde{v_1}(\alpha,y,\omega)&=&  \sum_{n=1}^{\infty}
\dfrac{\phi_n^{(\alpha)}(y)\int_{-1}^1\left[\mathcal{L}-i\omega\mathcal{M} 
\right]f \xi^*_n dy}{(\omega^2-\omega_0^2)(\omega-\omega_n^{OS}) 
K_n},\tag{2.8} \label{eq8} 
\end{eqnarray}
\end{fleqn}
\noindent where $K_n$ is given by
\begin{eqnarray}
  \int_{-1}^{1} {\xi}_k^* (k^2 -
  D^2) \phi_j \ dy  = K_j \delta_{jk}.\tag{2.9} \label{eq9}
 \end{eqnarray} \\
 
Inverting (\ref{eq8}) from the $(\alpha,y,\omega)$ to the physical $(x,y,t)$ plane by Laplace and Fourier inversions yields
\begin{fleqn}
\begin{eqnarray}
 v_1(x,y,z,t)=
\frac{1}{4 \pi} \sum_{n=1}^{\infty}\int_{-\infty}^{\infty}
\phi_n^{(\alpha)}(y) \frac{e^{i(\alpha x -\omega_0 t)}
I_{-}(\alpha) - e^{i(\alpha x -\omega_n t)} I_{n}^{OS}(\alpha)}{
(\omega_n- \omega_0) K_n } d \alpha \nonumber \\
- \frac{1}{4 \pi} \sum_{n=1}^{\infty}\int_{-\infty}^{\infty}
\phi_n^{(\alpha)}(y) \frac{e^{i(\alpha x +\omega_0 t)}
I_{+}(\alpha) - e^{i(\alpha x
-\omega_n t)} I_{n}^{OS}(\alpha)}{
(\omega_n+ \omega_0) K_n } d \alpha,\tag{2.10} \label{eq10} 
\end{eqnarray}
\end{fleqn}
 \begin{fleqn}
 \begin{align}
I_{\pm}(\alpha) = \int_{-1}^{1} [ \mathcal{L} \pm i \omega_0 \mathcal{M}] f
\xi_n^* dy, \,\, I_{n}(\alpha) = \int_{-1}^{1} [\mathcal{L} - i 
\omega_n \mathcal{M}]f \xi_n^* dy. \tag{2.11}
\label{eq11}
\end{align}
\end{fleqn}

\noindent The explicit forms of the integrals $I_{\pm}(\alpha)$ and $I_{n}(\alpha)$ are given in Appendix \ref{sec:appC}. The second and the fourth integrals are periodic in time with frequency $\omega_0$. They are similar to the solution obtained by \cite{tum}; the coefficients in the series are the receptivity coefficients.\\

The horizontal velocity $u(x,y,z,t)$ can be obtained, by using the continuity equation. The part of $u$ corresponding to $v_1$ can be obtained as\\
\noindent (\citealt{schmid})
\begin{fleqn}
 \begin{align}
u_1(x,y,z,t) = \frac{i}{4 \pi} \sum_{n=1}^{\infty}\int_{-\infty}^{\infty}
\frac{1}{\alpha} \frac{d \phi_n^{(\alpha)}(y)}{dy} \frac{e^{i(\alpha x -\omega_0 t)}I_{-}(\alpha) - e^{i(\alpha x -\omega_n t)} I_{n}^{OS}(\alpha)}{
(\omega_n- \omega_0) K_n } d \alpha \nonumber \\
- \frac{ i}{4 \pi} \sum_{n=1}^{\infty}\int_{-\infty}^{\infty}
\frac{1}{\alpha} \frac{d \phi_n^{(\alpha)}(y)}{dy} \frac{e^{i(\alpha x +\omega_0 t)}I_{+}(\alpha) - e^{i(\alpha x -\omega_n t)} I_{n}^{OS}(\alpha)}{(\omega_n+ \omega_0) K_n } d \alpha\tag{2.12}
\label{eq12}
\end{align}
\end{fleqn}
It is not difficult to see that the total velocity $v(x,y,z,t)$, given by (\ref{eq4}) is independent of the lifting function $f(y).$ Since $u$ is derived from $v,$ it follows that it is independent of $f$ as well.
 The parts of $u$ corresponding to the second term in $v$ (\ref{eq4}) is obtained from continuity equation as  
\begin{fleqn}
\begin{align*}
\frac{\partial u_2}{\partial x}=-h(t)f^\prime(y) \delta(x)
\end{align*}
\end{fleqn}
Applying the the far-field zero conditions, $u_2=0$ for $x\lessgtr 0$.
\subsection{Features of analytic solution}\label{sec:section2p3}
$v_1$ (\ref{eq10}) consists of four integral contributions, the first and third of which are
standard integrals that can be evaluated by closing the LOI with a Jordan arc
in the UHP (resp. LHP) for $x > 0$ (resp. $x < 0$). 
For simplicity, we consider the Fourier inversion of only the
$j^{th}$ term of the infinite sums, without the multiplicative factor.
Assuming all the spatial modes of the forcing frequency $\omega_0$ to be 
distinct, and that $\omega_j(\alpha)=\omega_0$ at only one point $\alpha =
\alpha_j$ in the complex plane, for $x>0$, the first integral becomes
\begin{fleqn}
\begin{equation}
v_{11j} = 2 \pi i \frac{e^{i(\alpha_j x - \omega_0 t)}\phi_j^{(\alpha_j)}(y) I_{-}(\alpha_j)}{\frac{d\omega_j}{
d\alpha}|_{\alpha=\alpha_j} K_j}.\tag{2.13}	\label{eq13} 
\end{equation}
\end{fleqn}
Since $\omega_n^{OS}(\alpha) = - \bar{\omega}_n^{OS}(-\bar{\alpha}),$ we have,
the poles for the third integral located at $\alpha = - \bar{\alpha}_j $ i.e.
in the same half plane as the poles for the first integral. Hence, for
$x > 0,$ the third integral becomes
\begin{fleqn}
\begin{equation}
v_{13j} = -2 \pi i \frac{e^{-i(\bar{\alpha}_j x + \omega_0 t)}\phi_j^{(-\bar{
\alpha}_j)}(y) I_{-}(-\bar{\alpha}_j)}{\frac{d\omega_j}{
d\alpha}|_{\alpha=-\bar{\alpha}_j} K_j}.	\tag{2.14}\label{eq14}
\end{equation}
\end{fleqn}
Thus the poles $\alpha_j$ lying in the UHP ($Im(\alpha_j) >0$)
contribute to the downstream development of the disturbance. A similar analysis for $x < 0$ shows the poles in the LHP contribute to the disturbance development upstream of the source. \\

We now consider the second and fourth integrals in (\ref{eq10}). These integrals 
cannot be evaluated by closures  of the contour using Jordan arcs, since 
$\omega_j(\alpha)$ is not linear in $\alpha$. We now consider each of these in turn.
Several approaches for large time asymptotic analysis of similar Fourier 
integrals exist in the literature; we use the steepest descent method. This involves locating the saddle points of the phase function $p_j(\alpha) = i\left[ \frac{x}{t} \alpha - \omega_j^{OS}(\alpha)\right]$.  
 $\omega_j^{OS}$ is a complicated function of $\alpha$, possessing in general more than one saddle and a host of branch points whose number increases with increasing $j$, and moreover is only
numerically known for pPf; an evaluation of $v_{12j}$ and $v_{14j}$ {\`a} la $v_{11j}$ and $v_{13j}$ is thus impossible. We present a first approximation to these integrals here.\\ 

The asymptotic solution of an integral of Laplace type  as found in \eqref{eq10} or \eqref{eq12} is given in Appendix \ref{sec:appD} (\ref{final_sol}). It is assumed in the following asymptotic analysis of these integrals that the phase function has many saddle points and a pole such that the steepest descent path from only one of the saddles passes through the pole; addition of more poles will result in addition of similar terms in the expression for $u$ or $v$. In a later section, the global topography of the phase function will be consulted before applying the formula  \eqref{final_sol}  to \eqref{eq10} or \eqref{eq12}.\\

 We define a saddle path to be the path traced by a saddle point as $v_d=\frac{x}{t}$ varies. $v_d$ varies from $-\infty$ to $\infty$.
Approximating the second integral by the saddle point contribution 
 at each $v_d$, we get the large-time saddle contribution for 
$v_{12}(x,y,z,t)$ to be

\begin{fleqn}
\begin{eqnarray}
 v_{12sj} = - \sqrt{\frac{2 \pi}{t |d^2 \omega_j^{OS} / d \alpha^2 |_{\alpha = 
\alpha_s}}} \frac{\phi_j^{(\alpha_s)}(y) I_j^{OS}(\alpha_s) 
e^{-i \gamma/2}}{\omega_j^{OS}(
\alpha_s) - \omega_0} e^{p_j(\alpha_s)\ t}\tag{2.15}\label{eq15}
\end{eqnarray}
\end{fleqn}

\noindent where $\gamma = arg(d^2 w_j^{OS} / d \alpha^2 |_{\alpha = \alpha_s}).$
If the pole arising from $\omega_j^{OS}=\omega_0$ lies between the SDP and the real axis for some interval $X=(v_{d_{min}},v_{d_{max}})$,  its contribution, given by
\begin{fleqn}
\begin{eqnarray}
v_{12pj} &=& - 2 \pi i \frac{e^{i(\alpha_j x - \omega_0 t)}
\phi_j^{(\alpha_j)}(y) I_{-}(\alpha_j)}{\frac{d\omega_j}{
d\alpha}|_{\alpha=\alpha_j} K_j} \quad v_d \in X	\nonumber\\
&=& 0 \quad \hbox{otherwise}
\tag{2.16}\label{eq16} 
\end{eqnarray}
\end{fleqn}
\noindent has to be included as well, with $v_{12j}$ given by $v_{12j} = v_{12sj} + 
v_{12pj}+$ a smoothing term. The smoothing term, whose analytic form is presented in  Appendix \ref{sec:appD}, arises when the steepest descent path through the saddle point crosses the pole, at  $v_{d_p}= v_{d_{min}}$ and/or $v_{d_p}= v_{d_{max}}$. It smooths the discontinuity arising from the inclusion and exclusion of pole residues. This term can become large when the saddle point approaches the pole. An increase in the number of relevant saddles can lead to an increase in the number of such smoothing terms as the number of possible descent paths crossing the pole also increases.  Let us consider the least stable mode here.
For the subcritical case, $v_{d_{min}}$ is positive and $v_{d_{max}}$ is $\infty$; $v_{d_p}= v_{d_{min}}$. Similarly, for the supercritical case, $v_{d_{min}}$ is $- \infty$ and $v_{d_{max}}$ is finite and positive; $v_{d_p}= v_{d_{max}}$. \\

For a supercritical Reynolds number, $v_{12pj}$ grows for some $v_d\in X.$ There may exist a range $Y$ of positive $v_d$ over which the real part of the phase function $p_j(\alpha)$ can become positive. In this range, $v_{12sj}$ gives rise to a temporally growing wave packet which is convected away from the source. If there is overlap between $X$ and $Y,$ for any $v_d \in X \cap Y,$ both contributions add  and a distinct growing TS wave and wavepacket cannot be seen. If the forcing frequency lies outside the neutral stability curve, $v_{12pj}$ decays but $v_{12sj}$ still grows in the interval $Y$, showing a distinct wavepacket. It is possible that the temporal growth rate of $v_{12sj}$ is small, in which case the wavepacket will decay transiently as $1 / \sqrt{t}$, but will grow exponentially as $t \rightarrow \infty$; here too, the wavepacket from  $v_{12pj}$ will be prominently seen in the asymptotic limit. \\

So far, the individual parts (\ref{eq13})-(\ref{eq15}) of the solution \eqref{eq10} have been discussed. In the following subsection, various asymptotic limits of $v_1$  will be obtained by studying the behavior of their sum. The asymptotic limits of the smoothing terms have not been
 studied here since it has to lie between the saddle and pole contributions.
\subsection{Asymptotic limits}
We will focus on $v_{11j}$ + $v_{12j}$; similar conclusions can be drawn for $v_{13j}$ + $v_{14j}.$\\

\begin{enumerate}
\item {$t \rightarrow \infty.$ Most of the description will be for downstream locations i.e. $x > 0$. Occasionally, the upstream evolution will also be mentioned.}\\
\begin{enumerate}
\item $x$ fixed i.e. $v_d \rightarrow 0.$\\
	\begin{enumerate}
\item Subcritical case.
$v_{12pj}=0$ for  $v_d < v_{d_{min}}$ where the latter is positive. Since $v_{12sj}$ decays, and $v_{11j}\ne 0$ for all $v_d>0$, 
 the large time asymptotic state is that of a sinusoidal oscillation at the input frequency. 
\item Supercritical case. $v_{12pj}\ne 0$ as $v_d\rightarrow 0$ and $v_{11j}=0$ for $v_d>0$.  $v_{12sj}$ decays as $v_d \rightarrow 0$; otherwise the flow would be absolutely unstable. The asymptotic field is, therefore, a sinusoidal oscillation at the input frequency, which however grows spatially downstream. $v_{12pj}$ would cancel $v_{11j}$ for upstream locations ($v_d<0$)
 leaving a decaying wavepacket to propagate upstream.\\
	\end{enumerate}

\item $v_d$ fixed. \\

	\begin{enumerate}
\item Subcritical case.
For $v_d \notin X,$ $v_{11j}\ne 0$ while $v_{12j}=v_{12sj}$, which is  a decaying wavepacket, so that the asymptotic solution is a decaying TS wave at the signal frequency. For $v_d \in X,$ $v_{12pj}$ cancels $v_{11j}$ with the result that only $v_{12sj}$ remains and gives rise to a decaying wavepacket.
\item Supercritical case. For $v_d \notin X $ and $v_d \in Y$, $v_{11j}=0$ and $v_{12j}=v_{12sj}$ which is a growing wavepacket over $Y,$ which is missing in previous asymptotic solutions published in literature. For $v_d \in X\cap(0,\infty)$ and $v_d \notin Y,\ v_{12j}\ (= v_{12pj}+v_{12sj})$ grows spatially, but remains periodic in time. For $v_d \in X \cap Y,$ the wavepacket $v_{12sj}$ grows in time while $v_{12pj}$ is periodic in time.\\
	\end{enumerate}
\end{enumerate}
\item Fixed t.\\
	\begin{enumerate}
\item $x \rightarrow \infty.$ 
For the subcritical case, the pole contributions from $v_{11j}$ and $v_{12pj}$ cancel, leaving a decaying wavepacket due to $v_{12sj}$. For the supercritical case, $v_{11j}=0$ for $v_d>0$ whereas $v_{12pj}=0$  for $v_d > v_{d_{max}}.$ Since $v_{12sj}$ decays outside the finite range $Y$, $v_{12j}$ decays to zero. 
\item $x \rightarrow - \infty.$
In the subcritical case, the upstream state is again one of a decaying sinusoidal oscillation at the input frequency, whereas in the supercritical case, the asymptotic upstream state is that of a decaying wavepacket.\\
	\end{enumerate}
\end{enumerate}

Unlike the previous studies, the asymptotic solution presented here is spatially bounded at all times for all Reynolds numbers. Earlier similar stability studies  invoked Sommerfeld conditions for a periodic signaling problem  or Brigg's criterion for an impulsive disturbance; hence, the disturbances were always studied in the frame $v_d\rightarrow 0$.  \cite{ling1}, for the first time, obtained
the time-asymptotic solutions for impulsive disturbances in rotating disk boundary layer at a non-zero $v_d$; the critical points of the dispersion equation such as the saddle and branch points had to be used.  \cite{ling1}, however, considered only one saddle point that varies with  $v_d$, thus avoiding the need for computing the global topography of the eigenmodes.  The periodic signaling problem, on the other hand, gives rise to a pole in the integrand (as shown above) and hence there is a need for determining the global topography of the Orr-Sommerfeld modes in the application of the steepest descent method.
\section{Eigenmodal maps}\label{sec:maps}
The Orr-Sommerfeld dispersion equation for pPf has coalescing temporal modes \citep{jones}; the corresponding wavenumbers are branch points in the  $\alpha$ plane. Fixing the
 branch cuts defines the eigenmaps.     Here, 
 in order to clearly separate out the temporal eigenmodes,  the eigenvalues, $\omega_j$ are first sorted into modes at the origin $\alpha = 0,$ based on their imaginary parts. From these, the neighboring values of each temporal mode along the real and imaginary axes are sorted by using discretised Cauchy-Riemann (C-R) equations; the details of the sorting procedure for OS temporal eigenvalues are given in Appendix \ref{sec:appA}.    This works in general because each $\omega_j(\alpha)$ is analytic except at the branch points. Marching in the $\alpha$ plane is done parallel to the imaginary axis, both into the UHP and the LHP, from the points on the real axis.  Even though the C-R equations are not satisfied at a branch point, the algorithm can still educe the $\omega$ branches at that point.  The vertical marching results in vertical branch cuts away from the real axis in both half planes of $\alpha$. This vertical mode tracing works as long as there is no double root for any real $\alpha$ in the wavenumber domain considered. We present here the first two dominant modes of the OS-even, OS-odd families for $Re = 5000$ in figure 1.\\
\begin{figure}
\centering
\subfloat[]{
\includegraphics[width=2.5in]{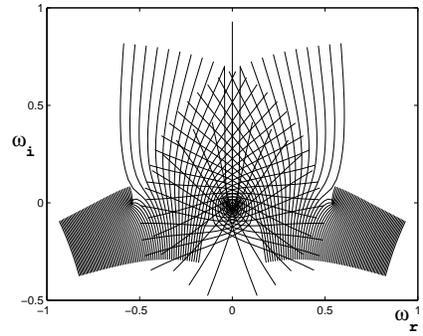}}
\hspace{0.1in}
\subfloat[]{
\includegraphics[width=2.5in]{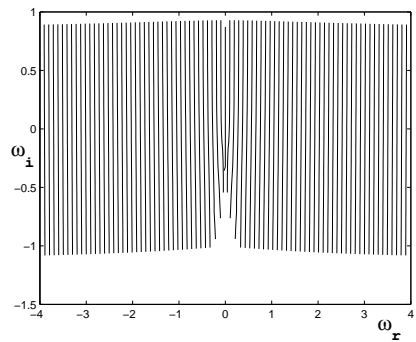}}
\hspace{0.1in}
\subfloat[]{
\includegraphics[width=2.5in]{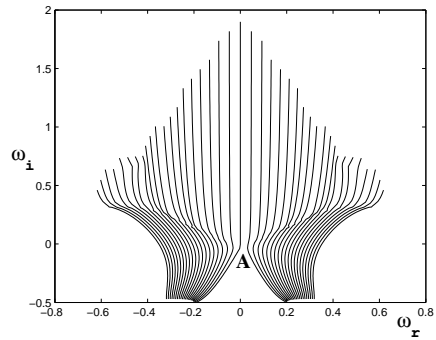}}
\hspace{0.1in}
\subfloat[]{
\includegraphics[width=2.5in]{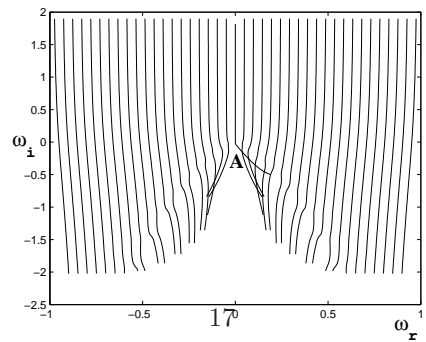}}
\caption{Temporal maps of the region $\textit{R} = [-4,4] \times [-1,1]$ in the $\alpha$ plane. Dominant modes of a) OS-even Mode 1 b) OS-even Mode 2 c) OS-odd Mode 1 d) OS-odd Mode 2. } 
\end{figure}

The understanding of the maps is enhanced by a knowledge of their branch points, saddles and poles. The maps are analytic at almost all points, the only exceptional points being those at which branching occurs. In a given window, these are seen to be finite in number. The double roots appear as half-saddles in the $\omega$ plane (figures 1 a-d) and as branch points (BPs) in the $\alpha$ plane. A saddle in the $\alpha$ plane appears as a cusp in the $\omega$ plane (figure 1a). Alternatively, one can plot level contours of $\omega_i$ (or $\omega_r$)in the $\alpha$ plane; the BPs appear with associated branch cuts (BCs). In order to save space, we present details of these points only for the dominant OS even mode.
\begin{figure}
\centering
\subfloat[]{
\includegraphics[width=2.5in]{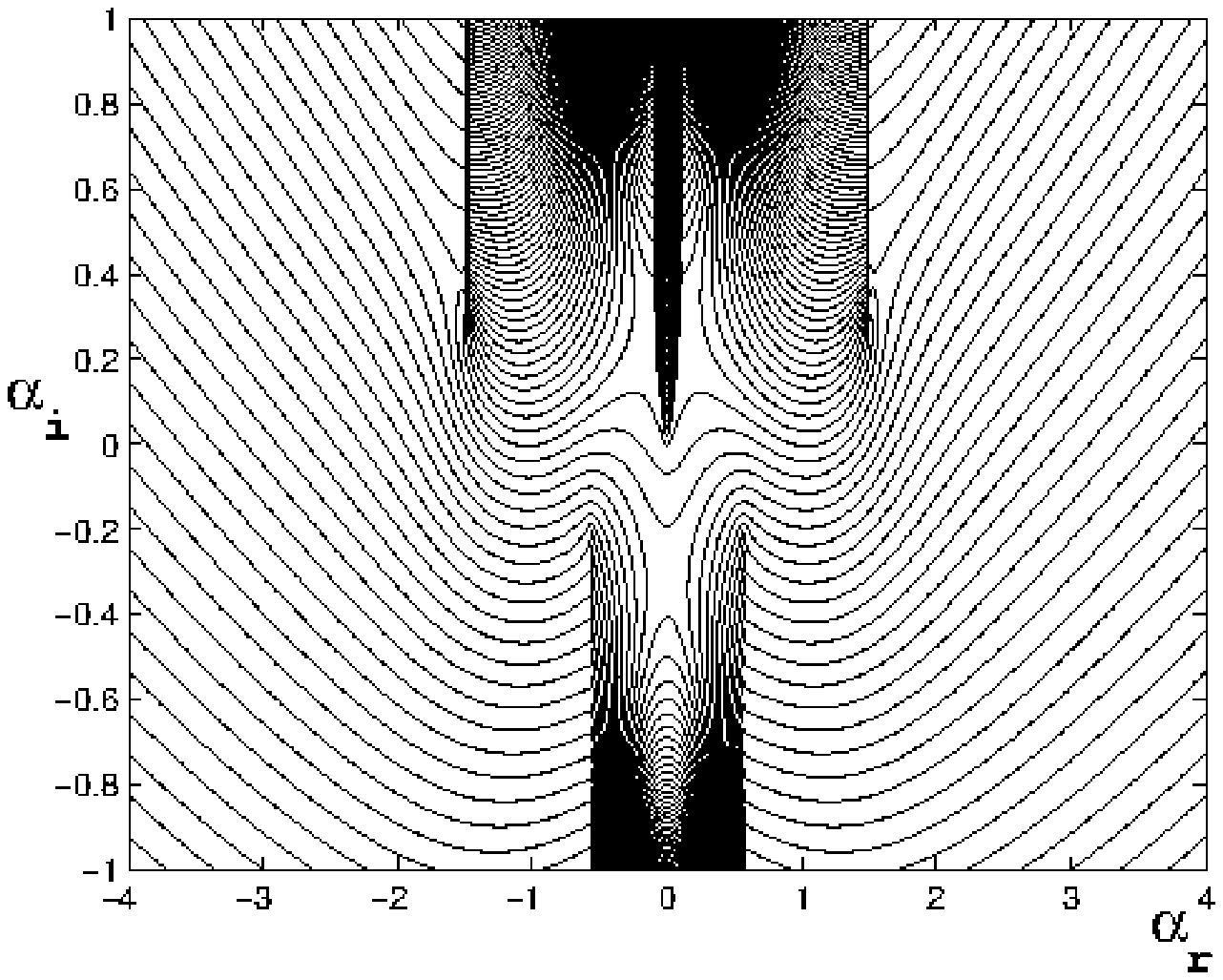}}
\hspace{0.1in}
\subfloat[]{
\includegraphics[width=2.5in]{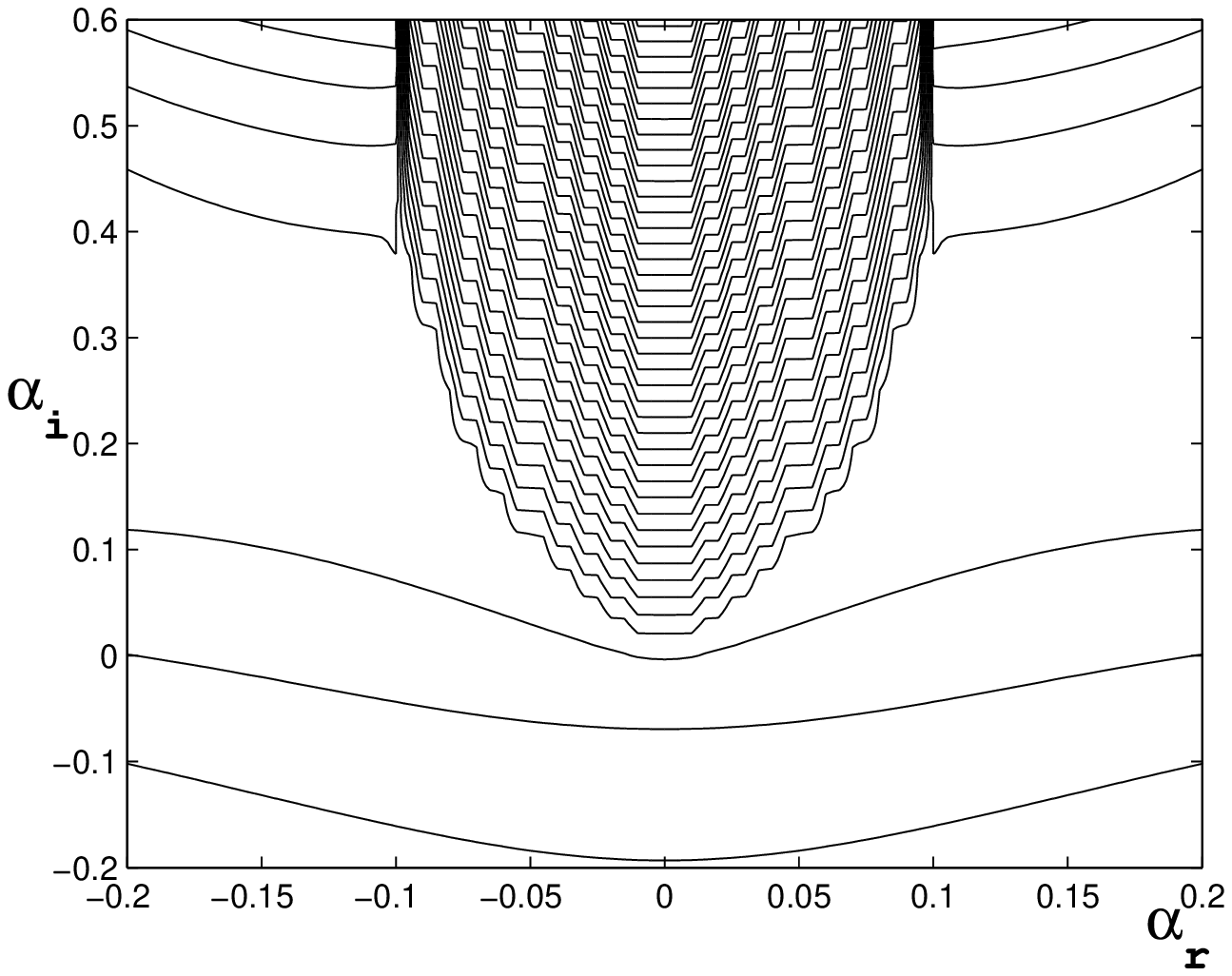} }
\caption{Contour plot of $\omega_i$ of mode 1 in the $\alpha$ plane. The branch points and the associated branch cuts in the rectangular domain can be seen. The central pillar in the UHP is actually a cluster of BCs; a close-up view is shown in (b).}
\end{figure}
Figure 2 shows the level contours of $\omega_i$; the bunching of these contours is indicative of BCs (in this case, vertical), emanating from the BPs. 
Plots can only give a rough indication of the BPs. A procedure to accurately locate them is given in Appendix \ref{sec:appB}.\\

Saddles
of the quantities $ p(\alpha) = \alpha v_d - \omega_n$ are the other entities that are important for computing disturbance wave integrals. The number and type of saddles depends on the mode number and $v_d$; in particular, saddles can appear and disappear from a modal map as $ v_d$ varies. We track the
saddles of the first dominant OS mode for $0 < v_d < 1;$ the saddle paths are found by solving $\partial \omega_i / \partial \alpha_r = 0$ numerically; the value of $\partial \omega_r / \partial \alpha_r$ at that point gives the $v_d$ value
for the corresponding saddle. In general, there could be an infinite number of saddles in each mode; however only a few of them make significant contributions to the integral. We present the saddle paths in figure 3. The modal topography is for a fixed value of $v_d = 0.5$; the saddle paths for varying $v_d$ are overlaid on this background. We track saddles only in the RHP, including the imaginary axis; by symmetry the corresponding LHP saddles can be inferred. Mode 1 has three saddles (colored lines in figure 3a) and they move upward with increasing $v_d$. Since an eventual goal would be to describe the disturbance evolution in terms of the dynamics of the critical points (BPs, saddles and poles), it is important to study the quantities $q = Im(\omega_{n*} - \alpha_* v_d)$, the real part of the phase function, that govern the growth rate of the
disturbance, the asterisk referring to the location of the critical point. Hence we parallelly plot this value at the
relevant critical points like the saddles and branch points as a function of $v_d$ in figure 3(b). This will give some indication of which critical points and modes contribute where, as a function of $v_d$.
\begin{figure}
\centering
\subfloat[]{
\includegraphics[width=2.5in]{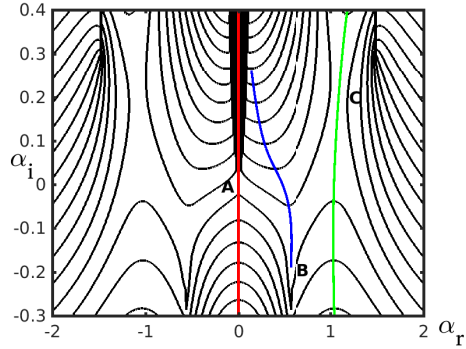}}
\hspace{0.1in}
\subfloat[]{
\includegraphics[width=2.5in]{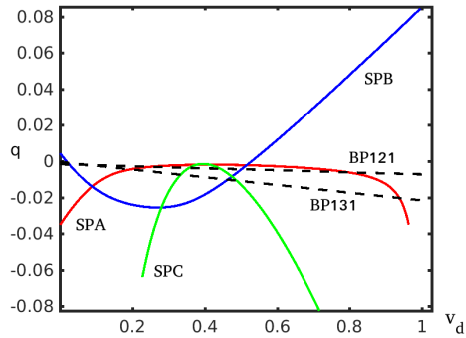}}
\caption{ a) Movement of three saddle points, A, B and C of the dominant even OS mode in the  $\alpha$ plane with varying $v_d.$ (b) Heights at saddle point A, B and C and at the branch points BP121 and BP131 (Appendix \ref{sec:appB}) as a function of $v_d.$}
\end{figure}
A third important entity is the pole $\alpha_p$ of the wave integral, given by $\omega_n(\alpha_p) - \omega_0 = 0$ where $\omega_0$ is the frequency of the disturbance source. A related quantity which is relevant to the residue calculation is $d \omega_n / d \alpha |_{\alpha_p}.$ These are straightforward to calculate by solving the OS equation.\\

 Higher modes are not shown here. However they possess interesting features   
worth mentioning here. The movement of a SP along the imaginary axis, as $v_d$ varies, is a common
feature of all modes. These SPs have zero phase (imaginary part) and decay slowly in the
neighborhood of $\alpha = 0$ and hence, can produce streamwise elongated structures. Collision of saddle points as $v_d$  varies also occurs frequently in the higher modes.  In the second mode, the central saddle increases in height till it collides with another central saddle point. Third mode  has two off-axis saddle points which collide at $v_d=0.4974$ forming a monkey saddle point.
 It will be interesting to study the disturbance velocity patterns corresponding to these saddle points; however, the associated decay rates are much higher than that of the primary mode and hence are not considered in the present study.
\subsection{Evaluation of Fourier integrals}
The results in N75 indicate, and an evaluation of the integrals in \S \ref{sec:solution} confirm, that the dominant mode is the two-dimensional first OS even mode. Hence we focus here only on the contributions of this mode to the velocity field. \\

We evaluate analytically the integral $v_{121}$, defined in \S \ref{sec:section2p3}. From figure 3 of \S \ref{sec:maps}, it is seen that one or more of the three saddles could contribute to the integral; we now determine which ones are relevant. The positive real axis ends in a valley of the right-most saddle point (C) and the valleys of the on-axis saddle point (A) connect the left and right half planes. Hence, both the saddle points must contribute to the real axis integral.\\

Before proceeding to the application of formulae in Appendix \ref{sec:appD}, we show (i) why  contribution from the middle saddle B is negligible and (ii) how only the SDP of the off-axis saddle point passes through the pole. Figure 4 shows the steepest descent paths of the saddle points corresponding 
to various $v_d$ values shown in the figure; the SDP of the on-axis saddle points are shown in 4(a) and those of the (third) off-axis saddle points are shown in 4(b). The least stable spatial mode corresponding to $\omega_0=0.28$ is located at $\alpha=(1.0302,0.0037)$; it is a pole in the alpha plane with leading contribution to the Fourier integrals. The SDP from the on-axis saddle point A never passes through this pole. It is interesting to note the Stokes phenomenon, when the SDP at $v_d=0.1586$ passes through saddle point B; however, it is not of any consequence as will be shown here. The SDP of the off-axis saddle point C crosses the pole at $v_d\approx0.4$.\\

\begin{figure}
\centering
\subfloat[]{
\includegraphics[width=3.2in]{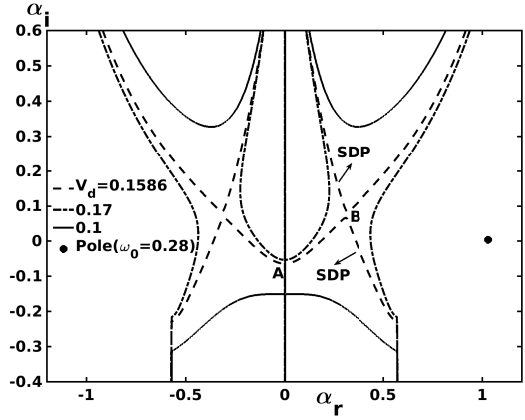}}
\hspace{0.1in}
\subfloat[]{
\includegraphics[width=3.2in]{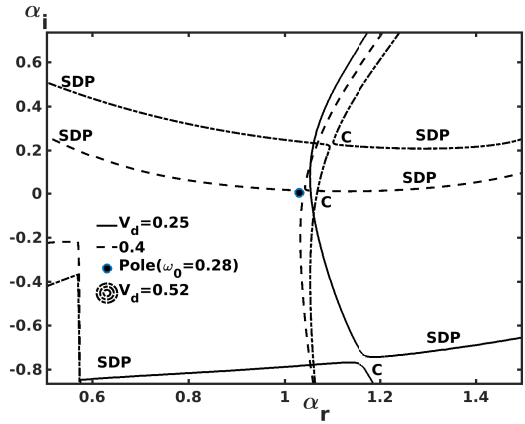}}
\caption{ SDP-Pole crossing; $\omega_0=0.28$; $\alpha_p=(1.032,0.0037)$  a) SDP of on-axis saddle point at $v_d=0.1, \ 0.1586, \ \hbox{and} \ 0.17$.  (b) SDP of off-axis saddle point $v_d=0.25, \ 0.4\ \hbox{and} \ 0.52$. The SDP for $v_d=0.4$ crosses the pole.}
\end{figure}

  In order to obtain asymptotic limits of the  $\alpha$-integrals for $v_1$ and $u_1$, we construct an integration path called Olver path, into which the real line is deformed;   by definition, an Olver path is the union of descent paths from the saddle points. The topography of the OS first mode shows a three-saddle cluster; such a cluster will always have four hills and four valleys such that all the three saddle points share a hill and a valley.  Hence,  a set of descent paths  from all the three saddles 
can only be linked through the common valley and hence only one valley of the middle saddle point will be used in the Olver path; whereas, both the valleys of the saddles A and C are involved.  Therefore, the middle saddle point B becomes an ordinary point of the Olver path and hence its contribution amounts to just the error term \citep{ough}. This SP transforms from an open point to an inadmissible SP, when $Re [p(\alpha) ]$ at this point is greater than the corresponding maximum value on the LOI;
if $B$ were admissible, we would obtain a growing mode which is convected downstream, an impossibility in subcritical pPf.\\

 In summary, the Fourier integral has two contributing saddle points and a pole which lies on the SDP of the off-axis saddle point at some $v_d$; hence the formula presented in Appendix \ref{sec:appD} is readily applicable. The validation of the formula presented in Appendix \ref{sec:appD} is based on the first temporal eigenmode described above.  The spatio-temporal solutions of the IBVP (2.10 and 2.11) using the formula are presented in the next section. \\
\section{Results I: Linear disturbance evolution}\label{sec:results1a}
The streamwise disturbance velocity component $u_1$ is computed using Olver method described in Appendix \ref{sec:appD}. It is important to note that the factor $\alpha$ in the denominator of (\ref{eq12}) cancels with the same in the numerator and hence is not a pole. This can be verified using the  expressions for $I_{\pm}(\alpha)$ and $I_n^{OS}$ given in Appendix \ref{sec:appC}.\\

The asymptotic solutions of $v_1$ (and $u_1$) in the previous section show a wavepacket arising from the saddle path along with the TS wave and sometimes distinct from it. In this section, the computed asymptotic spatio-temporal solutions for moderate times will be discussed.  The major part of the present study is for $Re=5000$. \\

   \begin{figure}
\centering
\includegraphics[width=3.9in]{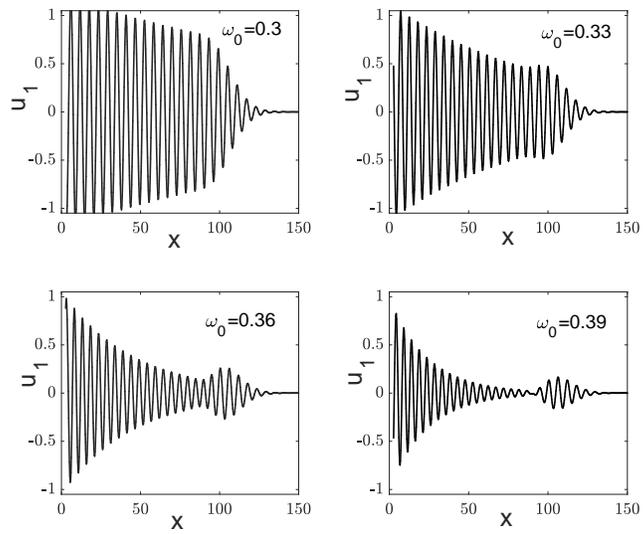}
\caption{IBVP solution at $t=268$ (large time) and $Re=5000 $ for four different 
forcing frequencies. (a) $\omega_0=0.3$: TS wave is dominant. (b) 
$\omega_0=0.33$: TS wave and wavepackets are of similar magnitudes and decay 
rates. (c) $\omega_0=0.36$: Clear wavepacket is formed. TS wave and wavepacket 
are of similar magnitude (d) $\omega_0=0.39$ TS wave decays rapidly in the 
neighborhood of the ribbon; wavepacket travels along the entire channel while 
decaying slowly.     }  
\end{figure}
The IBVP solutions for various values of $\omega_0$ are presented in figure 5. The TS wave and the
 wavepacket are indistinguishable for $\omega_0=0.3$, figure 5(a); in this case the disturbance state is a slowly decaying TS wave. At  higher frequencies, 
$0.33<\omega_0<0.39$, the decay rate of TS wave increases more steeply; the wavepacket can be clearly seen 
as in figure 5 (b and c). However, a large part of the wavepacket is still attached to the TS wave and
their amplitudes are comparable over a considerable length of the channel. This state is called a mixed state.  At 
still higher frequencies,
 $\omega_0\ge0.39$, as in figure 5(d), the TS wave decay rate is very high and a clear wavepacket is seen. Hence, in this case there are two distinct states of comparable magnitudes emerging from the wall disturbance, a TS wave and a wavepacket. \\
 
 The two states have very different decay rates and velocities of propagation.
 The TS wave decays (or grows) temporally in a reference frame which moves with its phase velocity $c_r$;    the phase velocity and  decay rate of a TS wave can be deduced from the OS dispersion equation. The wavepacket undergoes a slow spatio-temporal elongation. It is almost stationary in a reference frame moving with its group velocity $c_g$; the decay rate $\gamma_{wp}$ of the wavepacket in this reference frame varies slowly with time, unlike the TS wave. Neglecting these small spatio-temporal changes of the wavepacket, the group velocity and growth rate are calculated numerically from two  instantaneous solutions. 
  \subsection{Wavepacket characteristics}
The speed of the reference frame in which the wavepacket is stationary is computed as follows. It is determined  visually by noting the distance traveled by the centre of the wavepacket from the origin of a frame moving with a given velocity. When the centre of the wavepacket is almost stationary in a moving frame, $c$ is chosen as the velocity of that frame.  It is also an estimate of the group velocity of the wavepacket; this is a real quantity unlike the complex group velocity of the TS wave.\\

 This procedure is demonstrated in the two movies presented here for $\omega_0=0.45$; at this ribbon frequency, there is a distinct wavepacket in the test section. The movement of the wavepacket with respect to a frame moving with $c = 0.395$ is shown in Movie 1. For comparison, the movement of the wavepacket in a frame moving with $c = 0.35$ is shown in Movie 2. In these movies, the moving frame is represented by a box. The wavepacket stays in the box in the first movie while it moves slowly out of the box in the second one. \\
 
  \begin{figure}
\subfloat[Temporal decay of the wavepacket,indicated by the enveloping line]{
\includegraphics[width=5.in, height=2.in]{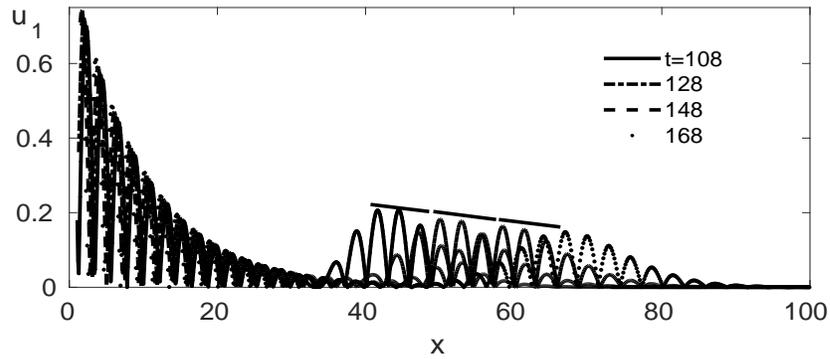}}
\newline
\subfloat[IBVP solutions multiplied by a factor of $e^{0.007t}$.]{
\includegraphics[width=5.in, height=2.in]{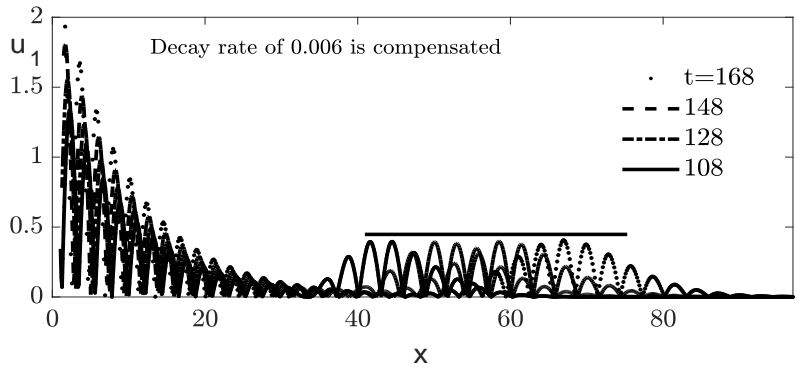}}
\caption{IBVP solutions for streamwise disturbance velocity at different instants for $Re=5000$; $\omega_0=0.45$.}
\end{figure}
   The temporal decay of the wavepacket is estimated by visually inspecting the temporal constancy of its amplitude when the IBVP solution is multiplied by a suitable exponential time factor. Figure 6(a) shows the IBVP solution at four different instants; the wavepacket decay is shown by the sloped envelope of the wavepacket. Figure 6(b) shows the IBVP solution at the same instants as in the previous figure multiplied by a factor of $e^{0.007 t}.$ Here, the envelope of the wavepacket is a horizontal line with the inference that the wavepacket decays at the rate of 0.007. The wavepacket arises from a saddle path that is traced as $v_d$ varies. Hence the temporal decay (or growth) of the wavepacket  is not purely exponential but also contains an algebraic factor $1 / \sqrt{t}.$ Here, the exponential decay rate of 0.007 roughly accounts for this algebraic decay too.\\

 Table 1 describes the characteristics of the TS wave and the wavepacket for 
different Reynolds number-frequency combinations. The third and fourth columns 
are the phase velocity and the spatial decay (or growth) rate of the TS wave. 
The spatial decay rate of the wavepacket, $\gamma_{wp}$ is given in the fifth 
column. It is defined as the spatial decay rate of the wavepacket peak.  The group 
velocity of the wavepacket, $c_g$ is given in the sixth column and the temporal decay rate of a wavepacket is the product of $c_g$ 
and $\gamma_{wp}$. The wavepacket elongates spatially (but slowly)  while 
moving downstream; the spatio-temporal variation of its group velocity and the decay rate  
is negligible. The values of $c_g$ and $\gamma_{wp}$ shown here are
hence computed from two instantaneous 
solutions.  The decay rates of the TS wave and the wavepacket are almost equal
for $0.3<\omega_0<0.33$ at $Re=5000$; the respective wavepacket group velocity is only 
slightly more than the TS phase velocity (Table 1). At  higher frequencies, 
$0.33<\omega_0<0.39$, the decay rate of TS wave increases more steeply compared 
to that of the wavepacket. At 
still higher frequencies,
 $0.39<\omega_0<0.45$, the TS wave decay rate is more than double the decay rate 
of the wavepacket; we choose this condition for identifying a clear wavepacket 
state as in figure 5(d). All the parameter combinations for $Re=6000$ shown in this 
table satisfy this condition. For $Re=4000$, the two decay rates are nearly same 
at $\omega_0=0.36$.  
\begin{table}
\begin{center}
 \begin{tabular}{|c|c|c|c|c|c|c|} \hline
$Re$&$\omega_0$ & $c_{r}$ & $-Im(\alpha_p)$  &$\gamma_{wp}$  & $c_g$& State\\ \hline
5000&0.3 &  0.2776 &-0.0046  & -0.0046    & - &-     \\
5000&0.32  & 0.281  &-0.00791  & -0.0074  & 0.34& TS         \\ 
5000&0.34  & 0.288 &-0.0139 & -0.0088& 0.37& Mixed    \\
5000&0.35 & 0.29  &-0.0179  & -0.0101& 0.375 &Mixed   \\
5000&0.39  &0.2996 &-0.0416 & -0.0138& 0.39  &TS, WP     \\
5000&0.45 & 0.312  &-0.1044 & -0.0176 & 0.395 & TS, WP  \\ \hline
6000&0.34&0.2815&-0.019&-0.01&0.38&TS, WP\\
6000&0.36&0.286&-0.032&-0.01&0.39&TS, WP\\
6000&0.39&0.2927&-0.0595&-0.0109&0.405&TS, WP\\ \hline
4000&0.36&0.3&-0.01782&-0.018&0.4&Mixed\\
4000&0.4&0.31&-0.0345&-0.016&0.4&TS, WP\\ 
4000&0.425&0.3159&-0.05&-0.0155&0.4&TS, WP\\ \hline
\end{tabular}
\caption{TS wave and wavepacket parameters for $Re=5000$} over a range of 
forcing frequencies $\omega_0$.
\end{center}
\end{table}
 \subsection{Comparison with N75 experiments - Linear stage}
     We start off with a brief description of the experiment in N75. A pPf was 
established in a long, quiet (turbulence level $< 0.01 \%$) channel, with a 
demonstration of the parabolic profile to a large degree in N75F3. A sinusoidal disturbance was introduced in this flow through a phosphor bronze ribbon, stretched close to the lower wall and vibrating, at a frequency 
$\omega_0$, in a direction normal to it. The 
test section, where the measurements were made, ranged from $\sim 44$ to $\sim 
78$ units downstream of the ribbon. \\

 For small disturbance amplitudes, less than 1\%, it was established that the disturbance appears in the flow as sinusoidal in time (N75F4) and antisymmetric in the wall normal direction $y$ (N75F5). Similar measurements at various streamwise locations confirmed that the disturbance was indeed a traveling wave, 
whose wavelength $\lambda$ was estimated. N75F7 shows the spatial evolution of the maximum disturbance value with $x$ for 
a variety of $Re$ and $\omega_0.$ N75F6 and N75F9 present the disturbance 
wavelengths as a function of $Re$ and $\omega_0$; we will consider only N75F6.  
N75F10 shows the amplification rate $\alpha_i = - h \,  d \, ln \, u_m^{'} \,/ 
\,dx$ vs. angular frequency. N75F11 shows the experimental stability boundary 
which is a little different from theory. \\

The digitized data from N75F6, for $Re = 3000, 4000, 5000$ are presented in 
Table 2, where we have also shown the $\alpha$ of the most dominant spatial 
mode, obtained by solving the spatial eigenvalue problem. It can be noted that 
the $\alpha_e$ are higher, in general, than the real part of $\alpha_t$ with a maximum 
discrepancy of up to $5 \%$. \\

\begin{table}
\begin{center}
\begin{tabular}{|c|c|c|c|c|c|} \hline
Re & f(Hz) & $\omega_0$ & $\lambda$(cm) & $\alpha_e$ & $\alpha_t$  \\ \hline
&  33 & 0.2597 & 4.919 & 0.9325 & 0.9275 + i 0.031  \\
&  39 & 0.3069 & 4.208 & 1.09 & 1.0348 + i 0.021 \\
3000 & 43 & 0.3384 & 3.898 & 1.1767 & 1.1056 + i 0.01915 \\
& 47 & 0.3699 & 3.697 & 1.2407 & 1.1757 + i 0.0214 \\ \hline
& 32.82 & 0.194 & 5.814 & 0.789 & 0.7935 + i 0.0389 \\
& 38.86 & 0.229 & 4.993 & 0.919 & 0.8802 + i 0.0243 \\
4000 & 50.34 & 0.297 & 4.117 & 1.114 & 1.0451 + i 0.0103 \\
& 60.42 & 0.357 & 3.661 & 1.253 & 1.1876 + i 0.017 \\
& 72 & 0.425 & 3.241 & 1.4152 & 1.3482 + i 0.052 \\ \hline
& 38.64 & 0.18 & 5.978 & 0.767 & 0.7732 + i 0.0325 \\
& 50.41 & 0.238 & 4.628 & 0.99 & 0.9234 + i 0.0097 \\  
5000 & 60.49 & 0.286 & 4.099 & 1.12 & 1.0454 + i 0.00375 \\ 
& 72.03 & 0.34 & 3.734 & 1.228 & 1.179 + i 0.014 \\ \hline 
\end{tabular}
\caption{Data from N75F6. Theoretical values are in column 6.}
\end{center}
\end{table}
N75F7 and N75F10 pertain to the damping rate of the disturbance; the former 
plots the disturbance maximum as a function of downstream distance whereas 
N75F10 synthesises this information into a single number at each $\omega_0$ and 
$Re.$
The data from N75F7(a), for an $f = 72 Hz$, are shown in Table 3, where we have 
also included the value obtained from a linear stability calculation. For example, for $Re = 5300, 72 Hz$ 
corresponds to $\omega_0 = 0.321$ which in turns produces a dominant spatial 
eigenvalue $\alpha = 1.141 + i 0.00831,$ the imaginary part of which is used in 
producing the respective values in the last column in Table 3.\\

\begin{table}
\begin{center}
\begin{tabular}{|c|c|c|c|} \hline
Re & $x - x_0 $ & $(u_m^{'} / u_{m,0}^{'})_e$  & $(u_m^{'} / u_{m,0}^{'})_t$    
\\ \hline
&  6 & 0.751 & 0.732  \\
&  14 & 0.538 & 0.483 \\
4000 & 20 & 0.391 & 0.354 \\
& 27 & 0.269 & 0.246 \\
& 34 & 0.194 & 0.171 \\ \hline
& 6 & 0.962 & 0.951\\
& 14 & 0.988 & 0.890\\
5300 & 20 & 0.885 & 0.847\\
& 27 & 0.732 & 0.799\\
& 34 & 0.641 & 0.754\\ \hline
& 6 & 1.086 & 1.015\\
& 14 & 1.155 & 1.034\\  
6400 & 20 & 1.101 & 1.049 \\ 
& 27 & 0.973 & 1.067\\ 
& 34 & 0.857 & 1.085 \\ \hline 
\end{tabular}
\caption{Data from N75F7. Theoretical values, from linear stability 
calculations, are in column 4.}
\end{center}
\end{table}
Several things can be noted from the table. For the lowest $Re = 4000,$ the 
amplitude decreases with increasing distance more or less in accordance with 
linear theory.
For $Re = 5300,$ there is a slight initial increase in experimental amplitude 
and later, a precipitous decline, which trends, the linear theory is unable to 
capture, producing as it does a constantly decreasing amplitude, given that the 
$Re$ is subcritical.  Further, the amplitude curve presented in this figure has 
a wavy pattern.
There are problems for the supercritical $Re = 6400$ as well; the experimental 
amplitude initially grows faster than what linear theory predicts and 
astonishingly, decreases after a certain distance, which the linear theory can 
never predict.\\

We turn to the solution of the IBVP for a clue as to what might be producing these 
behaviors. For $\omega_0 = 0.425$ and $Re = 4000,$ the wavepacket  has higher amplitude
in the test section due to high decay rate of the TS wave (Table 1) and it arrives 
much earlier than the 2D TS wave. Figure 7(a) shows the overlap of 
envelopes of several instantaneous solutions for this case in the time interval 
$[38,\ 208]$. The envelope of the wavepacket, shown in red, is above that of the 
TS wave (black), reflecting that the wavepacket decays at a slower rate.  
However, the measured decay rate matches closely that of the TS wave, as noted 
above. The seeming incongruity in the experimental observation of the faster 
decaying TS wave can be resolved if it is noted (N75) that `some distance from 
the ribbon was required for the disturbances to establish a structure which did 
not change downstream.' Thus, in this case, the experimenter can wait for the 
wavepacket to pass beyond the test section and for the unchanging TS wave to be 
established.
As the ribbon frequency approaches the neutral stability curve (or is in the 
unstable region) the wavepacket is indistinguishable from the TS wave due to its 
low group velocity as well as a comparatively higher decay rate than the TS 
wave. Hence, for such cases,  waiting does not amount to any difference in the 
envelope.\\

Figure 7(b) shows the overlap of the instantaneous IBVP solutions for the time 
interval $[198, \ 208]$.  The wavepacket has moved out of the test section 
before $t=198$, showing only the TS envelope. The black symbols are the suitably 
scaled experimental values obtained from N75F7a corresponding to $Re=4000$. The 
instantaneous solution for $Re=6000$ with a ribbon frequency of 72Hz is shown in 
figure 7(c) in the time interval $[38,\ 208]$. In this case, the wavepacket, due to its low group velocity, is not 
distinguishable from the growing TS wave. The envelope 
of the instantaneous solutions is shown as dashed line. Clearly, the envelope 
lies between the measured values for $Re=6400$ shown in triangles and for 
$Re=5300$ (squares); the slight nonlinearity and the 
apparent disturbance decay  farther from the ribbon is also captured well. Thus, 
the spatial decay in a growing mode ($Re=6400$) is probably more due to  
the choice of the time interval than a manifestation of any inherent flow 
physics.
\begin{figure}
\centering
\includegraphics[width=3.5in]{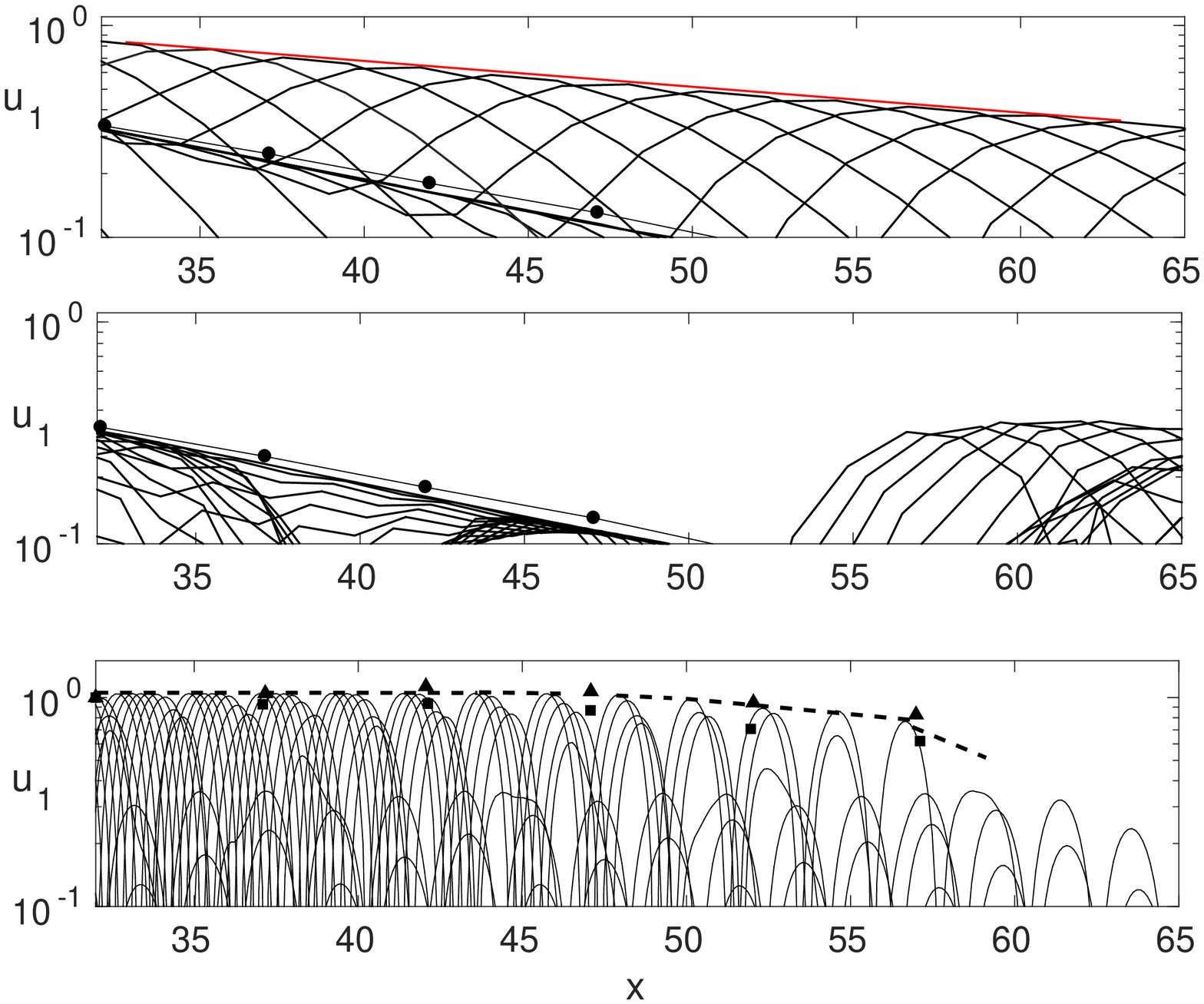}
\caption{Overlap of instantaneous IBVP solution envelopes  for $Re=4000 $ and 
$\omega_0=0.425$: The horizontal axis denotes the distance from the ribbon in 
centimeters. (a)   $38\le\ t \ \le 208$,  (b) $198\le \ t \ \le 208$; Filled 
circles denoted scaled measurements from N75F7a. \\
 (c) Instantaneous IBVP solutions for $Re=6000$ and  $\omega_0=0.28$; Dashed 
line denote the solution envelope. Triangles: Measurements for $Re=6400$; 
Squares: Measurements for $Re=5300$ from N75F7a.   }  
\end{figure} 
\subsection{Comparison with N75F15}
N75F15 records the downstream evolution of the disturbance maximum $u_m^{'}$, 
being over the channel height. The first recording station is 32 cm ($\approx 
44$ units) downstream from the ribbon and the last one, about 57cm ($\approx 78$ 
units). Experimental points, corresponding to six different initial intensities, 
are plotted. Six curves are drawn, one through each set of points. Curves (i) - 
(iii), for initial intensities $< 1 \%$, seem to show that $u_m^{'}$ increases 
slightly downstream of the initial station before decreasing continuously. Curve 
(iv) shows, after the initial rise, a constant disturbance for a considerable 
downstream distance, before again rising steeply.  We will be concerned in this 
study only with curves (i)-(iii) and the earlier part of (iv); (v) and (vi) 
depict evolution where the higher initial intensities means nonlinearity plays a 
role and is beyond the scope of the linear analysis of this paper.\\

\begin{figure}
\centering
\includegraphics[width=2.75in, height=2.25in]{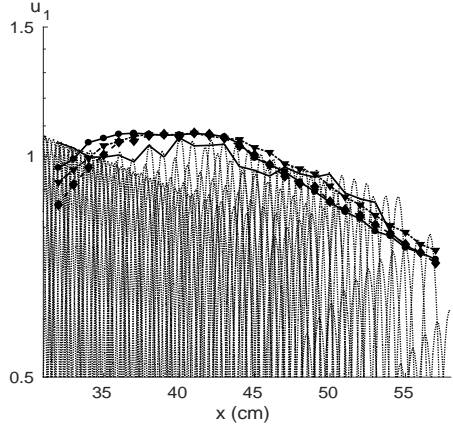}
\caption{ Instantaneous IBVP solutions in the test section for Re=5000 and 
$\omega_0=0.34$ over two periods of the vibrating ribbon (dotted lines). The 
three symbols correspond to three lower curves of N75F15 normalized to 1.1 at 
the peak. Solid line represents the computed maximum amplitudes at the 
experimental points.   }  
\end{figure}
We will attempt to explain the `apparent' spatial growth 
and subsequent decay of the disturbances (curves i - iii, N75F15)by examining 
the envelope of the instantaneous solutions.
N75's sampling rate seems to be 10 per time period (for example, N75F4) and we 
have used the same sampling to produce the envelope in figure 8. This 
figure shows the instantaneous IBVP solutions for $Re=5000$ and  $\omega_0=0.34$ 
approximately over two ribbon periods ($158\le \ t \ \le 198 $) in steps of two 
non-dimensional time units, the approximate sampling rate. These times 
correspond roughly to the residence time of the wavepacket in the test section. 
The solid line is the envelope formed by marking the maximum amplitude at the 
experimental points over all these time steps in the interval. The discrete 
nature of the spatial locations and the time steps results in the irregular 
shape of the envelope.  Also shown in the figure are the lower three curves of 
N75F15, normalized to a peak value of 1.1 in order to match with the computed 
peak. The matching between the experimental curves and the theoretical curve  is 
very good; in particular, the initial rise and subsequent decay are demonstrated. \\

\begin{figure}
\centering
\subfloat[]{
\includegraphics[width=2.5in]{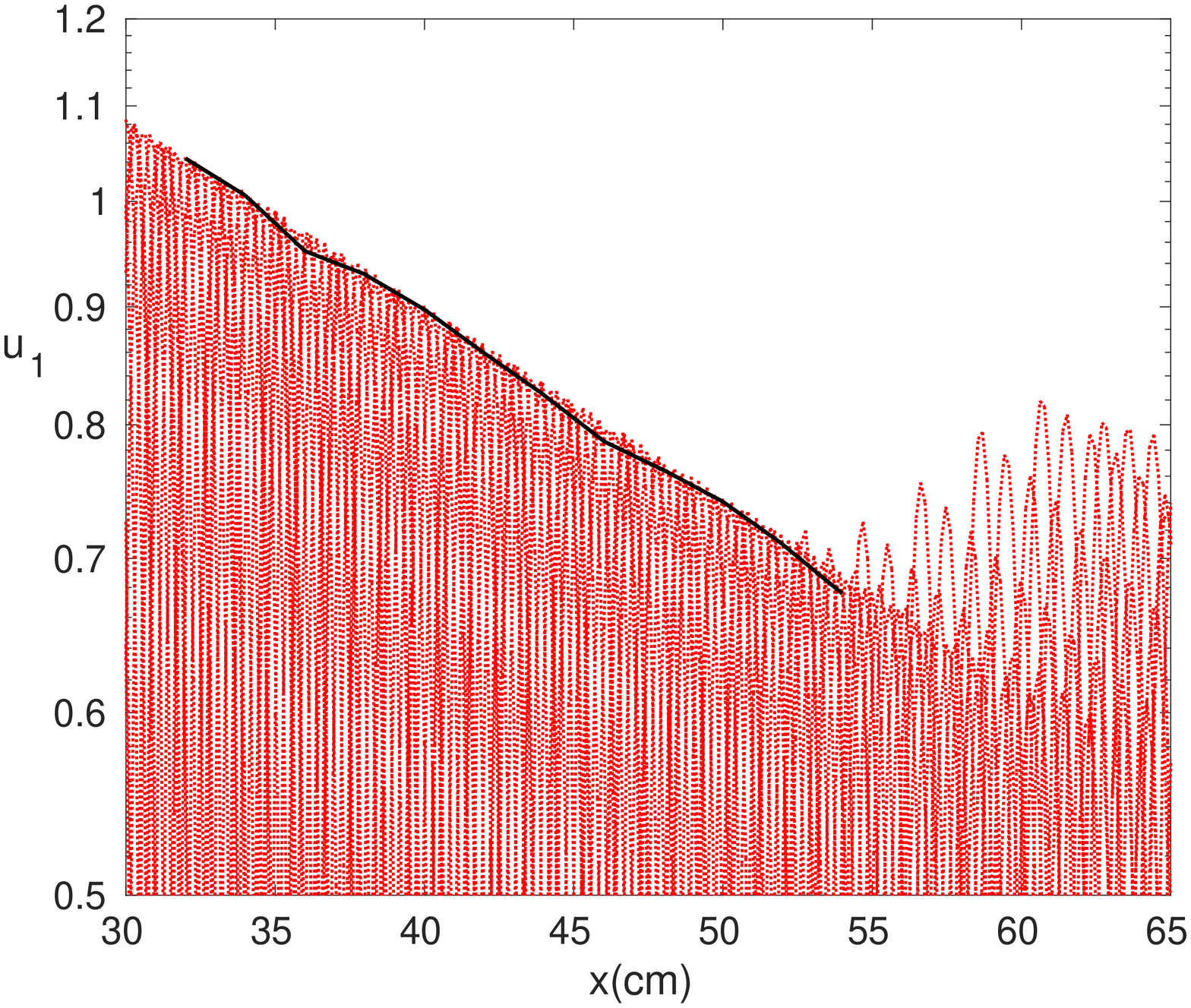}}
\subfloat[]{
\includegraphics[width=2.5in]{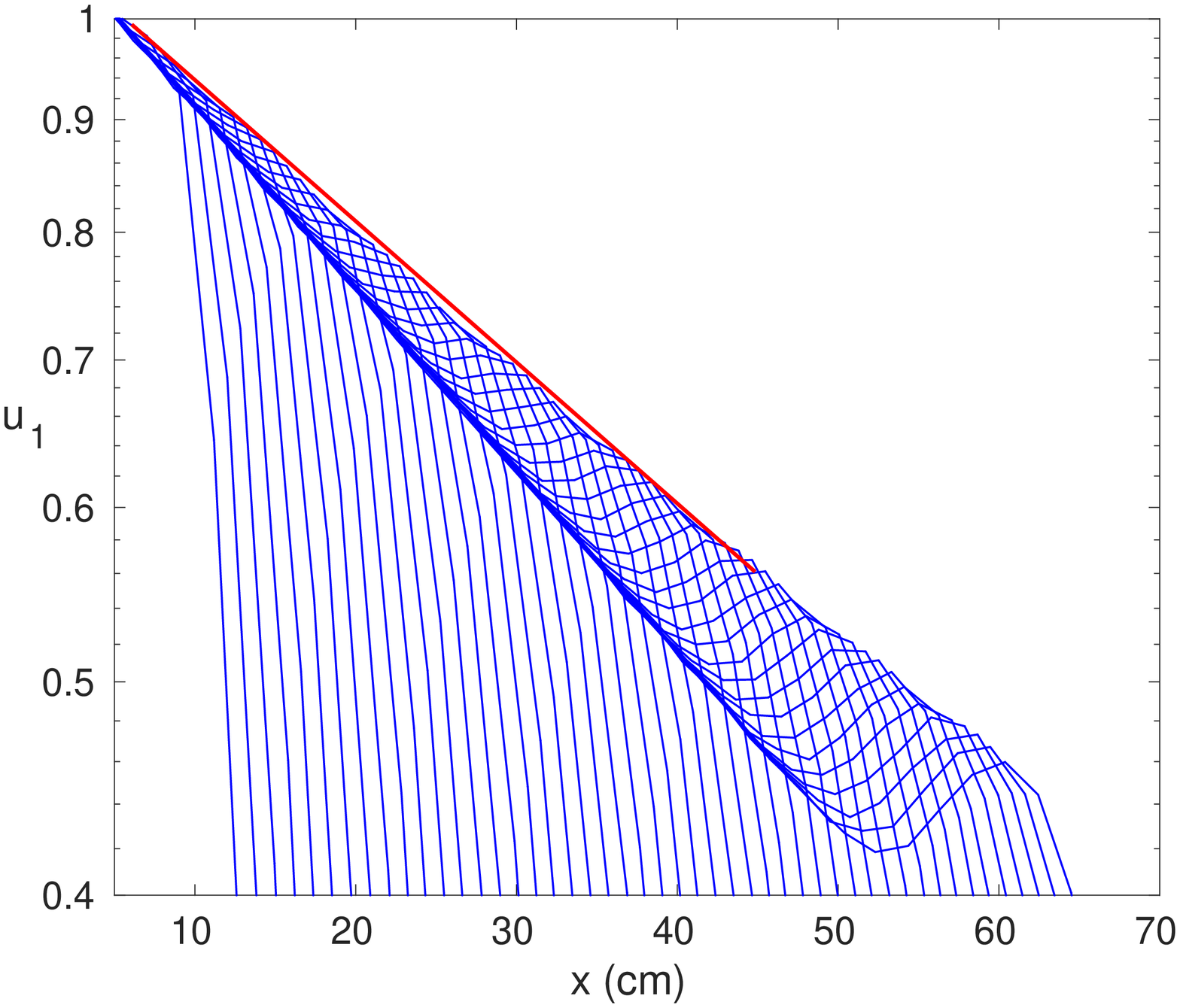}}
\caption{ Instantaneous IBVP solutions in the test section for Re=5000 ans 
$\omega_0=0.34$ over two periods of the vibrating ribbon (a) $200 
\le \ t \ \le \ 240$. Solid line represents the computed maximum amplitudes at 
the experimental points. (b) $38\le\ t \ \le 228$.  Envelope of the TS wave is shown by 
 bunching of blue lines (inner). Red line shows the envelope of the wavepacket.  }
\end{figure}
If the experimental observations were made at a later time, when the wavepacket moves out of the test section, the envelope is expected to show the TS wave. 
Figure 9(a) shows the resulting envelope. Note that the irregularity in this case 
is much less and the envelope follows the rate of decay of the TS wave very 
well. Hence, the initial rise as well as the different decay rate shown by the 
lower three curves of N75F15 are due to the combination of the choice of spatial 
and time steps and the passage of the wavepacket.
 On the other hand, if the experimental sampling rate was higher and also if more 
recording stations were located upstream, the envelope will be closer to that 
shown in figure 9(b); it will show a linear decay with rate different from the 
decay rate of the TS wave which is shown by the bunched blue lines.\\

N75 and N81 obtained only instantaneous hot-wire measurements at fixed locations and 
N75F15 is the maximum disturbance over many such instantaneous measurements at these locations.
Hence, these measurements cannot directly show the passage of a wavepacket at an earlier instant.
 We infer its existence in the experiments from the comparisons of IBVP solution with N75 measurements shown above. Since the wavepacket in the test section is of comparable size, or even bigger, than the TS wave at some ribbon frequencies, it  can equally well be considered as a base state for  secondary instability.
 \section{Secondary instability analysis}
 Secondary instability due to three dimensional background disturbances has long been considered a key mechanism in explaining 
subcritical transition in wall-bounded shear flows \citep{herb3}. A variety of 
base states have been considered for the linearisation - the dominant TS mode 
with the damping neglected (\citealt{herb1};
H83  hereafter), nonlinear equilibria and quasi-equilibria \citep{ors} and 
streamwise vortices and streaks\\
\noindent (\citealt{schmid}). Typically, in all such studies, 
the base state has to be considered in a reference frame moving with an 
appropriate velocity. 
This renders the coefficients of the disturbance equations periodic in the frame 
variable, with the implication that Floquet modes, in that variable, can be 
sought.  The three-dimensional background disturbances are represented by spanwise wavenumbers, $\beta$.\\

The general strategy is to study temporal secondary instability; a basic 
 traveling wave of wavenumber $\alpha$ is considered and the secondary temporal growth rate 
determined from the solution of an eigenvalue problem.  
The traditional secondary instability analyses of H83 and others stop at computing TS threshold amplitude for a neutral Floquet mode at a given $\beta$, which we term neutral threshold amplitude, for easy reference. 
The neutral threshold amplitude is merely the lowest one for a possible secondary growth and cannot be directly compared with experiments duch as N75 or N81, as (a) the 3-D disturbance amplitude is not accounted for and (b) the mild decay of the TS wave is uncompensated for. The experimental studies on flow stability reported in literature, for example, N75 and N81 not only present the wavenumbers of the background three-dimensional disturbances but also the corresponding initial amplitudes $\epsilon_z$. Their threshold amplitude measurements are closely linked to $\epsilon_z$ as is evident from N81. Apart from these experiments, a few other measurements  of pPf, (for e.g. \citealt{nish}, \citealt{ramaz}), have also quantified three-dimensionality of the experimental set-up.\\
 
 We have taken into account both factors in our computation of the threshold amplitude. The amplitude and decay rate of the base state and the magnitude of the three-dimensional background disturbances  have been combined into a formula  for net growth or decay of the total disturbance over one time-period of the vibrating ribbon; the formula is discussed in the following subsection. This combination is  similar to how the 
primary state is formed by superposition of the TS wave onto pPf.
\subsection{Threshold amplitudes}
For a small ribbon velocity amplitude $A_R$ the total streamwise velocity is given by
\begin{fleqn}
\begin{equation}
u_2(x,y,t)= U(y) + A_R\ e^{i(\alpha x -\omega t)}u_{TS}(y)
\end{equation}
\end{fleqn}
  \noindent where $u_{TS}$  is the normalized TS  eigenfunction. For a given wavenumber $\alpha$, when $Im(\omega)=\omega_i $ is very small, $u_2$ can be a  secondary base state which may be unstable to three dimensional disturbances.  \\

The total disturbance function $\bar{u}(x,y,t)$  is composed of 
the secondary base state and the corresponding Floquet modes. 
 For simplicity, we define $\bar{u}(x,y,z,t)$  
as:
\begin{fleqn}
\begin{equation}
 \bar{u}(x,y,z,t)=U(y)\ +\ A_R\ e^{i(\alpha x -\omega_0 t)}\ \left\{ e^{\omega_i t} 
u_{TS}(y)+\frac{\epsilon_z \ e^{i\beta z}}{A_R}\  Max\left [ (e^{-i\sigma t}-1, 0\right] \ 
u_f(y)\right\} 
\end{equation}
\end{fleqn}

 \noindent where $Re(\omega)=\omega_0$ and $\sigma$ is the least stable Floquet mode and $u_f$ is the eigenfunction; for sufficiently small $A_R$, $\sigma_i < 0$.  
 The second term on the R.H.S within the brackets is the three dimensional secondary growth and is modeled such that for either $\epsilon_z=0$ or $\sigma_i \le 0$, only the secondary base state remains. The maximum criterion has been used to ensure this happens in the latter case.\\
 
 The quantities $\beta$ and $\epsilon_z$ are inputs from the measurements. These are presented in N75 and N81 as the wavenumber and amplitude of spanwise variations in the centerline velocity $U_c$. 
    N81F5 and N81F6 show, for  some small ribbon amplitudes, that the spanwise percentage variation of TS amplitudes is also roughly the same as that of $U_c$. However, in these cases, the ribbon amplitudes are in the neighborhood of the threshold values and hence  secondary growth is already taking place, even though it may not be large enough to compensate the base state decay. 
    The form of three dimensional disturbances  for very small ribbon amplitudes is not known. In the absence of this knowledge, the formula shown above is a simple way of incorporating 
 the developing three dimensionality while establishing the two dimensional base state in the absence of secondary growth.\\
 
 For the Floquet expansion to be valid, 
 it is only necessary that $\epsilon_z \ll u_2$ and hence it can be of the order of  $A_R$. 
 Here, we consider only the most unstable 
fundamental mode $\sigma$ whose real part is often negligibly small. As we are interested only in obtaining the threshold amplitudes, we further assume that
 there exists a $y=y_1$ such that $u_{TS}(y_1)= u_f(y_1) = 1$ which would maximize 
 $\bar{u}$ across the channel; hence, the threshold amplitudes for $\bar{u}$ for growth under these assumptions 
 will give the minimum threshold amplitude for secondary growth.
 At the spanwise peaks ($z=2n\pi/\beta$), the equation given above 
simplifies to :
\begin{fleqn}
\begin{equation}
 \bar{u}(x,y_{1},z_{peak},t)=U(y)+A_R\ e^{i(\alpha x -\omega_0 t)}\ \left\{  
e^{\omega_i t}+\frac{\epsilon_z }{A_R}\ Max\left[ e^{-i\sigma t}\ 
-1,0\right]\right\}
\end{equation} 
\end{fleqn}
 The expression within the curly brackets models the total 
growth or decay of the input disturbance. We now describe the two methods of 
determining the threshold amplitude $A_T.$\\

In amplitude plateauing, which is used in N75, $A_T$ is determined by requiring 
the average growth/decay of this term over one time-period, $T$ ($= 
2\pi/\omega_0$) to be zero, i.e. 
\begin{fleqn}
\begin{equation}
  \ \ \frac{1}{T}\left[ \frac{e^{\omega_i T}-1}{\omega_i} \ + \ 
\frac{\epsilon_z}{A_T}\frac{e^{\sigma_i(A_T) T}-1}{\sigma_i(A_T)} \right] \ - \ 
\frac{\epsilon_z}{A_T}\ = \ 1. 
\end{equation}
\end{fleqn}
In case of peak-valley splitting, we first note that the leading 
component in the Floquet eigenfunction series is symmetric while the TS 
eigenfunction is antisymmetric. Hence, their sum, as in the formula for 
$\bar{u}$, will have sharper peaks and shallow valleys. The time-averaged  disturbance amplitude at the valley is given by 
\begin{fleqn}
\begin{equation}
  \bar{u}(z_{valley})= \frac{A_T}{T}\left[ \frac{e^{\omega_i T}-1}{\omega_i} \ 
- \ \frac{\epsilon_z}{A_T}\frac{e^{\sigma_i(A_T) T}-1}{\sigma_i(A_T)} \right] \ 
+ \ \frac{\epsilon_z}{A_T} 
\end{equation}
\end{fleqn}
while that at the peak continues to be given by the LHS of (5.4).\\
 
For decaying Floquet modes, by the present definition, the peak and valley 
amplitudes are identical. They start splitting when $\sigma_i = 0$. For nearly 
two-dimensional disturbances, these amplitudes may not differ noticeably up to a 
certain $\sigma_i >0$ and that is why these have not been used in N75; the 
splitting is more rapid in the case of 3D disturbances as shown in figure 10. 
The solid and dashed lines refer to the variation of peak and valley amplitudes 
w.r.t ribbon velocity amplitude respectively; the splitting is clearly seen. The 
intersection of the peak amplitude with the identity line (dotted) in figure 10 
indicates the plateauing amplitude.  \\

The threshold  amplitudes in N75  were measured based on amplitude plateauing as 
shown in N75F15 while peak-valley splitting was chosen in N81. Following the 
measurements, we have used (average) amplitude plateauing given by (a) for small 
$\beta$ and the peak-valley splitting amplitude for $\beta=1.76$ as in N81. \\

\begin{figure}
\centering
\includegraphics[width=3.2in, height=2.5in]{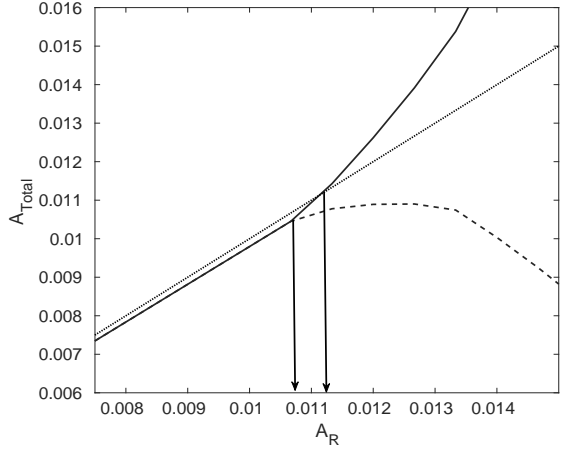}
\caption{Peak-Valley splitting.  $Re = 5000$; $\omega_0=0.34$ (72Hz) 
and   $\beta=1.76$.  }
\end{figure}
We have considered only the growing fundamental mode, which is (or nearly) always in phase with the TS wave. The decaying Floquet modes are not important as they lead to net decay for subcritical Reynolds numbers.  
The TS amplitude and the secondary disturbance amplitude  are simply added here along with their decay and growth rates respectively; this situation is possible only if the corresponding eigenfunctions have peaks at the same $y$ location. 
Therefore, the computed threshold values are the lowest possible estimates under the one-ribbon period averaging.
 \subsection{Background disturbances in N75 and N81}
The spanwise distribution of the laminar centreline velocity $U_c$ was
found to be wavy for $Re > 3500$ (N75F2), with the authors suggesting that it was due to a
slight warping of the upper channel wall. Warping can induce a variety of spanwise velocity distortions over a range of wavenumbers $\beta$; the smallest value is zero.
 The mid-third of the 40 cm wide channel is warped which produces a variation of 1.5\% of the mean
channel depth (N75); the velocity on either side of the warped portion is not known. In the absence of velocity
data across the entire channel, we assume a spanwise mean flow distortion, of wavelength equal
to warping width,  ($\beta = 0.35$); the corresponding distortion amplitude is assumed to be 1\% ($\epsilon_z = 0.01$ )
based on the given mean channel depth variation of 1.5\%. This set of 
parameters is a typical one for a mildly three-dimensional background disturbance.  \\

  Another set of values for ($\beta,\ \epsilon_z$) can be obtained
from the velocity distortions within the warped section as shown in N75F2. Figure 11 shows the Fourier transform of this data.
A peak at a 
wavenumber of $\beta=1.5$ can be seen at all Reynolds numbers. The largest 
amplitude deviations from the mean centerline velocity are roughly 0.0037 
and 0.003 for $Re=6000$, $5000$ and $4000$ respectively. 
Another peak of similar amplitude occurs at $\beta=2.2$ for Re=6000; however, it 
is not considered in the present analysis. For 
comparison with N75, we hence consider two  sets of parameters, ($\beta$, $\epsilon_z$)$=(0.35,\ 0.01)$ and $(1.5,\ 0.0037)$, arising from the warping on the top channel wall. These spanwise amplitudes are much smaller than that of the mildly three-dimensional disturbance presented above. However, they are still an order of magnitude higher than the freestream
disturbance amplitudes.\\

   \begin{figure}
\centering
\includegraphics[width=3.5in]{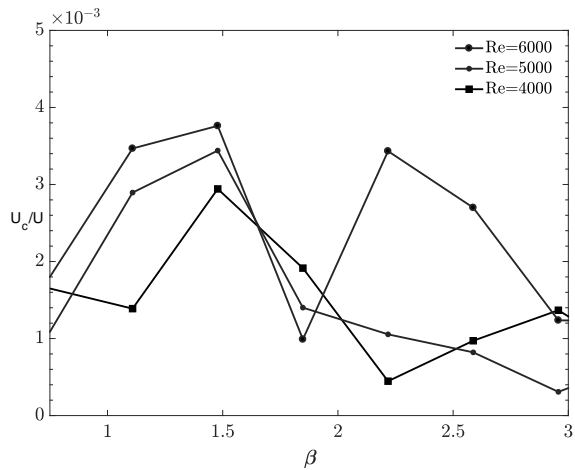}
\caption{Fast Fourier transform of data presented in N75F2 for Re=5000 and 6000. 
The maximum percentage deviation from the mean velocity is approximately 0.4 
($\epsilon_z=0.0037$). The maximum occurs at $\beta=1.5$.  }
\end{figure}
The measurements of N81 are for highly three-dimensional disturbances both in 
terms of the spanwise wavenumbers and the corresponding percentage variation in 
the mean velocity. A periodic spanwise variation of the base flow was achieved 
with the help of a damping screen with the wavelength and variation in the 
centerline velocity being roughly 25 mm ($\beta=1.76$) and 5\% respectively. 
Unlike N75, the threshold amplitudes in N81 were measured based on peak-valley 
splitting and are presented in N81F15.  Following the experiments, we have 
computed the threshold amplitudes  for $\beta =1.76$, using a spanwise amplitude 
of 0.05; the  results for peak-valley splitting and amplitude plateauing are 
shown in Figure 10. Plateauing occurs at a higher amplitude than the 
peak-valley splitting since, theoretically, peak-valley splitting occurs for any 
$\sigma_i >0$, whereas, as shown in subsection 5.1, time-averaged plateauing of the 
peak amplitude  (similar to N75F15), occurs at a positive $\sigma_i$. Even for 
nearly two dimensional disturbances as in N75, the peak-valley splitting will 
occur at lower amplitudes compared to N75F16. However, the splitting may not be 
significant up to some amplitude and hence would  not be a convenient criterion  
for threshold amplitudes in that case.
\subsection{Floquet analysis of base states}
It is clear from Table 1 that distinct TS and wavepacket states and mixed states  exist within the ribbon frequency ranges considered in N75 and N81. H83 pioneered the
secondary instability analysis with the TS wave as base state; some questions regarding the fundamental and subharmonic instabilities have been reconsidered in \cite{kidambi}.
\subsection{Secondary instability of wavepacket state} 
  It may be noted that only one wavepacket emerges in the solution of the IBVP, while the Floquet framework necessarily implies a periodic system of wavepackets.  For this purpose, we construct a periodic wavepacket system based on the IBVP wavepacket,  padding with zero on either side so as to control the separations of the packets. The procedure for wavepacket reconstruction using Fourier coefficients is described in Appendix \ref{sec:appE}. \\
  
The governing equations for secondary disturbance evolution and their discretized forms are given in Appendix \ref{sec:appF}; the discretised equations have been written for the wavepacket for the first time and reduce to the known form for the TS wave state. The number of Fourier modes in the Floquet expansion  depends on the number of significant coefficients, $N_f$, in the Fourier expansion of the wavepacket; smaller the base $\alpha$, larger the index $N_f$, which in turn increases the  size of the resulting Floquet matrix. Hence in this analysis, we choose $N_f$ to be 22 at the maximum. \\

 For the present analysis to have any relevance to the original problem, it is important to know what effect the separation between the wavepackets has on the secondary growth rates. Two different convergence tests have been performed: (i) the wavepacket at different times have been considered and  (ii) the number $N_f$ is varied from  11 up to 20.  The convergence of the least stable / most unstable fundamental mode at different amplitudes $A$ of the wavepacket, for the two times and various $M$ and $N_f$ is demonstrated in Table 4 for $Re=5000$, $\omega_0=0.45$ and the spanwise wavenumber $\beta=1.84$. Most of the computations in this paper are done using $N_f=11$ Fourier coefficients. 
\begin{table}
\fontsize{8}{11}\selectfont
\begin{center}
\begin{tabular}{|c|c|c|c|c|} \hline
A & $N_f$ & M & $\sigma_r$ & $\sigma_i$    \\ \hline 
 & 11 & 15 & 0.00227 & -0.00048  \\ 
0.0022 & 17 & 20 & -0.00226 & -0.00042 \\
 & 22 & 25 & 0.00226 & -0.00048 \\ 
(t = 128) & 22 & 25 & 0.00226 & -0.00077 \\ \hline
& 11 & 15 & 0.00235 & 0.00122 \\
0.0024 & 17 & 20 & 0.00235 & 0.00127 \\
 & 22 & 25 & 0.00234 & 0.00122 \\ 
(t = 128) & 22 & 25 & -0.00233 & 0.00093 \\ \hline
 & 11 & 15 & -0.00243 & 0.00278  \\
0.0026 & 17 & 20 & -0.00242 & 0.00282 \\
 & 22 & 25 & 0.00242 & 0.00276 \\ 
(t = 128) & 22 & 25 & 0.00241 & 0.00247 \\ \hline
 & 11 & 15 & 0.0025 & 0.00421 \\
0.0028 & 17 & 20 & -0.00250 & 0.00425 \\
 & 22 & 25 & -0.00250 & 0.00419 \\ 
(t = 128) & 22 & 25 & 0.00248 & 0.0039 \\ \hline
\end{tabular}
\caption{Least stable / most unstable fundamental Floquet eigenvalue for various amplitudes $A$ of the wavepacket.  $\omega_0 = 0.45, Re = 5000.$  }
\end{center}
\end{table}
\section{Results II: Comparison with N75F15 and N81F16}\label{sec:comparison}
We now present the threshold amplitudes for several drive frequencies $\omega_0\in 
(0.25,0.45)$. From the IBVP solution (for e.g. figure 1), relatively clear base 
states of TS wave and wavepacket can be established for the lower and upper ends 
of the frequency range. It is for these ranges that a secondary analysis can be 
performed and the threshold amplitudes obtained.
We have chosen the wavenumber-amplitude combinations, based on the data 
presented in the introduction, viz. (1) $\beta=1.5$, $\epsilon_z=0.0037$ and (2) $\beta= 0.35$, $\epsilon_z=0.01$,  in order to meaningfully compare with N75F16. \\

\begin{figure}
\centering
\subfloat[]{
\includegraphics[width=2.75in]{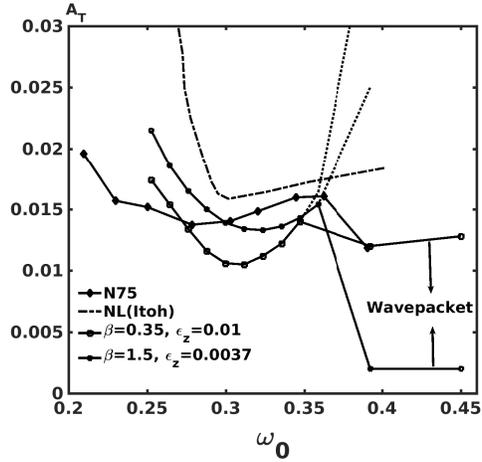}}
\hspace{0.2in}
\subfloat[]{
\includegraphics[width=2.75in]{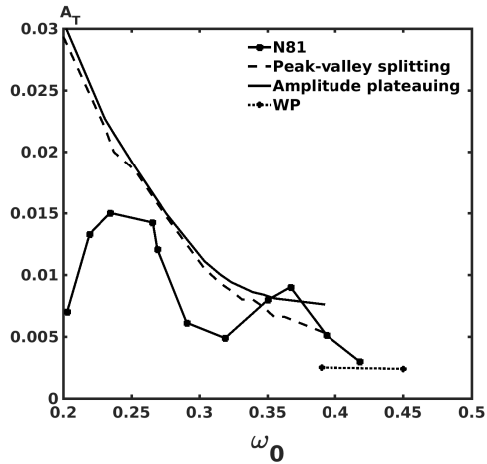}}
\caption{Threshold amplitudes as a function of ribbon frequency 
$\omega_0.$  $Re = 5000$.(a) Filled diamond : experimental values from N75; 
  Dash-Dot lines: Itoh's (1974) nonlinear calculations. Present threshold 
computations for secondary instability for $\beta=0.35; \ \epsilon_z=0.01 \ 
\hbox{and} \ \beta=1.5; \epsilon_z=0.0037$ are shown. The wavepacket base state 
computations are indicated. (b)  Filled square : experimental values from N81 
 (strongly three-dimensional) $\beta=1.76$, $\epsilon_z=0.05$.  }
\end{figure}
We plot the computed threshold amplitudes $A_T$ for these two sets in figure 12(a), for $Re=5000$ as a function of $\omega_0$; experimental data from N75F16 are also shown.
The experimental first 
minimum $Mi_1$  occurs at $\omega_0=0.28$ with an amplitude of 0.0135.
The 
computed $Mi_1$ occurs at $\omega_0=0.3$ and $0.32$ for 
$(\beta,\ \epsilon_z)=(0.35,\ 0.01)$ and $(1.5,\ 0.0037)$
respectively; their corresponding amplitudes are 0.01 and 0.013. 
At higher  $\omega_0$ $>0.34$, the computations for the TS base state show 
increasing threshold amplitudes in both calculations, as indicated by dotted 
lines in the figure. The nonlinear calculations of Itoh (1974) (Dash-dot) also 
show the same trend. The 
minimum threshold amplitude of Itoh (1974) occurs at the same frequency as the present 
calculation for $(0.35, \ 0.01)$.
The rate of increase in the threshold amplitude from Itoh (1974) is somewhat lower than the present calculations, however the second minimum $Mi_2$ is not 
shown by the nonlinear analysis.
We recall that at these higher frequencies, the base 
state for the secondary analysis is not a pure TS wave but a mixed state or even a wavepacket.\\

The Floquet analysis for $\omega_0> 0.39$  is performed on the wavepacket state.
The wavepacket threshold amplitude for $(1.5,\ 0.0037)$ is much lower compared to that of $(0.35,\ 0.01)$  even though its $\epsilon_z$ is very low.
 For both sets of $(\beta,\ \epsilon_z)$, computed threshold amplitudes for $\omega_0=0.39$ and $0.45$ are almost equal with 
 $A_T=0.0124$ and 0.0025 respectively; the values at $\omega_0=0.45$ are slightly higher.  The second minimum $Mi_2$ of the present calculations, hence occurs at $\omega_0=0.39$. The experimental value at 
$\omega_0=0.39$ is roughly 0.012, which is close to the wavepacket threshold 
amplitude for $(0.35,\ 0.01)$.    \\
    
The N81 measurements show three minima at $\omega_0=0.2, \ 0.32, \ \hbox{and} \ 
0.425$. The present computations show a monotonically decreasing threshold amplitude 
for the TS wave. The base state is in fact a mixed one for $\omega_0\ge 0.34$ and hence 
the present threshold computations are not applicable in this range; they are shown in the figure 
only to indicate what numbers would be obtained with such an analysis.
 The computed threshold amplitude for the wavepacket at $0.39$ is roughly 0.0025, which does not vary till 
 $\omega_0 = 0.45$.
  One difference from the N75 case is that the intermediate peak, 
demonstrated by experiment in $\omega_0 \in (0.34,0.39)$ cannot be deduced from 
the present computations and a separate analysis is required for the mixed 
object in this range. The computed threshold values at $\omega_0=0.34$ and 0.39 
are 0.0075 and 0.0025, the first of which is higher than the corresponding 
experimental minimum of 0.005 at $\omega_0=0.32$ whereas the second matches well 
with the measured minimum.  Unsurprisingly, the two-dimensional nonlinear 
threshold calculations of Itoh are very high compared to both  the present 
computations and the measurements of N81 and neither capture the minima nor 
their location.\\

The third minimum $Mi_L$ at $\omega_0=0.2$ is not shown by the present 
computations. The IBVP solution is a mixed state at $\omega_0=0.2$. For $\omega_0 < 0.2$, a
 clear wavepacket emerges in the test section, with a decay rate higher than those corresponding
  to $\omega_0 \ge 0.34$. In addition to this, the least stable OS mode for $\beta=1.76$ has a 
  decay rate comparable (or even lower) to that of the TS wave. The high initial amplitude  $\epsilon_z$ (=0.05)  at this $\beta$ will also affect the receptivity of the three-dimensional primary mode for this $\beta$. Hence, in this range of ribbon
  frequencies, the secondary base state cannot be deduced from the IBVP using the least stable
  OS mode alone.  We have not considered the resulting compound base state in the present study.     \\
  
The comparison of the computed threshold amplitudes for Re=6000 with the 
corresponding data of N75F16 is shown in Figure 13(a). Two sets of calculations for 
the combinations ($0.35,\ 0.01$) and 
($1.5,\ 0.0037$) have been done. N75F16 shows  $Mi_1$ and $Mi_2$ at 
$\omega_0=0.27$ and at $\omega_0=0.345$ respectively. The  computed minimum $Mi_1$   
for the TS base state is at $\omega_0=0.28$. 
At $\omega_0=0.34$, the threshold amplitudes for TS wave and that of the wavepacket are the same for ($0.35, \ 0.01$); on the other hand, for  ($1.5, \ 0.0037$)   
the threshold value for the wavepacket base state drops to a very low value of 0.002.  
For $\omega_0 >0.34$, the TS base state threshold amplitude continues to grow (not shown here) while the wavepacket threshold amplitudes plateau to 0.0096 and 0.002 
respectively for the parameters ($0.35,\ 0.01$) and ($1.5,\ 0.0037$) respectively. The 
experimental value at $Mi_2$ lies between these two values. 
The nonlinear threshold amplitudes increase with increasing
ribbon frequencies. \\

The threshold computations for Re=4000 at $\beta=0.35$ is shown in Figure 13(b). 
The computed values agree very well with the measured values in the frequency 
range $(0.32,\ 0.34)$, at which the TS decay rates are the lowest. For all other 
frequencies, the application of Floquet analysis is more and more in error as the damping rate is no longer negligible. For 
$\beta=0.35$,  the threshold amplitudes of the wavepacket 
corresponding to $\omega_0=0.4, \ 0.425$ are less than the minimum threshold of 
the TS waves. The measured $Mi_2$ matches with the computation at $\omega_o=0.4$.
The wavepacket at $\omega_0=0.425$, 
however, shows a very low threshold amplitude of 0.003 when $\beta=1.5$. 
Threshold amplitude computations were not done for $\beta=1.5$ and 
$\omega_o=0.4$.\\

Interestingly, the criterion for threshold amplitude presented in subsection 5.1
 is not satisfied at all for TS wave at $\beta=1.5$ and $\epsilon_z=0.003$,
  which indicates that a plateauing similar to N75F15  does not occur at all.
   By increasing $\epsilon_z$ to $0.004$, the threshold amplitude condition 
   can be satisfied  over a small range of $\omega_0$. This verification, however, 
   is not shown here.\\

For $Re=5000$ and 6000, the threshold amplitude at $Mi_1$ is much lower for the 
$\beta=0.35$ than for $\beta=1.5$; this may be due to the higher value of 
$\epsilon_z$ assumed at $\beta=0.35$. For the wavepacket state, the growth rate 
of Floquet modes increases quite rapidly with its amplitude and hence the 
threshold amplitudes are insensitive to the variation in $\epsilon_z$. The 
wavepacket thresholds decrease with increasing $\beta$.
\begin{figure}
\centering
\subfloat[]{
\includegraphics[width=2.7in,height=2.66in]{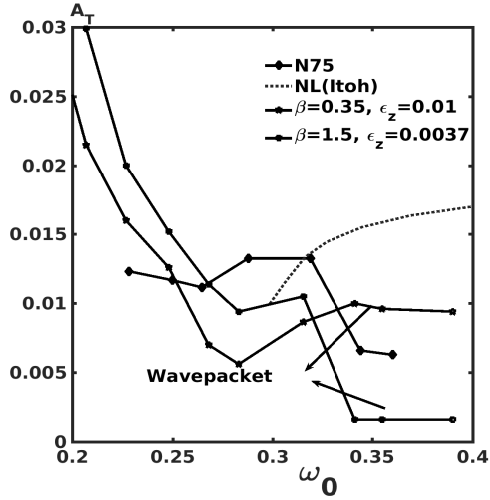}}
\hspace{0.15in}
\subfloat[]{
\includegraphics[width=2.7in]{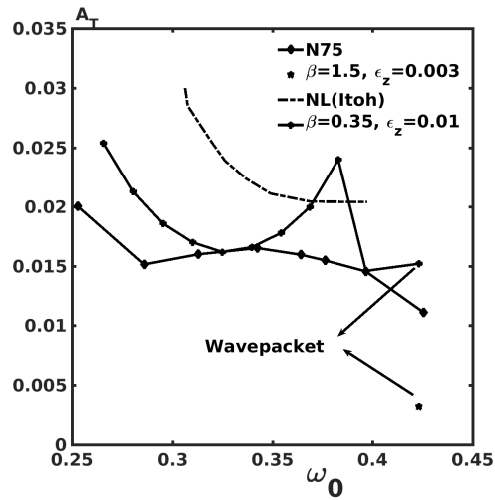}}
\caption{(a) Threshold amplitudes as a function of ribbon frequency 
$\omega_0.$  $Re = $ 6000. Filled diamonds represent experimental values from 
N75.  Dash-Dot lines represent Itoh's (1974) nonlinear calculations. Present 
threshold computations for secondary instability for $\beta=0.35; \ 
\epsilon_z=0.01 \ \  \hbox{and} \ \beta=1.5; \epsilon_z=0.0037$ are shown. The 
wavepacket base state computations are indicated. (b)Threshold amplitudes as a function of ribbon frequency  $\omega_0.$  $Re = $ 4000. Filled diamonds represent experimental values from 
 N75.  Dash-Dot lines represent Itoh's (1974) nonlinear calculations. Present 
 threshold computations for secondary instability for $\beta=0.35; \ 
 \epsilon_z=0.01 \ \  \hbox{and} \ \beta=1.5; \epsilon_z=0.003$ are shown. The 
 wavepacket base state computations are indicated.}
\end{figure}
\subsection{Discussion}
As pointed out in the previous section, most of the presented experimental data (for e.g. N75F7 and N75F15) do not represent a constant-rate decay, as required by linear stability analysis; in fact there are regions of spatial growth and plateauing followed by decay even for very small disturbance amplitudes at subcritical Reynolds numbers. At first glance, these features may seem attributable to transient growth, at least two manifestations of which have been long studied - (a) due to streamwise independent structures \citep{ell} and (b) due to the interaction of at least two non-normal modes (\citealt{schmid}). From the results that have been presented earlier, we argue that both these mechanisms are not in play here. Instead, we have shown that the curves (i) - (iii) of N75F15 are a reflection of the spatio-temporal nature of the interaction between the TS wave and the associated wavepacket. Non-normality of the underlying operator is not directly relevant, as these features are shown by the primary mode itself. As mentioned in the Introduction, Trefethen et al (1993) demonstrated similar behavior with a $2 \times 2$ nonlinear, non-normal model; however, its relevance to an experimental situation like N75 has not been established. The marginal role played by nonlinearity in explaining the behavior of the data considered here is further illustrated by the following facts - (a) Itoh's (1974) non-linear threshold amplitudes are higher than the experimental values
for all Reynolds numbers and (b) Even for a high initial amplitude of 2 \%, the initial amplitude and growth of the first harmonic  is very small (N75F17).  The fact that our results, computed using a secondary instability analysis based on a linear solution, can explain the experimental observations to a large extent, further confirms this.\\

As is well-known, Floquet analysis allows detuned modes as solutions, the 
fundamental (resp. subharmonic) being not detuned at all (resp. being the most 
detuned).   The computed thresholds should correspond to whatever detuning 
produces the lowest values. However, the thresholds presented in figures 12 and 13  
are based only on the fundamental secondary mode. One reason for this is the 
lack of experimental observations of signatures of the detuned modes, despite  
sometimes having lower thresholds, detailed explanations for which have been 
advanced ( 
for example, in  \citealp{kim}, \citealp{zang}, \citealp{kidambi}). Also, the threshold amplitude in the present scenario cannot be read 
off as the value at which the secondary mode begins to grow but rather has to 
be 
computed, taking into account the slight decay of the base state (be it a TS 
wave or a wavepacket), as detailed in section 5.1. This computation gives 
unambiguous results for the case of the fundamental mode as it is phase-locked 
with the primary wave but would have to be further modified to produce sensible 
results for experimental comparison, if one were to consider detuned modes. In 
view of the aforementioned lack of experimental observations of such modes, we 
have considered only the fundamental modes in this study.\\    
  
 The IBVP solution shows only one wavepacket downstream of the TS wave. The wavepacket travels downstream with a group velocity, $c_g$, much higher than the phase velocity of the TS wave.
 Even though the wavepacket evolves spatio-temporally, it is nearly steady in the reference frame moving 
 with the group velocity $c_g$. Given the nearly constant nature of the group velocity and size, 
 secondary instability of the wavepacket is as much a 
possibility as that of the TS wave.  
Though wavepackets have been objects of study in the stability community since 
at least \cite{gas1} which considered three-dimensional wavepacket development 
in a boundary layer, and the TS wave for even longer, the two have not been 
considered together in the vibrating ribbon problem. This is possibly (for e.g. 
\citealt{gasd}) because the vibrating ribbon was seen as producing a TS wave and a 
pulsed point source as producing a wavepacket.
The secondary instability of the single wavepacket arising from the IBVP solution  is 
examined by considering a periodic train of wavepackets as the base state; each wavepacket constituting this train is identical to the wavepacket state. The wavepackets are  sufficiently separated from each other spatio-temporally; the larger the separation, closer its secondary stability characteristics will be to those of a single wavepacket.   \\

We now discuss the sensitivity of the computed threshold amplitudes to the primary wave characteristics.
The computed threshold amplitude $A_T$ depends on the reference frame velocity $c.$ Everything else remaining same, an increasing $c$ leads to a decreasing $A_T,$ to a certain extent. A representative variation is shown in Table 5 for the 
fundamental mode at $\alpha = 1.12, \beta = 2, Re = 5000.$ $A_T$ attains a minimum for $c \approx 0.8$. It is well-known (for e.g. \citealt{crosw}) that the transfer of energy from the mean flow to the secondary disturbance is the key instability mechanism and is represented by the term $T_{30} = - \epsilon_z^2 
\int_{\Omega} u_3 \, v_3 \, \frac{dU}{dy} \, d \Omega$; $u_3$ and $v_3$ are the velocity eigenfunctions in the streamwise and normal directions and $\Omega$ is the channel volume over one wavelength of the TS wave. For the large TS amplitudes considered in H83 and \cite{crosw}, the eigenfunctions $u_3$ and 
$v_3$ (for both fundamental and subharmonic) are peaked in the neighborhood of the critical layer and hence a large contribution to $T_{30}$ happens in that 
neighborhood. However, for smaller amplitudes  closer to the threshold $A_T$, $u_3$ and $v_3$ can be dramatically different, as shown in Figure 14. Though they still have maxima in the vicinity of the critical layers, these are more broad and a larger section around the critical layer contributes. An increasing $c$ implies that the critical layer moves towards 
the centre of the channel but the gradient $d U / dy$ goes to zero at the centre, with the net effect that the maximum transfer of energy occurs for some $c \in (0,\ 1)$ and this also corresponds to the lowest threshold $A_T$. We recall here that the wavepackets are  stationary in frames moving with group velocities $c_g$ 
which are higher than the phase velocities of the TS waves considered here; correspondingly, their threshold amplitudes are
much lower than those of TS amplitudes as seen from figures 12 and 13. \\
\begin{table}
\begin{center}
\begin{tabular}{|c|c|c|c|c|c|c|c|c|c|c|c|c|} \hline
$c$ & 0.2  & 0.3 & 0.4 & 0.5 & 0.6 & 0.7 & 0.75 & 0.8  & 0.9 & 0.95  \\ \hline
$A_T$ & 0.0362  & 0.0108 & 0.0066 & 0.0052 & 0.0046 & 0.0042 & 0.0041 & 0.004  & 
0.0042 & 0.005 \\ \hline
\end{tabular}
\caption{Threshold amplitude of the fundamental mode as a function of the 
reference frame velocity $c$. $Re = 5000, \alpha = 1.12$ and $\beta = 2.$}
\end{center}
\end{table}
\begin{figure}
\centering
\includegraphics[width=5.25in]{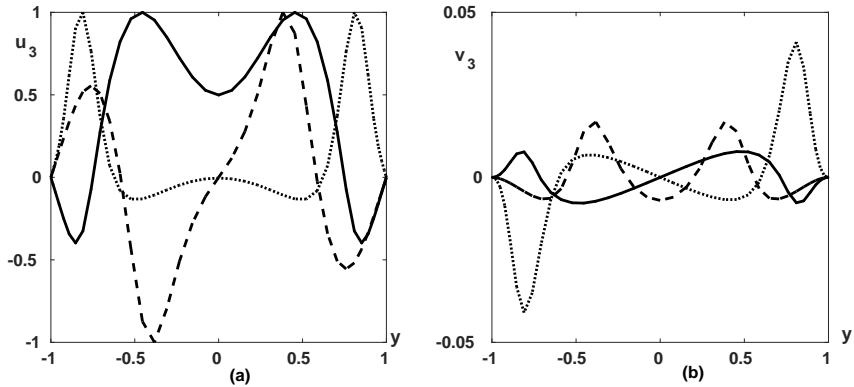}
\caption{The dominant component of a) $u_3$ and b) $v_3$ of the fundamental mode 
as a function of $y.$ Solid and dashed lines correspond to threshold amplitudes 
for $c = 0.3$ and 0.8 respectively. Dotted line is for an amplitude of 0.0248 
and $c = 0.2818.$ These are the values similar to the ones in the experiment of 
Nishioka \& Asai (1984). Arrows indicate approximate location of the critical 
layers.}
\end{figure}

  The threshold amplitude $A_T$, as computed from the formula,  also depends on the temporal decay rate of the base state.
The measured and
 computed threshold curves show a parabolic variation with respect to $\omega_0$ which follows closely the parabolic curve which the TS spatial decay rate $\alpha_i$ traces w.r.t $\omega_0$. The first minimum, $Mi_1$, in N75F16 occurs at the $\omega_0$ corresponding to the least decaying spatial TS wave. 
In the computations presented in figures 12 and 13, $Mi_1$ is attained  at  slightly higher forcing frequencies for 
all the three Reynolds numbers. The reason for this shift is, the TS decay rates in a small  neighborhood of $M_{i1}$ do not vary as rapidly as the phase velocity; as $\omega_0$  increases the phase velocity increases and hence the threshold amplitude decreases. $Mi_1$ occurs at higher and higher frequency with  increasing $\beta$ as can also be seen from the N75 and N81 experiments; the same behavior is shown 
 by the computations presented in figure 12(a) for $Re=5000$. $Mi_1$  is not sensitive to $\epsilon_z$ but 
the corresponding threshold amplitude is mildly sensitive; a larger value of 
$\epsilon_z$ lowers the minimum threshold amplitude slightly. At higher 
forcing frequencies where a wavepacket state exists, the threshold amplitude
 does not show a parabolic variation around $Mi_2$ but increases mildly
 beyond $Mi_2$.  For example, for $Re=5000$, wavepacket states are seen 
for $\omega_0>0.39$. It can be seen from Figure 12(a) that the threshold amplitude for $\omega_0=0.45$ is higher than that at 0.39, even though the group velocity $c_g$ for $\omega_0=0.45$ is greater than the $c_g$ for $\omega_0=0.39$ (Table 1). But, the decay rate of the former is higher which causes the mild increase in the threshold amplitude at $\omega_0=0.45$.  \\

Unlike the effect of frame velocity, which comes directly from the Floquet analysis as discussed in the preceding paragraph, the decay rate of the base state is artificially 
 incorporated in  the  equation (5.4). We have also computed threshold values by  using  another method (not presented here) where the primary decay rate is compensated by simply adding the computed Floquet growth rate to it; thus, it is a multiplicative compensation as against an additive one for the present method. However,  both methods correctly reproduce the parabolic part of the threshold curves, for the TS base state. The two sets of threshold values differ only a little as long as the primary decay rates are small even though the second method does not include  $\epsilon_z$ at all. This is not surprising if we recall  the weak dependence, on $\epsilon_z$,  of threshold amplitudes computed using the formula. The sudden drop 
  in the threshold amplitude when the base state changes is mainly due to the higher frame velocity  and hence happens irrespective of how the primary and secondary disturbances
  are combined. In summary, the two major features of the threshold curves of N75F16
   and N81F15, the parabolic nature and the sudden drop in the threshold at higher frequencies, are due to the arrest of primary decay by secondary growth and the change in the base state respectively; in particular, they are not artifacts of the formula that is used for computing the threshold amplitude.\\

Threshold amplitudes have sometimes been obtained from DNS studies, for a fixed set of wavenumbers $(\alpha, \beta)$. We present findings from one such study \citep{red}, where different types of disturbances such as TS waves, three dimensional noise 
(N), oblique waves (OW) and two-dimensional optimal disturbances (2DOPT) were considered, alongside the current results and results from N75 in figure 15. It is evident from the present computations that the threshold amplitudes of the two different base states vary differently with Reynolds numbers. Hence, it is interesting to compare the nature of these variations with the similar results from DNS for other disturbance types such as 2D optimal disturbances and oblique waves etc. 
The first minima in N75F16 at three Reynolds numbers (4000, 5000 and 6000) are  shown by dashed line with diamond symbols. The computed minima at these Reynolds numbers for TS state at $\beta=0.35$ are shown as solid line with square symbols. The discrepancy between these two curves is the highest at $Re=6000$; 
the computations show a smaller threshold amplitude at this supercritical Reynolds number. The amplitudes at $Mi_2$ of N75F16, which are not necessarily the minimum values, are shown by solid line, triangles. The computed minima for wavepackets at $\beta=0.35$ are shown by dashed line, triangles. All 
these threshold amplitudes lie close to the DNS of secondary instability of TS waves but are slightly higher. It has to be noted that the DNS for the TS state was performed for $\alpha =1$ and $\beta=1.0$.  The computed minima for TS state at 
$\beta = 1.5$ are much higher than those for $\beta=0.35$ and hence not shown in this figure. The minimum threshold values for wavepacket at $\beta=1.5$ are shown by solid line with star symbols. These values are much lower than the  group of values 
for $\beta=0.35$. These values are comparable to the threshold amplitudes for 2DOPT and random noise.   It is evident from figures 12 - 13 that the wavepacket threshold values for $\beta=2$ are not very different from those of $\beta=1.5$. 
The figure also shows that the threshold amplitudes in the 2D vibrating ribbon experiments are closer to the wavepacket thresholds than those of oblique waves and streamwise vortices strengthening our claim that this is indeed the operative mechanism for these parameter values. 
\begin{figure}
\centering
\includegraphics[width=3.5in]{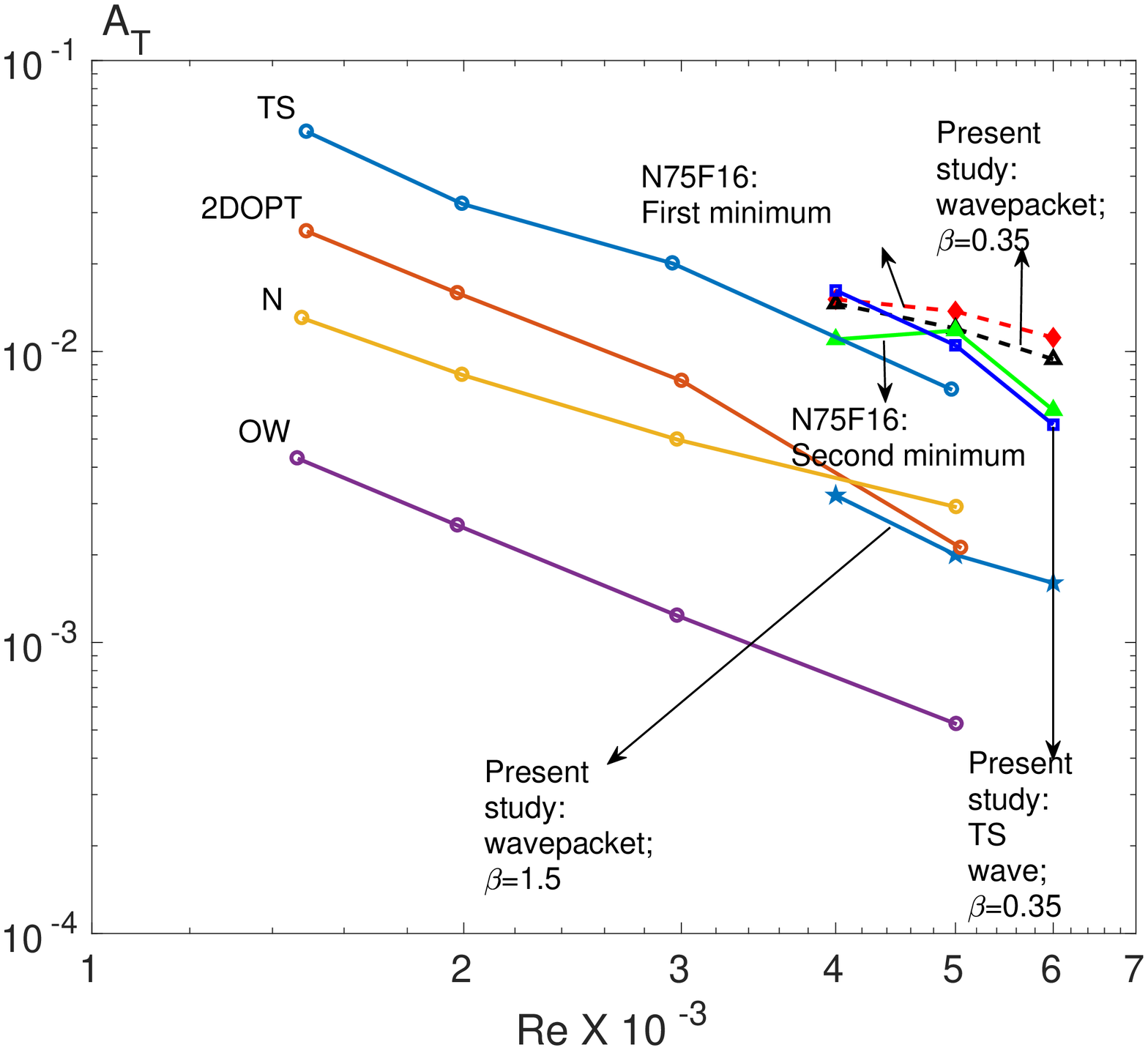}
\caption{Comparison of present computations with threshold amplitudes \cite{red}. Circles denote threshold amplitudes computations of \cite{red} for various states.  }
\end{figure}
\section{Concluding remarks}\label{sec:conclude}
A semi-analytic solution for the IBVP of a vibrating ribbon in pPf has been provided and clearly delineates the distinct states of the TS wave and the wavepacket. The solution is largely made possible by using an algorithm (Appendix A) to sort complex temporal eigenvalues into modal families, identifying salient features like saddles, poles and branch points  and then incorporating these features to properly evaluate the relevant disturbance integrals. These states are then used to provide a novel explanation for the incipient stages of subcritical transition in pPf.\\

 This involves not only the well-known secondary instability 
of the primary TS wave but also a seemingly overlooked secondary instability of 
a wavepacket that often dominates for higher drive frequencies. To this end, a secondary instability analysis of a wavepacket state has been provided for the first time. This framework 
can not only explain the behavior of the lower curves in N75F15, but also the 
reason for the maximum in N75F16, something previous theories based solely on 
primary linear stability, nonlinearity or transient growth have failed to do. 
The current model also provides a counterview to the widely accepted route to 
transition, involving the secondary instability of nonlinear TS states and their 
subsequent breakdown, for controlled disturbance environments. N75 claims to 
have seen spot-like fluctuations directly triggering transition for higher drive 
frequencies. It is tempting to speculate that these spots are further evolution 
of the wavepackets that have been observed and documented in this study. In 
fact this kind of speculation is quite old, though in a different context; we 
find, for example, in the Introduction of \cite{gasd} - `Natural 
transition often occurs through the formation and growth of turbulent spots 
which are presumably initiated by these linear wavepackets.' Wavepacket 
dynamics in the wingtips of turbulent spots has also been investigated 
(\citealt{henn1}, \citealt{li} ). A fair amount of print, mostly for boundary later flows, in the form of 
DNS studies, has been devoted to evolution of a wavepacket into a spot; a recent example is \citealt{cher}. 
Post the secondary instability analysis, we have employed heuristic methods, motivated by the experimental ones, to estimate the threshold amplitudes. It may be worthwhile to examine if these methods can be endowed with more rigor.
Other interesting and difficult problems would involve an analytic 
exploration of the nonlinear development of the structures identified in this 
paper, so as to obtain a 
better analytic description of the later stages of the transition process.\\

\noindent The authors acknowledge financial support from
National Board of Higher Mathematics, Department of Atomic Energy, India through\\
Project No. 2/48(3)/2013/NBHM(RP)/R\&DII/685.

\appendix\section{SPECRE: Sorting Procedure for Eigenvalues based on Cauchy - Riemann Equations }\label{sec:appA}
SPEC-RE is based on a well-
known result of function theory for polynomials; the roots of such polynomials are then analytic functions of the same parameter {\sl{with only algebraic singularities}}  \citep{Kato-1995}.
 whose coefficients are analytic functions of a parameter.
Thus, the eigenmode $\omega_j$ is an analytic function of the complex wavenumber $\alpha$ except at isolated branch
points. At any such point of analyticity $\alpha = \alpha_p$ , the quantity

 $$\displaystyle  F=\left|\frac{\partial{\mathcal{R}e}(\omega_j)}{\partial
\alpha_r}-
\frac{\partial{\mathcal{I}m}(\omega_j)}{\partial\alpha_i} \right|+
 \left|\frac{\partial{\mathcal{R}e}(\omega_j)}{\partial\alpha_i}+\frac{\partial{
\mathcal{I}m}
(\omega_j)}{\partial\alpha_r} \right|$$ has to be negligible
 since Cauchy-Riemann conditions for analytic functions have to be satisfied. In
 what follows, this quantity $F(\omega_j;\ \alpha_p)$ is called the CR residue.

The primary task of SPEC-RE is to sort each eigenvalue of the spectrum
from a given initial point in the $\alpha$ plane by minimization of the CR
residue at the points of analyticity.
The algorithm also makes use of the negation of the CR criterion at the branch
points rather than describing a `method' to identify branch points.\\

The computational domain is a rectangular patch in the 
the $ \alpha$ plane with edges parallel to the axes. The grid points are equally
 spaced along both the axes;
 however, the grid size in these directions may be different. The sorting
algorithm is implemented on a 4-point stencil of this grid (dashed line in figure 16(a)); at any given pivot
point $\alpha_{n,m}=(\alpha_r,\alpha_i)$
 the stencil consists of the neighbouring points $\alpha_{n-1,m} = (\alpha_r - \delta \alpha_r,\alpha_i), \alpha_{n+1,m}= (\alpha_r+
\delta \alpha_r,\alpha_i)$ and $\alpha_{n,m+1}=(\alpha_r,\alpha_i+\delta
\alpha_i)$. The inclusion of the upstream point $\alpha_{n-1,m}$ ensures the continuity of slope
in the sweeping direction, while the other three points ensure orthogonality (or local
harmonicity). Given a particular eigenvalue
 $\omega_j$ at $\alpha_{n,m}$, the algorithm is designed to pick one (and only
one) of the eigenvalues from the spectrum at two neighbouring points $\alpha_{n+1,m}$ and $\alpha_{n,m+1}$
 such that the CR condition at $\alpha_{n,m}$ is satisfied. Equivalently, the
relevant complex derivatives at $\alpha_{n,m}$ must make the CR residue $F$
 to be negligible.  In the numerical procedure,
  these derivatives are replaced by the central and forward differences
 $$\frac{\partial\omega_j}{\partial\alpha_r}=\frac{\omega_l(\alpha_{n+1,m})-
\omega_j(\alpha_{n-1,m})}{2 \delta\alpha_r},\   l=1,2,3,...$$
 $$\frac{\partial\omega_j}{\partial\alpha_i}=\frac{\omega_k(\alpha_{n,m+1})-
\omega_j(\alpha_{n,m})}{i\delta\alpha_i}, \  k=1,2,3,...$$
 \noindent The CR residue $F(\omega_j;\ \alpha_{n,m})$ is defined using these
central-forward differences and is actually a set of numbers
$F_{l
k}(\omega_j;\alpha_{n,m}) ; \ l=1,2,3,... \
 k=1,2,3,... $ . The indices $k_p$ and $l_p$  that correspond to the minimum of
these numbers for a given $j$, which is expected to be a negligible quantity,
are picked.  As the analyticity condition
  for $\omega_j$ at $\alpha_{n,m}$ is numerically satisfied between $\omega_j,
\omega_{l_p}$ and $\omega_{k_p}$, all three $\omega$s belong to the same
analytic function.
  In other words,
  \begin{gather}
  \omega_j(\alpha_{n+1,m})=\omega_{l_p}, \tag{a} \\
  \omega_j(\alpha_{n,m+1})=\omega_{k_p}.  \tag{b}
 \end{gather}
 \noindent The pivot point can then be moved to one of the two adjacent points
either in the horizontal direction or the vertical
direction and the sorting procedure can be repeated for the new stencil.
Hence, starting from an initial point $\alpha_0$, the sorting procedure
picks one and only one value from the spectrum at each grid point and assigns
it to the $j$th collection
so that an analytic function $\omega_j(\alpha)$ is constructed, on the entire rectangular patch in the $\alpha$ plane.
\subsection{Sweep direction}
In a horizontal sweep, the pivot point $\alpha_{n,m}$ moves along the direction
of increasing $\mathcal{R}e(\alpha)$, keeping $\mathcal{I}m(\alpha)$ constant.
After reaching the right-most point of the grid, the pivot point is moved to
 $\alpha_{1,m+1}$.
Further computations are performed on stencils containing $\alpha_{n,m+1}, \alpha_{n+1,m+1}$ and $\alpha_{n,m+2}$ starting from
n = 1. It may be noted that the eigenvalues at this level have already been sorted from the computation
at the m-th level, as shown in equation (b). Hence, using the eigenvalues at the m+1-st level, either (i)$\omega_j$
at the m + 2-nd level may be sorted, or (ii) re-sorting may be done afresh at the m + 1-st level. Method
(ii) will not produce any new arrangement 
of eigenvalues at the m + 1-st level unless a branch point lies
between the m-th and the m + 1-st levels. Eduction of a branch cut along the sweep direction (horizontal)
by Method (ii) will be explained in the following subsection.
The sweep direction is not rigidly fixed. A vertical sweep, for instance, will produce a different modal
map, with vertical branch cuts. One could indeed sweep even along any family of parametric curves; the
C-R equations would then have to be satisfied in the appropriate coordinates.
 \subsection{Mode sorting around a branch point}
Assume that there exists a branch point between $\omega_j$ and $\omega_{j+k}$
located in the box formed by the $m$-th, $m+1$-st,  $n$-th and $n+1$-st lines as shown in figure 16 (
i.e. $\omega_j(\alpha)$ and $\omega_{j+k}(\alpha)$ intersect at some
$\alpha_b$). 
 By design, the sorting algorithm produces an analytic $\omega_j$ not only up to the $m$-th
 line, but also up to the point $\alpha_{n+1,m+1}$ on the $m+1$-st line.
 At the stencil formed by $\alpha_{n+1,m} $, $\alpha_{n+2,m}
$ and
$\alpha_{n+1,m+1} $, application of CR condition forces analyticity of $\omega_j$
 at both edges of the stencil and hence, does not allow the BC to cut the
$\alpha_{n+1,m} $ - $\alpha_{n+1,m+1} $ edge. The forcing of analyticity
 on the lower and left edges of the box by the previous stencil leads to the BC cutting the
$\alpha_{n,m+1} $ - $\alpha_{n+1,m+1}$ edge, as shown
 in figure 16(a).
 If further computations were to be done using Method (i) to sort eigenvalues at
 $m+2$-nd level, application
of CR condition for the stencil at $\alpha_{n,m+1}$ will be erroneous due
to the aforementioned  non-analyticity  at the $\alpha_{n,m+1} $ - $\alpha_{n
+1,m+1} $ edge. By Method (ii), $\omega_j$ values along
that line are rearranged and analytic sorting between $\omega_j(\alpha_{n,m+1})$
and $\omega_j(\alpha_{n+1,m+1})$ is ensured. Analyticity along this edge forces non-analyticity of $\omega_j$
along the $ \alpha_{n+1,m}$ -
$ \alpha_{n+1,m+1}$ edge, which is equivalent to the BC being horizontal in that
 grid box as shown in figure 16(b).  By continuation of the horizontal sweep at the $m+1$-st level, a horizontal
BC evolves naturally. A vertical sweep, together with the application of
Method (ii) in the vertical direction would produce a vertical BC. It should, in principle,
be possible to modify the algorithm to obtain a branch cut along a suitable complex curve from the
branch point by allowing non-anlayticity at suitable edges of the stencils while sweeping.
 \begin{figure}
\centering
\includegraphics[width=3.25in]{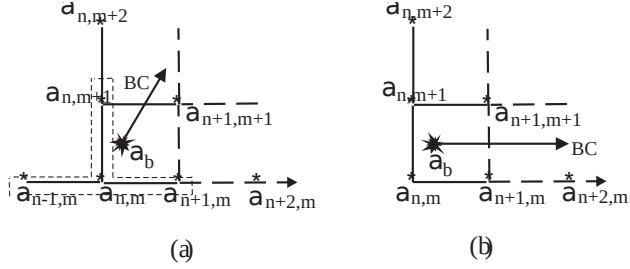}
\caption{Schematic to explain mode sorting around a BP $\alpha_b$ by the
algorithm SPECRE.}
\end{figure}
\section{ Branch point identification}\label{sec:appB}
 The coordinates for which $\norm{ \omega_i - \omega_j}$ is the smallest will locate a BP between the $i^{th}$ and the $j^{th}$ modes. This BP location is further verified by mapping a circle,
of suitably small radius (typically a value of 0.001 has been used here) and centred around the suspected BP, under the relevant modal maps; it is well-known that the modal maps will generate open curves if a BP is being circled and that the open curves together form a closed curve. For example, the
paths traced by modes 1 and 3 when a circle is traced in the $\alpha$ plane around the BP (0.0226,0.0213) are shown in figure 17; each of these paths is an open curve but together they form a closed curve showing that a square root BP is indeed enclosed. \\

For ease of reference, we label the branch points with 3 digit numbers - the first two digits are the mode numbers sharing the branch point and the last is its serial number in the list of branch points between those modes. The branch points are listed in the order they occur from
top to bottom in the complex plane. Thus BP342 is the second of the branch points between modes 3 and 4. Though we refer primarily to BPs in the RHP, it is understood that the images of these in the left half plane are also BPs and the number, in general, refers to both. Also, BPxyn and BPyxn refer to the same BP.
\begin{figure}
\centering
\includegraphics[width=2.in]{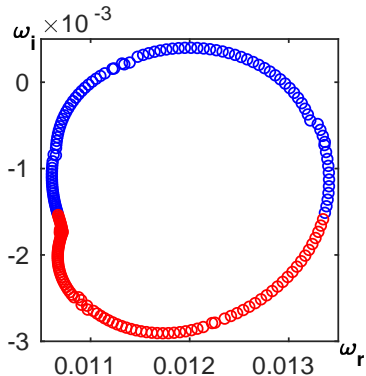}
\caption{Tracks of $\omega_1$ (blue circles) and $\omega_3$ (red circles) when a circle of radius 0.001 is traced around the 131 BP (0.0226,0.0213) in the $\alpha$ plane. The open tracks 
indicate the presence of a branch point, which can be inferred to be a square root as the circles of both colours form a closed curve.}
\end{figure}
\section{Explicit forms of integrals in \eqref{eq11}}\label{sec:appC}
We give explicit forms for the integrals in  \ref{eq11} that involve even and odd adjoint OS and Squire eigenfunctions.
We recall that these eigenfunctions are first computed on the half-domain [0,1] and reflected in the appropriate manner. The function and second derivative are required to vanish at $y = 0$ for an odd eigenfunction whereas vanishing of the first and third derivatives lead to an even eigenfunction. \\

We now consider one part of the integral $I_{+}$ viz.
\begin{fleqn}
\begin{gather*}
I_L = \int_{-1}^{1} \mathcal{L} f \xi_n^{*} dy.
\nonumber
\end{gather*}
\end{fleqn}

Splitting $\mathcal{L} f$ into an even and odd part, we have
\begin{equation}
\mathcal{L} f = f_{Le} + f_{Lo}
\end{equation}

For an even $\xi_n$, $I_L$ can be written as
\begin{equation}
I_{L} \equiv I_{Le}^n = 2 \int_{0}^{1} f_{Le} \xi_n^{*}(y) dy
\end{equation}

\noindent Since $\hat{\xi_n^{*}}(\hat{y}), \hat{y} \in [-1,1]$ is what is actually computed as a Chebyshev expansion, we need to express (C 2) in terms of $\hat{y}.$ The relation between the hatted and unhatted $y$ is given by $ y = (1 + \hat{y}) / 2.$ Dropping the hat, with the understanding that the integrand is a function of $\hat{y}$, we have
\begin{gather}
I_{Le}^n = \int_{-1}^{1} f_{Le} \xi_n^{*} d \hat{y}. 
\end{gather}
A similar expression obtains for the integral $I_{Lo}$ involving the odd eigenfunctions, with $f_{Lo}$ in place of $f_{Le}.$
After some algebra, $f_{Le}$ and $f_{Lo}$ can be shown to be
\begin{gather}
f_{Le} = (G_0 + \frac{G_2}{2}) T_0 + G_1 T_1 + \frac{G_2}{2} T_2\\
{\rm{where}} \, G_0 = \frac{1}{2} \bigg{(}2 i \alpha - \frac{3}{4} i \alpha^3 - \frac{\alpha^4}{Re} \bigg{)}, \,\,
G_1 = \frac{i \alpha^3}{4} \, {\rm{and}} \, G_2 = \frac{i \alpha^3}{8}. \nonumber
\end{gather}
The $T_n(\hat{y})$ are Chebyshev polynomials of the first kind.
\begin{gather}
f_{Lo} = H_0 T_0 + H_1 T_1 + H_2 T_2 + H_3 T_3 + H_4 T_4 + H_5 T_5  \\
{\rm{where}} \, H_0 = a_0 + \frac{a_2}{2} + \frac{3}{8} a_4, \, H_1 = a_1 +
 \frac{3}{4} a_3 + \frac{5}{8} a_5, \, H_2 = \frac{1}{2} (a_2 + a_4), \nonumber
 \\ H_3 = \frac{a_3}{4} + \frac{a_5}{2}, \, H_4 = \frac{a_4}{8},  {\rm{and}}
 \, H_5 = \frac{a_5}{16} \, {\rm{with}}, \,
a_0 = \frac{33}{128} i \alpha^3 - \frac{1}{8} i \alpha + \frac{3}{2} 
\frac{\alpha^2}{Re} + \frac{11}{32} \frac{\alpha^4}{Re}, \nonumber \\
 a_1 = \frac{5}{128} i \alpha^3 - \frac{3}{8} i \alpha + \frac{3}{2} 
\frac{\alpha^2}{Re} + \frac{9}{32} \frac{\alpha^4}{Re}, \, a_2 = - (
\frac{19}{64} i \alpha^3 + \frac{3}{8} i \alpha + \frac{3}{32} 
\frac{\alpha^4}{Re}), \nonumber \\ a_3 = - (\frac{3}{64} i \alpha^3 + 
\frac{i \alpha}{8} + \frac{1}{32} \frac{\alpha^4}{Re}), \, a_4 = \frac{5}{128}
 i \alpha^3, \, a_5 = \frac{i \alpha^3}{128}. \nonumber
\end{gather}
Similarly we have
\begin{equation}
\mathcal{M} f = f_{Me} + f_{Mo}
\end{equation}
with
\begin{gather}
f_{Me} = - \frac{\alpha^2}{2} T_0, \, {\rm{and}} \, f_{Mo} = (F_0 + \frac{F_2}{2}) T_0 + (F_1 + \frac{3}{4} F_3) T_1 + \frac{F_2}{2} T_2 + \frac{F_3}{4} T_3,  \\
{\rm{with}} \, F_0 = \frac{3}{4} + \frac{11}{32} \alpha^2, \, F_1 = \frac{3}{4} + \frac{9}{32} \alpha^2, \, F_2 = - \frac{3}{32} \alpha^2, \, F_3 = - \frac{\alpha^2}{32}. \nonumber
\end{gather}
Using the fact that $\xi_n^{*}(\hat{y})$ is given by the Chebyshev series
\begin{gather*}
\xi_n^{*} = \sum_{i=0}^{\infty} b_{ni}^e T_i(\hat{y})
\end{gather*}
and the fact that
$$
\int_{-1}^{1} T_m(x) T_n(x) dx = \begin{cases} \frac{1}{1 - (m-n)^2} + \frac{1}{1 - (m+n)^2} & \mbox{if} \, m + n \, {\rm{is \, even}} \\ 0 & \mbox{if} \, m + n \, {\rm{is \, odd}} \end{cases},
$$
(C 3) becomes
\begin{gather}
I_{Le}^n = 2 \sum_{k = 0}^{\infty} \bigg{[} \bigg{(} \frac{a_0}{1 - 4 k^2} + \frac{4 k^2 + 3}{(9 - 4 k^2)(4 k^2 - 1)} a_2 \bigg{)} b_{n,2k}^e + \frac{1}{(1 - 2k)(2k + 3)} a_1  b_{n,2k+1}^e \bigg{]}. 
\end{gather}
In this expression, 
\begin{gather*}
a_0 = G_0 + \frac{G_2}{2}, \, a_1 = G_1, \, a_2 = \frac{G_2}{2}.
\end{gather*}
and $b_{ni}^e$ are the Chebyshev coefficients determining the even eigenfunction.
Similarly,
\begin{gather}
I_{Lo}^n = 2 \sum_{k = 0}^{\infty} \bigg{[} \bigg{(} \frac{H_0}{1 - 4 k^2} + \frac{4 k^2 + 3}{(9 - 4 k^2)(4 k^2 - 1)} H_2 + \frac{4 k^2 + 15}{(25 - 4 k^2)(4 k^2 - 9)} H_4 \bigg{)} b_{n,2k}^o \nonumber \\ + \bigg{(} \frac{1}{(1 - 2k)(2k + 3)} H_1  + \frac{10 k + 9}{(9 - 4k^2)(2 k + 1)(2 k + 5)} H_3 + \frac{4 k^2 + 4 k + 25}{(25 - 4 k^2)(2 k - 3)(2 k + 7)} H_5 \bigg{)} b_{n,2k+1}^o \bigg{]}. 
\end{gather}
The $b_{ni}^o$ are the Chebyshev coefficients determining the odd eigenfunction.

As for the second part of the integral, we have
\begin{gather}
I_{Me}^n = - 2 \sum_{k=0}^{\infty} \frac{\alpha^2}{1 - 4 k^2} b_{n,2k}^e,  \\
I_{Mo}^n = 2 \sum_{k = 0}^{\infty} \bigg{[} \bigg{(} \frac{a_0}{1 - 4 k^2} + 
\frac{4 k^2 + 3}{(9 - 4 k^2)(4 k^2 - 1)} a_2 \bigg{)} b_{n,2k}^o \nonumber \\ + \bigg{(} 
\frac{1}{(1 - 2k)(2k + 3)} a_1  + \frac{10 k + 9}{(9 - 4k^2)(2 k +1)(2 k + 5)}
 a_3 \bigg{)} b_{n,2k+1}^o \bigg{]}. 
\end{gather}
In the above expression, 
\begin{gather*}
a_0 = F_0 + \frac{F_2}{2}, \, a_1 = F_1 + \frac{3}{4} F_3, \,
a_2 = \frac{F_2}{2}, \, a_3 = \frac{F_3}{4}.
\end{gather*}
\section{Integral asymptotics by Olver method }\label{sec:appD}
We collect here asymptotic expansions for the case of interacting saddle and pole. Most of the material is sourced from Oughstun (2009) which has the original references. \\

The formulae are presented for the case of $N$ isolated saddles $\alpha_{si}, i = 1, \cdots, N$ interacting with a pole $\alpha_{p1}.$ It is also assumed that the steepest descent path from only one of the saddles crosses the pole with varying $ v_d,$ a real parameter. For the case considered in the text, $N = 2.$
\begin{fleqn}
\begin{eqnarray}
I_{sp}(t;v_d) &\approx& \sum_{i=1}^{2} q(\alpha_{si}) \bigg{(} - \frac{2 \pi}{t p^{''}(\alpha_{si})}
\bigg{)}^{1/2} e^{t p(\alpha_{si})} \nonumber \\
&+& \gamma_1 \bigg{[} \pm i \pi erfc(\mp i \Delta_1 \sqrt{t}) 
e^{t p(\alpha_{p1})} + \sqrt{\frac{\pi}{t}} \frac{e^{t p(\alpha_{s1})}}{\Delta_1} \bigg{]}, \label{A1}
\end{eqnarray}
\end{fleqn}
where
\begin{fleqn}
\begin{eqnarray*}
\Delta_1 &=& [p(\alpha_{s1}) - p(\alpha_{p1}) ]^{1/2},\nonumber \\ \gamma_1 &=& \lim{\alpha \rightarrow \alpha_{p1}} [(\alpha - \alpha_{p1})] q(\alpha), \nonumber 
\end{eqnarray*}
\end{fleqn}
and
\begin{fleqn}
\begin{eqnarray*}
 erfc(z) &=& 1 - \frac{2}{\sqrt{\pi}} \int_{0}^{z} e^{- \xi^2} \, d \xi.\hspace{0.8in}
\end{eqnarray*}
\end{fleqn}
A crucial point in the computation is that the correct branch be chosen for $\Delta_1.$ For this, we make use of the following -
\begin{fleqn}
\begin{align*}\sqrt{\Delta_1^2} = \begin{cases} \Delta_1 & \mbox{if} \,\, - \frac{\pi}{2} <  \theta_1 \le \frac{\pi}{2} \\
- \Delta_1 & \mbox{if} \,\, - \pi < \theta_1 \le \frac{-\pi}{2} \, {\rm{or}} \, \frac{\pi}{2} < \theta_1 < \pi,
\end{cases} \end{align*}
\end{fleqn}
where $ \theta_1 = arg(\Delta_1),$ in turn is computed by eq.(10.91) of Oughstun (2009) as
\begin{fleqn}
\begin{align*} \lim{\alpha_{p1} \rightarrow \alpha_{s1}} \theta_1 = \theta_c + arg\{[-p^{''}(\alpha_{s1}]^{1/2}\} + 2 n \pi  \end{align*}
\end{fleqn}
 with $\theta_c$ being the angle made by the vector from $\alpha_{s1}$ to $\alpha_{p1}$ and $n$ an integer chosen such that $\theta_1$ lies in the principal range $(-\pi,\pi)$.
The upper (lower) signs are to be used when $Im(\Delta_1) > (<) 0.$ When $Im(\Delta_1) = 0$ but $\Delta_1 \ne 0,$ we have
\begin{fleqn}
\begin{eqnarray}
I_{sp}(t;\theta) &\approx& 
\sum_{i=1}^{2} q(\alpha_{si}) \bigg{(} - \frac{2 \pi}{t p^{''}(\alpha_{si})}
\bigg{)}^{1/2} e^{t p(\alpha_{si})} \nonumber \\
&+& \gamma_1 \bigg{[} i \pi erfc(- i \Delta_1(v_d) \sqrt{t}) 
e^{t p(\alpha_{p1})} \nonumber \\ &+& \sqrt{\frac{\pi}{t}} \frac{e^{t p(\alpha_{s1})}}{\Delta_1} - i \pi e^{t p(\alpha_{p1})} \bigg{]}. \label{A2}
\end{eqnarray}
\end{fleqn}
For $\Delta_1 = 0,$ we have
\begin{fleqn}
\begin{eqnarray}
&I_{sp}(t;v_d) &\approx \sum_{i=1}^{2} q(\alpha_{si}) \bigg{(} - \frac{2 \pi}{t p^{''}(\alpha_{si})}
\bigg{)}^{1/2} e^{t p(\alpha_{si})}  \nonumber \\
&- \gamma_1& \bigg{(} - \frac{2 \pi}{t p^{''}(\alpha_{s1})}
\bigg{)}^{1/2} e^{t p(\alpha_{s1})} 
\bigg{[} \frac{1}{\alpha_{s1} - \alpha_{p1}} + \frac{p^{'''}(\alpha_{s1})}{6 p^{''}(\alpha_{s1})} \bigg{]}. \label{A3}
\end{eqnarray}
\end{fleqn}\\

The total integral $I$ is then given as below. The assumption is that the pole is fixed and the saddle moves upward with increasing $v_d.$  There are two cases to consider. If, for small $v_d,$ the LOI is in between the SDP and the pole, $I$ is given by 
 \begin{fleqn}
\begin{align*}
 I = \begin{cases} I_{sp} & \mbox{if} \,\, v_d < (v_d)_p  \\
I_{sp} + i \pi \gamma_1 e^{t p(\alpha_{p1})} & \mbox{if} \,\,  v_d = (v_d)_p \\
I_{sp} + 2 i \pi \gamma_1 e^{t p(\alpha_{p1})} & \mbox{if} \,\,  v_d > (v_d)_p.
\end{cases} 
 \end{align*}
\end{fleqn}
In this case, the pole is encircled in an anticlockwise manner when the LOI is deformed into the SDP. $(v_d)_p$ is the value of $v_d$ at which the pole and saddle collide.

On the other hand, if, for small $v_d,$ the pole is between the LOI and the SDP, then
we have
 \begin{fleqn}
\begin{align*}
 I = \begin{cases} I_{sp} & \mbox{if} \,\, v_d > v_{d_p}  \\
I_{sp} - i \pi \gamma_1 e^{t p(\alpha_{p1})} & \mbox{if} \,\,  v_d = v_{d_p} \\
I_{sp} - 2 i \pi \gamma_1 e^{t p(\alpha_{p1})} & \mbox{if} \,\,  v_d < (v_{d_p}.
\end{cases} 
 \end{align*}
\end{fleqn}
In this case, the pole is encircled in a clockwise manner when the LOI is deformed into the SDP.\\

We now consider the numerical implementation of formulae \ref{A1}-\ref{A3}. If $\Delta_1$ is bounded away from zero, the implementation is straightforward. This is the most likely scenario when an off-axis pole and an off-axis saddle interact; though $Im(\Delta_1)$ passes through zero when the SDP passes through the pole, $Re(\Delta_1)$ and hence $\Delta_1$ itself remain non-zero, in general. However, when $\alpha_{s1}$ and $\alpha_{p1}$ lie on the imaginary axis, it is inevitable that $\Delta_1 =0$ for some $v_d$; this happens when the pole and saddle collide. For other $v_d,\Delta_1$ is pure imaginary. Formulae \ref{A1} and \ref{A3} are to be used in this case. From \ref{A1}, it appears that $I_{sp} \rightarrow \infty$ as $\Delta_1 \rightarrow 0.$ However, the first term also tends to infinity and indeed the resultant cancellations result in \ref{A3}, which is valid for $\Delta = 0.$ For $\Delta \ne 0$ but small, large errors can result if \ref{A1} is used as is. For numerical purposes, we adopt the following procedure -

Choose an $\epsilon > 0.$
For $0 <  Im(\Delta_1) < \epsilon,$ we write
\begin{fleqn}
\begin{eqnarray}
I_{sp}(t;v_d) &\approx& \sum_{i=1}^{2} q(\alpha_{si}) \bigg{(} - \frac{2 \pi}{t p^{''}(\alpha_{si})}
\bigg{)}^{1/2} e^{t p(\alpha_{si})}  \nonumber \\
&-& \gamma_1 \bigg{(} - \frac{2 \pi}{t p^{''}(\alpha_{s1})}
\bigg{)}^{1/2} e^{t p(\alpha_{s1})} 
\bigg{[} \frac{1}{\alpha_{s1} - \alpha_{p1}} + \frac{p^{'''}(\alpha_{s1})}{6 p^{''}(\alpha_{s1})} \bigg{]}\nonumber \\
&-& \gamma_1 i \pi ( 1 - erfc(- i \Delta_1 \sqrt{t}) ) 
e^{t p(\alpha_{p1})}, \label{A4} \\
I_p& =& i \pi \gamma_1 e^{t p(\alpha_{p1})}.  \nonumber
\end{eqnarray}
\end{fleqn}
and for $- \epsilon < Im(\Delta_1) < 0,$ we write
\begin{fleqn}
\begin{eqnarray}
I_{sp}(t;v_d) &\approx& \sum_{i=1}^{2} q(\alpha_{si}) \bigg{(} - \frac{2 \pi}{t p^{''}(\alpha_{si})}
\bigg{)}^{1/2} e^{t p(\alpha_{si})}  \nonumber \\
&-& \gamma_1 \bigg{(} - \frac{2 \pi}{t p^{''}(\alpha_{s1})}
\bigg{)}^{1/2} e^{t p(\alpha_{s1})} \nonumber 
\bigg{[} \frac{1}{\alpha_{s1} - \alpha_{p1}} + \frac{p^{'''}(\alpha_{s1})}{6 p^{''}(\alpha_{s1})} \bigg{]}
\nonumber \\ &-& \gamma_1 i \pi erfc(- i \Delta_1 \sqrt{t}) 
e^{t p(\alpha_{p1})}. \label{A5}
\end{eqnarray}
\end{fleqn}
Typically, we take $\epsilon = 0.01.$ We now check that \ref{A3}-\ref{A5}, result in $I$ being a continuous function of $v_d$, even though the constituents of $I$ viz. $I_{sp}$ and $I_p$ are discontinuous. To avoid clutter, we denote the sum in \ref{A4} as $S$ and the next term as $T.$ Note that $T$ is the limit of the second term in the square bracket of \ref{A1} (which we denote by $U$) as the pole approaches the saddle. Assume the pole does not contribute when $Im(\Delta_1) > 0.$ Then, we have,
\begin{fleqn}
\begin{align}
 I = \begin{cases} S + U + i \pi \gamma_1 erfc(-i \Delta_1 \sqrt{t}) & \mbox{if} \,\, Im(\Delta_1) > \epsilon  \\
S + T  + \gamma_1 i \pi erfc(- i \Delta_1 \sqrt{t}) )& \mbox{if} \,\,  0 < Im(\Delta_1) < \epsilon \\
S + T  - \gamma_1 i \pi erfc( i \Delta_1 \sqrt{t}) ) + 2 \pi i \gamma_1 e^{t p(\alpha_{p1})} & \mbox{if} \,\,  - \epsilon < Im(\Delta_1) < 0 \\
S + U - \gamma_1 i \pi erfc( i \Delta_1 \sqrt{t}) ) + 2 i \pi \gamma_1 e^{t p(\alpha_{p1})} & \mbox{if} \,\,  Im(\Delta_1) < - \epsilon.
\end{cases} 
\label{final_sol}
 \end{align}
\end{fleqn}
The first line is just \ref{A1}, the second, \ref{A4} plus the pole contribution, the third, \ref{A5} plus the pole contribution and the fourth again \ref{A1} with the appropriate sign. It can be checked that $I$ is continuous at a) $Im(\Delta_1) = \epsilon$ as $T$ approaches $U$ as $Im(\Delta_1) \rightarrow \epsilon,$ b) $\Delta_1 = 0,$ as the discontinuity in $I_{sp}$ exactly cancels the discontinuity in the pole contribution and finally at c) $Im(\Delta_1) = - \epsilon$ for the same reason as in (a).\\

Similar expressions can be obtained for the second case, when the pole is encircled in a clockwise manner; the signs of the residue and error function terms will be different.\\

For validation, we consider the integral $I$ defined by
$$\displaystyle I = \int_{-\infty}^{\infty}\frac{\alpha(1-\frac{5}{16}\alpha^2)\ Exp\left[ i(\alpha\ x \ -\omega\ t\right]}{\omega-\omega_0}\ d\alpha .$$ where $\omega$ is the first Orr-Sommerfeld mode. 
We validate the analytic computation by integrating $I$ numerically. For the numerical integration, although the line of integration is the real line, it is more convenient, due to the slow decay rate of the exponent along the real axis, to integrate along a path such as shown in figure 18 where we have also shown the contour plot of $\omega_i-\alpha_i \frac{x}{t}$ to convey an impression of how fast the integrand decays. As can be seen, we have chosen the numerical integration path to follow the steepest descent path beyond $\norm {\alpha_r}  = 2$ (shown as dotted lines in figure 18). The contour levels at two points on the descent paths are indicated in the figure. The descent paths reach a level of $\approx -0.275$ within a rectangular domain $[-4, \ 4] \times [-4, \ 4]$. Hence, for $t=40$, for example, the magnitude of the integrand along the descent path decays to nearly  zero within this domain. This assures that the length of the integration path is sufficient to obtain a converged value.   The grid independence of the numerical integral has been checked by doubling the number of integration points along the integration path.  For $t<40$, numerical integration requires much longer descent paths.
\begin{figure}
\centering
\includegraphics[width=3.5in]{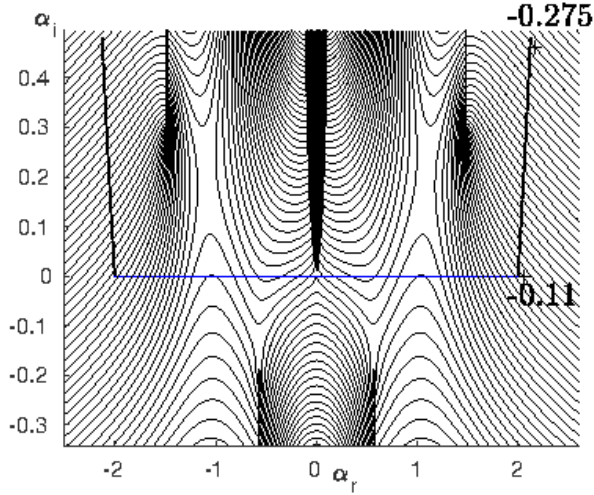} 
\caption{Contour plot of $(\omega_i-\alpha_i \frac{x}{t}); \ \frac{x}{t}=0.5$; Numerical integration path: $[-2,\ 2]\cup \ \hbox{SDP}$ from $\alpha=\pm 2$. }
\end{figure}
We show the real and imaginary parts of $I$, as a function of $v_d$ in figures 6 for $t = 100;$ the drive frequency $\omega_0 = 0.3$.\\

\begin{figure}
\centering
\subfloat[]{
\includegraphics[width=2.in]{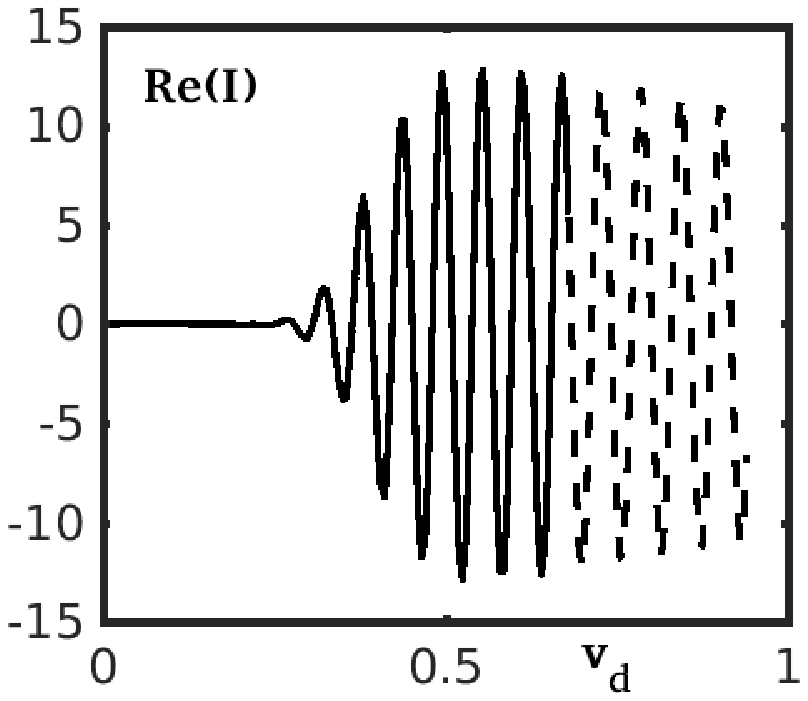}}
\subfloat[]{
\includegraphics[width=2.1in]{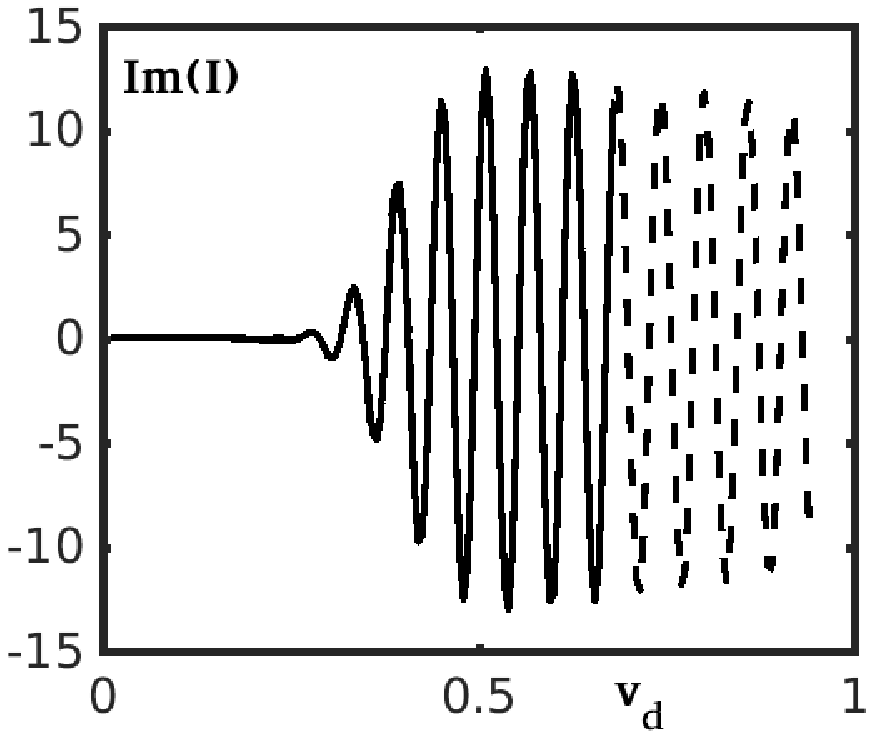}}
\caption{a) Real and b) Imaginary parts of $I_1^{OSE}$. Solid and dashed lines represent numerical and analytic computations.$ \omega_0 = 0.3,t = 100.$}
\end{figure}
 It can be discerned from the analytic computation that the initial transient is almost entirely due to the saddle family C (green line in figure 3a) while the final oscillatory part is due to the pole contribution, with the wavelength and damping related to the real and imaginary parts of the pole coordinates. Though the central saddle (red line in figure 3a) is at a similar height as the right saddle for a range of $v_d,$ (figure 3b) its contribution still turns out to be negligible because of the factor $\alpha$ in the integrand. Figures 19 (a) and (b) show the comparison of the numerical (solid lines) and analytical (dashed lines) values of $Real(I)$ and $Im(I)$ respectively.  The agreement between the numerical integration and the analytical values is very good. Validations were also done for many values of $\omega_0$, though not shown in this paper. \\
\begin{table}
\fontsize{8}{11}\selectfont
\begin{center}
\begin{tabular}{|c|c|c|c|c|c|} \hline

n & 0 & 1 & 2 & 3 & 4    \\ \hline 
$a_n$ & -0.0065 & -0.0056 + 0.0037 i & -0.0016 + 0.0068 i & 0.0128 + 0.0092 i & 0.0657 + 0.0416 i  \\ \hline \hline
n & 5 & 6 & 7 & 8 & 9  \\ \hline
$a_n$ & 0.0312 + 0.257 i & -0.4041 + 0.0926 i & -0.0586 - 0.2329 i & 0.0344 - 0.048 i & 0.0079 - 0.0127 i  \\ \hline \hline
$n$ & 10 & 11 & & & \\ \hline
$a_n$ & 0.0009 - 0.0057 i & -0.0008 - 0.0028 i & & & \\ \hline
\end{tabular}
\caption{Fourier coefficients of the wavepacket at $t = 108$ for $\omega_0 = 0.45. \ N = 11.$  }
\end{center}
\end{table}
 
\begin{table}
\fontsize{8}{11}\selectfont
\begin{center}
\begin{tabular}{|c|c|c|c|c|c|} \hline
n & 0 & 1 & 2 & 3 & 4    \\ \hline 
$a_n$ & -0.0033 & 0.0031 - 0.0009 i & -0.0028 + 0.0018 i & 0.0021 - 0.0026 i & -0.0008 + 0.0034 i \\ \hline \hline
n & 5 & 6 & 7 & 8 & 9  \\ \hline
$a_n$ & -0.0015 - 0.0039 i & 0.0064 + 0.0046 i & -0.016 - 0.0076 i & 0.0328 + 0.0208 i & -0.0462 - 0.0604 i \\ \hline \hline
n & 10 & 11 & 12 & 13 & 14  \\ \hline
$a_n$ & 0.0156 + 0.1285 i & 0.0982 - 0.1554 i & -0.2021 + 0.0463 i & 0.1509 + 0.0991 i & -0.0129 - 0.1164 i  \\ \hline \hline
n & 15 & 16 & 17 & &  \\ \hline
$a_n$ & -0.0197 + 0.0597 i & 0.0172 - 0.024 i & -0.0088 + 0.0106 i & & \\ \hline
\end{tabular}
\caption{Fourier coefficients of the wavepacket at $t = 108$ for $\omega_0 = 0.45. N = 17.$  }
\end{center}
\end{table}
\section{Wavepacket reconstruction}\label{sec:appE}
    It may be noted that only one wavepacket emerges in the solution of the IBVP, while the Floquet framework necessarily implies a periodic system of wavepackets. For the present analysis to have any relevance to the original problem, it is important to know what effect the separation between the wavepackets has on the secondary growth rates. For this purpose, we construct a periodic wavepacket system based on the IBVP wavepacket, with zero padding on either side so as to control the separations of the packets. Assuming the length of the padded wavepacket to be the wavelength $\lambda_b,$ the Fourier coefficients of the periodic wavepacket system are obtained. With $I$ denoting the index of the maximum of these fourier coefficients, $I \alpha$ gives the central wavenumber $\alpha_c$. The smaller the $\alpha,$ the larger the number of fourier modes $N$. Too small an $\alpha$ does lead to an increased separation of the wavepackets but will also entail large $N$ and consequently large computational times. In the procedure adopted here, we first find $\alpha_c$ accurately by choosing a very small base wavenumber $\alpha.$ The base wavenumber is then redefined to be an integral fraction of $\alpha_c$ and the fourier coefficients are recomputed. For the wavepacket at $t=108$, $\alpha_c$ is found to be $\approx 1.0802$ and we have calculated using $\alpha = \alpha_c / 6, \alpha_c / 12$ and $\alpha_c / 16$ (0.18004, 0.09002 and 0.06751 respectively).
The fourier coefficients for 
$\alpha_c/6$ $\alpha_c/12$ are tabulated in Tables 6 and 7. It may be noted that the seventh  and the thirteenth coefficient are the largest in these two representations respectively. \\

\begin{figure}
\includegraphics[width=5.1in, height=1.75in]{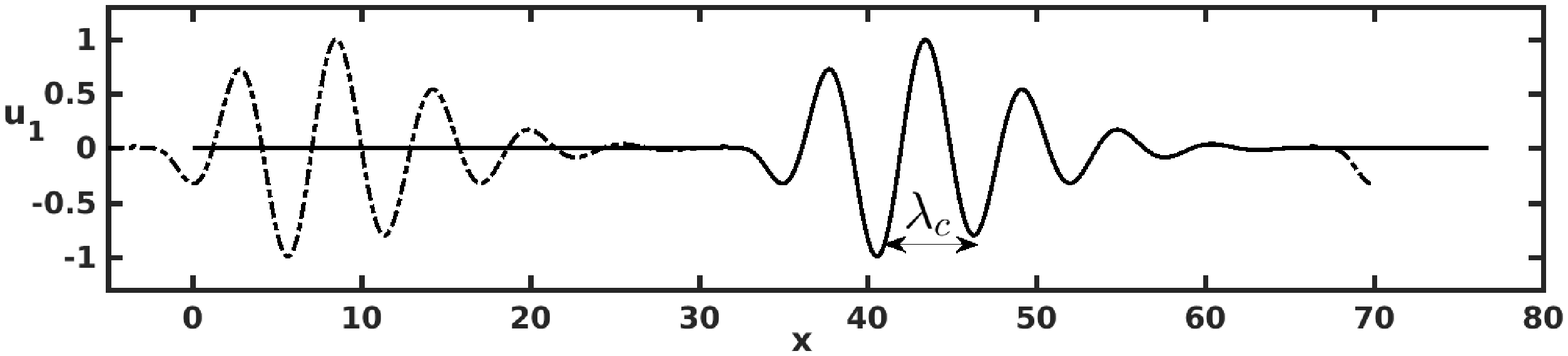}
\caption{Periodic wavepacket system for $Re=5000$, $\omega_0=0.45$ at $t=108$. $\alpha\ = \ \alpha_c/6$}
\end{figure}
\begin{figure}
\includegraphics[width=5.1in, height=1.75in]{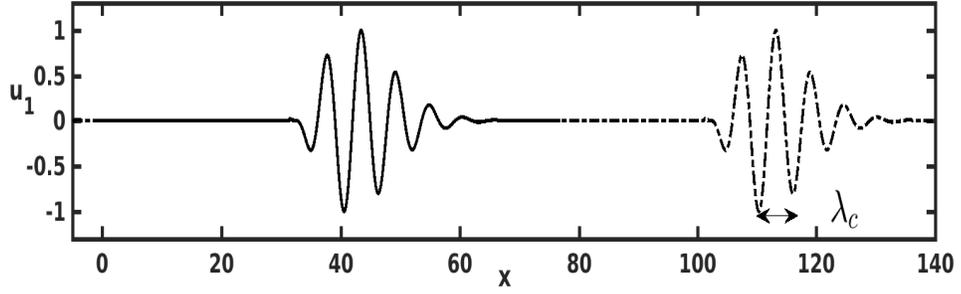}
\caption{Periodic wavepacket system for $Re=5000$, $\omega_0=0.45$ at $t=108$. $\alpha\ = \ \alpha_c/12$}
\end{figure}
Figures 20 and 21 illustrate the periodic wavepacket systems constructed with these $\alpha,$ the effect of which on the separation length can be clearly seen. The larger the separation between the wavepackets, the closer they may be assumed to approximate the actual situation of a single wavepacket. In these two figures, the solid line denotes the localized wavepacket from the IBVP solution; the dashed line, indistinguishable from the solid one, is the periodic reconstruction of the wavepacket. The magnitudes of the fourier coefficients corresponding to $\alpha = \alpha_c / 6$ and $\alpha_c / 12$ are shown in figure 22. 
 \begin{figure}
\centering
\includegraphics[width=5.6in, height=1.75in]{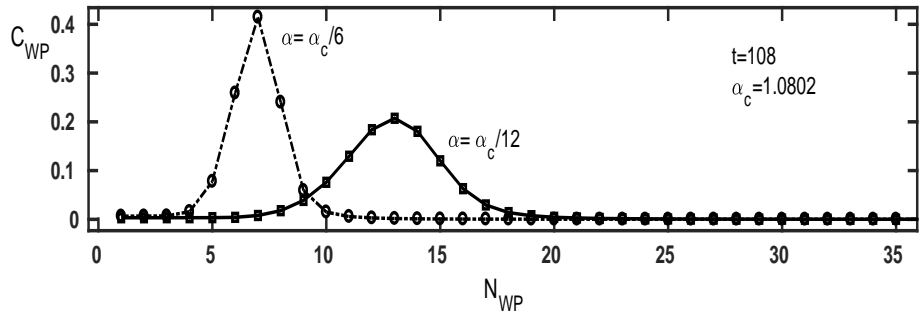}
\caption{Fourier coefficients of the periodic wavepacket systems corresponding to $\alpha=\alpha_c/6,\ \hbox{and} \ \alpha_c/12$ for $Re=5000$; $\omega_0=0.45$.}
\end{figure}
\section{Floquet equations }\label{sec:appF}
The secondary base state velocity is given by
\begin{fleqn}
\begin{align}
\vec{v}_2(\xi,y)=U(y)+A\ \vec{v}(\xi,y); \quad \xi=x-ct
\end{align}
\end{fleqn}
where $ \vec{v}(\xi,y)$ is the normalized TS wave or the wavepacket from the IBVP solution. The  secondary disturbances $\vec{v}_3(\xi,y,z,t)$ are defined  as perturbations in the total velocity:
\begin{fleqn}
\begin{align}
\vec{v}_T(\xi,y,z,t)= \vec{v}_2(\xi,y)+\epsilon\ \vec{v}_3(\xi,y,z,t)
\end{align}
\end{fleqn}
\noindent Substituting (F 2) into the N-S equations, linearising in $\epsilon$, eliminating the pressure and the spanwise velocity component $w_3$ by taking curl and using continuity respectively, a coupled PDE system for the streamwise and normal velocity components $u_3$ and $v_3$ is obtained as (Herbert et al 1987) -
\begin{fleqn}
\begin{align}
&\bigg{[} \frac{1}{Re} \nabla^2 - (U - c) \frac{\partial}{\partial x} -  \frac{\partial}{\partial t} \bigg{]}   \frac{\partial \eta_3}{\partial z} + \zeta_0 \frac{\partial^2 v_3}{\partial z^2} 
\nonumber \\ &+ A \bigg{[}( - \frac{\partial \psi_1}{\partial y} \frac{\partial}{\partial x} + 
\frac{\partial \psi_1}{\partial x} \frac{\partial}{\partial y} - 
\frac{\partial^2 \psi_1}{\partial x \partial y} \frac{\partial \eta_3}{\partial z} 
\nonumber \\ &+  \frac{\partial^2 \psi_1}{\partial x^2} 
( \frac{\partial^2 u_3}{\partial x \partial y} + \frac{\partial^2 v_3}{\partial y^2} ) 
- \frac{\partial^2 \psi_1}{\partial y^2} \frac{\partial^2 v_3}{\partial z^2} \bigg{]} = 0,  \\
&\bigg{[} \frac{1}{Re} \nabla^2 - (U - c) \frac{\partial}{\partial x} - \frac{\partial}{\partial t} \bigg{]} \nabla^2 v_3 - \frac{d \zeta_0}{dy} \frac{\partial v_3}{\partial x} \nonumber \\&+ A \bigg{[} (- \frac{\partial \psi_1}{\partial y} \frac{\partial}{\partial x} + \frac{\partial \psi_1}{\partial x} \frac{\partial}{\partial y} ) \nabla^2 v_3 + \frac{\partial^2 \psi_1}{\partial x^2} (\frac{\partial \zeta_3}{\partial y} + \frac{\partial \eta_3}{\partial z}) \nonumber \\ &- \frac{\partial^2 \psi_1}{\partial x \partial y} (\frac{\partial \zeta_3}{\partial x} + \frac{\partial \xi_3}{\partial z}) - \frac{\partial \zeta_1}{\partial x} (2 \frac{\partial u_3}{\partial x} + \frac{\partial v_3}{\partial y})\nonumber \\ &- \frac{\partial \zeta_1}{\partial y} \frac{\partial v_3}{\partial x} - (u_3 \frac{\partial}{\partial x} + v_3 \frac{\partial}{\partial y} ) \frac{\partial \zeta_1}{\partial x} \bigg{]} = 0. 
\end{align}
\end{fleqn}
{\noindent The boundary conditions are $u_3 = v_3 = \partial v_3 / \partial y = 0  \ \ \hbox{at} \ \ y = \pm 1$}. Standard Floquet theory (for e.g. Nayfeh \& Mook 1995) indicates that the disturbance equations, with periodic coefficients of period $\lambda = 2 \pi / \alpha$ admit solutions of the form 
\begin{fleqn}
\begin{gather}
\begin{bmatrix} u_3(\xi,y,z,t) \\ v_3(\xi,y,z,t) \end{bmatrix} =e^{\gamma \xi} e^{\sigma t} e^{i\beta z}\sum_{m=-\infty}^\infty \begin{bmatrix} u_{m}(y) \\ v_m(y) \end{bmatrix} e^{i m \alpha \xi}.  
\end{gather}
\end{fleqn}
Substituting (F 5) into (F 3) and (F 4) , and collecting coefficients of $e^{i m \alpha \xi}$, we have, for each $m,$ the following sets of equations -
\begin{fleqn}
\begin{align}
\bigg{[}& \frac{(\delta_m^2 - \beta^2)^2}{Re} + (U - c) \delta_m (\beta^2 - \delta_m^2) - A \delta_m (\delta_m^2 - \beta^2) (a_0 \phi^{'} + \overline{a}_0 {\overline{\phi}}^{'} ) \bigg{]} u_m 
+ \frac{\delta_m^2 - \beta^2}{Re} \frac{d^2 u_m}{d y^2}\nonumber \\
&+ \beta^2  \bigg{[}U^{'} + A (a_0 \phi^{''} + b_0 \overline{\phi}^{''}) \bigg{]} v_m 
+ \bigg{[} \frac{\delta_m (\delta_m^2 - \beta^2)}{Re} 
- (U - c) \delta_m^2 - A \delta_m^2 (a_0 \phi^{'} + \overline{a}_0 \overline{\phi}^{'}) \bigg{]} \frac{d v_m}{d y}\nonumber \\ 
&+ \frac{\delta_m}{Re} \frac{d^3 v_m}{d y^3} 
- A \phi^{'} \sum_{n=1}^{N} a_n (\delta_{m-n}^2 - \beta^2) (\delta_{m-n} + i \alpha n) u_{m-n}   \nonumber\\
&+ A \phi \alpha \sum_{n=1}^{N} \bigg{[} i n (\delta_{m-n}^2 - \beta^2) - \alpha n^2 \delta_{m-n} \bigg{]} a_n \frac{d u_{m-n}}{dy} \nonumber \\
&- A \overline{\phi}^{'} \sum_{n=1}^{N} \overline{a}_n (\delta_{m+n}^2 - \beta^2) (\delta_{m+n} - i \alpha n) u_{m+n} \nonumber \\\
&- A \overline{\phi} \alpha \sum_{n=1}^{N} \bigg{[} i n (\delta_{m+n}^2 - \beta^2) + \alpha n^2 \delta_{m+n} \bigg{]} \overline{a}_n \frac{d u_{m+n}}{dy} 
+ A \beta^2 \phi^{''} \sum_{n=1}^{N} a_n v_{m-n} \nonumber \\
&- A \phi^{'} \sum_{n=1}^{N} \delta_{m-n} (\delta_{m-n} + i \alpha n) a_n \frac{d v_{m-n}}{dy} + A \alpha \phi \sum_{n=1}^{N} (i \delta_{m-n} - \alpha n) n a_n \frac{d^2 v_{m-n}}{d y^2} \nonumber \\ 
&+ A \beta^2 \overline{\phi}^{''} \sum_{n=1}^{N} \overline{a}_n v_{m+n} - A \overline{\phi}^{'} \sum_{n=1}^{N} \delta_{m+n} (\delta_{m+n} - i \alpha n) \overline{a}_n \frac{d v_{m+n}}{dy} \nonumber \\
&- A \alpha \overline{\phi} \sum_{n=1}^{N} (i \delta_{m+n} + \alpha n) n \overline{a}_n \frac{d^2 v_{m+n}}{d y^2}  = \sigma \bigg{[} (\delta_m^2 - \beta^2) u_m + \delta_m \frac{d v_m}{d y} \bigg{]},  
\end{align}
\end{fleqn}
\begin{fleqn}
\begin{align}
&\bigg{[} \frac{(\delta_m^2 - \beta^2)^2}{Re} - (U - c) \delta_m (\delta_m^2 - \beta^2) - 2 \delta_m \nonumber\\
 &+ A \delta_m \bigg{(} a_0 \phi^{'''} + \overline{a}_0 \overline{\phi}^{'''} - (\delta_m^2 - \beta^2) * (a_0 \phi^{'} + \overline{a}_0 \overline{\phi}^{'}) \bigg{)} \bigg{]} v_m \nonumber \\
 &+ \bigg{[} \frac{2}{Re} (\delta_m^2 - \beta^2) - (U - c) \delta_m \bigg{]} \frac{d^2 v_m}{d y^2} + \frac{1}{Re} \frac{d^4 v_m}{d y^4} \nonumber \\
&+ A \sum_{n=1}^{N} \bigg{[} (n^2 \alpha^2 \phi - \phi^{''})(\alpha n - 2 i \delta_{m-n}) - \alpha n \phi (\delta_{m-n}^2 - \beta^2) \bigg{]} n \alpha a_n u_{m - n} \nonumber \\ 
&+ 2 i \alpha A \phi^{'} \sum_{n=1}^{N} n a_n \delta_{m - n} \frac{d u_{m-n}}{dy} 
+ A \alpha^2 \phi \sum_{n=1}^{N} n^2 a_n \frac{d^2 u_{m-n}}{dy^2} \nonumber \\
&+ A \sum_{n=1}^{N} \bigg{[} (n^2 \alpha^2 \overline{\phi} - \overline{\phi}^{''})(\alpha n 
+ 2 i \delta_{m+n}) - \alpha n \overline{\phi} (\delta_{m+n}^2 - \beta^2) \bigg{]} n \alpha \overline{a}_n u_{m + n} \nonumber \\ \nonumber
\end{align}
\end{fleqn}
\begin{fleqn}
\begin{align} 
&-  2 i \alpha A \overline{\phi}^{'} \sum_{n=1}^{N} n \overline{a}_n \delta_{m + n} \frac{d u_{m+n}}{dy} + A \alpha^2 \overline{\phi} \sum_{n=1}^{N} n^2 \overline{a}_n \frac{d^2 u_{m+n}}{dy^2}-A \nonumber \\
&\sum_{n=1}^{N} \bigg{[} \phi^{'} \bigg{(} \delta_{m-n} (\delta_{m-n}^2 - \beta^2) + i \alpha n (\delta_{m-n}^2 + \beta^2) \bigg{)} + (n^2 \alpha^2 \phi^{'} - \phi^{'''})(i \alpha n + \delta_{m-n}) \bigg{]} v_{m-n} \nonumber \\
&+ A \sum_{n=1}^{N} \bigg{[} i \alpha n \phi (\delta_{m-n}^2 - \beta^2) - 2 \alpha^2 n^2 \phi \delta_{m-n} - i \alpha n (n^2 \alpha^2 \phi - \phi^{''}) \bigg{]} a_n \frac{d v_{m-n}}{dy} \nonumber \\ 
&+ A \phi^{'} \sum_{n=1}^{N} (i \alpha n - \delta_{m-n}) a_n \frac{d^2 v_{m-n}}{d y^2} + A i \alpha \phi \sum_{n=1}^{N} n a_n \frac{d^3 v_{m-n}}{d y^3} -A\nonumber \\
\nonumber
&\sum_{n=1}^{N} \bigg{[} \overline{\phi}^{'} \bigg{(} \delta_{m+n} (\delta_{m+n}^2 - \beta^2) + i \alpha n (\delta_{m+n}^2 + \beta^2) \bigg{)} + (n^2 \alpha^2 \overline{\phi}^{'} - \overline{\phi}^{'''})(-i \alpha n + \delta_{m+n}) \bigg{]} v_{m+n} \nonumber \\
&- A \sum_{n=1}^{N} \bigg{[} i \alpha n \overline{\phi} (\delta_{m+n}^2 - \beta^2) - 2 \alpha^2 n^2 \phi \delta_{m+n} - i \alpha n (n^2 \alpha^2 \overline{\phi} - \overline{\phi}^{''}) \bigg{]} \overline{a}_n \frac{d v_{m+n}}{dy} \nonumber \\
 &- A \overline{\phi}^{'} \sum_{n=1}^{N} (i \alpha n + \delta_{m+n}) \overline{a}_n \frac{d^2 v_{m+n}}{d y^2} - A i \alpha \overline{\phi} \sum_{n=1}^{N} n \overline{a}_n \frac{d^3 v_{m+n}}{d y^3} = \sigma \bigg{[} (\delta_m^2 - \beta^2) v_m + \frac{d^2 v_m}{d y^2} \bigg{]}. 
 \end{align}
 \end{fleqn}
In the above, $\delta_m = \gamma + i m \alpha.$
As is evident, the equation for the $m^{th}$ fourier coefficient involves fourier coefficients from the $(m-N)^{th}$ to the $(m+N)^{th}$ levels. Even though $m$ ranges over the real line, for numerical purposes, we truncate to a maximum of $m = M$ where $M \ge N.$ Chebyshev collocation at $K + 1$ points in $y$ renders this ODE system into a matrix eigenvalue problem
\begin{fleqn}
\begin{eqnarray*}
S {\bf{V}} = \sigma T {\bf{V}}
\end{eqnarray*}
\end{fleqn}
where ${\bf{V}}$ is the $2 (K + 1)(2M + 1)$ dimensional vector with components \begin{fleqn} 
\begin{align*}(u_{-M,0},\cdots,u_{-M,K},v_{-M,0},\cdots, v_{-M,K},\cdots,u_{M,0},\cdots,u_{M,K},v_{M,0},\cdots,v_{M,K})\end{align*}
\end{fleqn} and $S$ and $T$ are square matrices of the same dimension. $u_{i,j}$ is the value of the fourier coefficient $u_i$ at the $j^{th}$ collocation point. $S$ and $T$ are both banded matrices, the former with varying bandwidth and the latter with a fixed width, as shown below -
\begin{fleqn}
\begin{gather}
S = \left(\begin{array}{cccccc}
s_{1,1}& \cdots& \cdots & s_{1,j_1}&\ddots &0 \\
\vdots & \ddots & \ddots & \ddots & \ddots &  \ddots\\
s_{i_1,1}& \cdots & \cdots & \cdots& \cdots& s_{i_1,i_2} \\
\vdots & \ddots & \ddots & \ddots & \ddots & \vdots \\
0& \ddots & \ddots &  s_{i_2,j_2}& \cdots & s_{i_2,i_2} \\
\end{array}\right),  \\
T = \left(\begin{array}{cccccc}
t_{1,1} & \cdots & t_{1,j_3} & 0 & 0 & 0\\
0 & \ddots & \ddots & \ddots & 0 & 0 \\
\vdots & \ddots & t_{i_1,i_1} & \cdots & t_{i_1,j_4} & \vdots \\
0 & 0 & \ddots & \ddots & \ddots & 0 \\
0 & 0 & 0 & t_{i_2,j_5}& \cdots& t_{i_2,i_2} \\
\end{array}\right). 
\end{gather}
\end{fleqn}
In the above,
\begin{fleqn} 
\begin{align*}
&i_1 = 2 M (K + 1) + 1, i_2 = 2 (2 M + 1)(K+1),\\
 &j_1 = 2(N + 1)(K + 1), j_2 = 2(2 M - N)(K + 1) + 1, \\
  &j_3 = 2 K + 2, j_4 = 2(M + 1)(K + 1), j_5 = 4 M (K + 1) + 1.
\end{align*}
\end{fleqn}

\begin{thebibliography}{99}
\bibitem[Ashpis \& Reshotko (1990)]{ash1}{\sc Ashpis, D.E.~ \& Reshotko, E.~} 1990 The vibrating ribbon problem revisited. {\em J.~Fluid Mech.\/} {\bf{213}}, 531--547.
\bibitem[Cherubini et al (2010)]{cher} {\sc Cherubini, S.~, Robinet, J.-C.~, Bottaro, A.~ \& De Palma, P.~} 2010 Optimal wave packets in a boundary layer and initial phases of a turbulent spot. {\em J.~Fluid Mech.\/}{\bf{656}}, 231 - 259.
\bibitem[Croswell (1985)]{crosw}{\sc Croswell, J.~ W.~} 1985 On the energetics of primary and secondary instabilities in plane Poiseuille flow, M.S Thesis, VPI \& SU. 
\bibitem [Dhanak (1983)]{dhan} {\sc Dhanak, M.~R.~} 1983 On certain aspects of three-dimensional 
instability of parallel flows. {\it Proc. Roy. Soc. Lond. A}{\bf{385}}, 53 - 84.
\bibitem[Di Prima \& Habetler (1969)]{prima} {\sc Di Prima, R.~C.~ \& Habetler, G.~J.~} 1969 A completeness theorem for non-selfadjoint problems in hydrodynamic stability, {\it Arch. Rat. Mech. \& Anal.} 34(3), 218 - 227.
\bibitem[Drazin \& Reid (1985)]{draz}{\sc Drazin, P.~G.~ \& Reid, W.~H.~} 1985 Hydrodynamic stability. {\em Camb. Univ. Press\/} 
\bibitem[Ellingsen \& Palm (1975)]{ell} {\sc Ellingsen, T.~ \& Palm, E.~ }1975 Stability of linear flow. {\em Phy. Fluids}, {\bf{18}}, 487 - 488.
\bibitem[Felsen \& Marcuvitz (1973)]{felsen}{\sc Felsen, L.~B.~ \& Marcuvitz, N.~ }1973 Radiation and scattering of waves, IEEE PRESS Series on Electromagnetic Waves, Wiley - Interscience.
\bibitem[Gaster (1965)]{gaso}{\sc Gaster, M.~} 1965 On the generation of spatially growing waves in a boundary layer. {\em J.~Fluid Mech.\/} {\bf{22(3)}}, 433--441.
\bibitem[Gaster (1968)]{gas1} {\sc Gaster, M.~ 1968} The development of three-dimensional wave packets in a boundary layer. {\em J.~Fluid Mech.\/}{\bf{32(1)}}, 173 - 184.
\bibitem[Gaster \& Davey (1968)]{gasd}{\sc Gaster, M.~ \& Davey, A.~} 1968 The development of three-dimensional wave-packets in unbounded parallel flows. {\em J.~Fluid Mech.\/} {\bf{32}}, 801-808.
\bibitem[Gordillo \& Perez-Saborid (2002)]{gor}{\sc Gordillo, J.~ M.~ \& Perez-Saborid, M.~} 2002 Transient effects in the signaling problem. {\em Phy. Fluids}{\bf{14(12)}}, 4329--4343.
\bibitem[Gustavsson (1979)]{gus}{\sc Gustavsson, L.~H.~} 1979 Initial value problem for boundary layer flows. {\em Phy. Fluids\/} {\bf{22}}, 1602--1605.
\bibitem[Henningson (1989)]{henn1} {\sc Henningson, D. S.~} 1989 Wave growth and spreading of a turbulent spot in plane Poiseuille flow. {\it Phys. Fluids A}, {\bf{1}}, 1876-1882.
\bibitem[Herbert(1983)]{herb1}{\sc Herbert, Th.~} 1983 Secondary instability of plane channel flow to subharmonic three-dimensional disturbances. {\em Phy. Fluids \/}{\bf{26(4)}}, 871--874. 
\bibitem[Herbert et al (1987)]{herb3}{\sc Herbert, Th.~, Bertolotti, F.~P.~ \& Santos, G.~R.~} 1987 Floquet analysis of secondary instability in shear flows.  {\em Stability of time-dependent and spatially varying flows.\/} {\sc Ed. Dwoyer, D.~L.~ \& Hussaini, M.~Y.~}, 43--57.
\bibitem[Herbert (1988)]{herb2}{\sc Herbert, Th.~} 1988 Secondary instability of  boundary layers. {\em Ann. Rev. Fluid Mech.\/} {\bf{20}}, 487--526.
\bibitem[Hill (1995)]{hill}{\sc Hill, D.~C.~} 1995 Adjoint systems and their role in the receptivity problem for boundary layers. {\em J.~Fluid Mech.\/}{\bf{292}}, 183--204.
\bibitem[Itoh(1974)]{ito}{\sc Itoh, N.~} 1974 Spatial growth of finite wave disturbances in parallel and nearly parallel flows. Part 1. The theoretical analysis and the numerical results for plane Poiseuille flow. {\em Trans. Japan Soc. Aero. Space Sci.} {\bf{17}}, 160--174.
\bibitem[Jones (1988)]{jones}{\sc Jones, C.~A.~} 1988 Multiple eigenvalues and mode classification in plane Poiseuille flow, {\it The Quart. J. Mech. App. Math.} 41(3),363 - 382.
\bibitem[Juniper (2006)]{juniper}{\sc Juniper, M.~ P.~} 2006 The effect of confinement on the stability of two-dimensional shear flows, {\em {J. Fluid Mech.}}  {\bf{565}}, 171 -- 195.
\bibitem[Kato(1995)]{Kato-1995} 
{\sc Kato, T.~} 1995 Perturbation theory for linear operators. Springer.
\bibitem[Kidambi \& Srinivasan (2018)]{kidambi}{\sc Kidambi, R.~ \& Srinivasan, U.~} 2018 Is the subharmonic threshold always lower than the fundamental one in plane Poiseuille flow?{\em Phy. Fluids\/}{\bf{30}(1)}
\bibitem[Kim \& Moser (1989)]{kim} {\sc Kim, J.~ \& Moser, R.~D.~} 1989 On the secondary instability in 
plane Poiseuille flow. {\it Phys. Fl. A} {\bf{1(5)}}, pp. 775 - 777.
{\it Theoret. Comput. Fluid Dynamics} {\bf{1}}, pp. 41 - 64.
\bibitem[Kleiser (1982)]{klei}{\sc Kleiser, L.~} 1982 Spectral simulations of laminar-turbulent transition in plane Poiseuille flow and comparison with experiments. {\em Lecture Notes in Physics} {\bf{170}}, 280--285.
\bibitem[Koch (1986)]{koch}{\sc Koch, W.~} 1986 Direct resonances in Orr-Sommerfeld problems. {\em Acta Mechanica\/} {\bf{59}}, 11--29.
\bibitem[Lanczos (1996)]{lanc}{\sc Lanczos, C.~} 1996 Linear differential operators. {\em Classics in App. Math.} {\bf{18}}, SIAM.
\bibitem[Li \& Widnall (1989)]{li} {\sc Li, F.~ \& Widnall, S.~}1989  Wave patterns in plane Poiseuille flow created by concentrated disturbances. {\it{JFM}}, {\bf{208}}, 639-656.
\bibitem[Lingwood (1997)]{ling1}{\sc Lingwood, R.~J.~} 1997 On the Application of the Briggs’ and
Steepest-Descent Methods to a Boundary-Layer Flow. {\em Stud. App. Math} {\bf{98:3}}, 213-254.
\bibitem[Ma et al (1999)]{ma}{\sc Ma, B.~, van Doorne, C.~W.~H.~, Zhang, Z.~ \& Nieuwstadt, F.~T.~} 1999 On the spatial evolution of a wall-imposed periodic disturbance in pipe Poiseuille flow at $Re = 3000.$ Part 1. Subcritical disturbance. {\em J.~Fluid Mech.\/}{\bf{398}}, 181--224.
\bibitem[Manuilovich (1992)]{man}{\sc Manuilovich, S.~V.~} 1992 Sensitivity of plane Poiseuille flow to vibration of the channel walls. {\em Izvestiya Nauk.\/} {\bf{4}}, 12--19.
\bibitem[Nayfeh \& Mook (1995)]{nay} {\sc Nayfeh, A.~H.~ \& Mook, D.~T.~} 1995 {\em Nonlinear oscillations.} John Wiley \& Sons.
\bibitem[Nishioka et al (1975)]{nish1}{\sc Nishioka, M.~, Iida, S.~ \& Ichikawa, Y.~} 1975 An experimental investigation of the stability of plane Poiseuille flow. {\em J.~Fluid Mech.\/}{\bf{72(4)}}, 731--751.
\bibitem[Nishioka et al (1981)] {nish1a}{\sc Nishioka, M.~, Iida, S.~ \& Kanbayashi, S.~}  1981 An experimental investigation of the subcritical instability in plane Poiseuille flow. NASA TM-75885\footnote{The names of the authors in this translated report are incorrect. The original version is in Japanse published in Proc. 10th Turbulence Symposium, Inst. Space Aeron. Sci., Tokyo Univ., 1978, p. 55-62}.
\bibitem[Nishioka \&  Asai (1984)]{nish} {\sc Nishioka, M.~ \& Asai, M.~} 1984 Three-dimensional 
wave-disturbances in plane Poiseuille flow, pp. 173 - 182 of {\it 
Laminar-Turbulent transition, Ed. Kozlov, V.~V.~} IUTAM Symposium, Novosibirsk, 
USSR.
 \bibitem[Orszag \& Patera (1983)]{ors} {\sc Orszag, S.~A.~ \& Patera, A.~T.~} 1983 Secondary instability of wall-bounded shear flows. {\em J.~Fluid Mech.\/}{\bf{128}}, 347--385.
\bibitem[Oughstun (2009)]{ough}{\sc Oughstun, K.~E.~} 2009 Electromagnetic and optical pulse propagation 2. {\em Springer Series in Optical Sciences}, Springer.
\bibitem[Ramazanov (1984) ]{ramaz} {\sc Ramazanov, M.~P.} 1984 Development of finite-amplitude 
disturbances in Poiseuille flow, pp. 183 - 190 of {\it Laminar-Turbulent 
transition}, Ed. Kozlov, V.~V.~ IUTAM Symposium, Novosibirsk, USSR.
\bibitem[Reddy et al (1998)]{red}{\sc Reddy, S.~C.~, Schmid, P.~J.~, Baggett, J.~S.~ \& Henningson, D.~S.~} 1998 On stability of streamwise streaks and transition thresholds in plane channel flows. {\em J.~Fluid Mech.\/}{\bf{365}}, 269--303.
\bibitem[Reynolds \& Potter (1967)]{rey}{ \sc Reynolds, W.~C.~ \& Potter, M.~C.~} 1967 Finite-amplitude instability of parallel shear flows. {\em J.~Fluid Mech.}{\bf{27}}, 465--492.
\bibitem[Schmid \& Henningson (2001)]{schmid}{\sc Schmid, P.~J.~ \& Henningson, D.~S.~} 2001 Stability and transition in shear flows.{\em App. Math. Sci. \/} {\bf{142}}, Springer.
\bibitem[Schot (1992)]{schot}{\sc Schot, S.~ H.~} 1992 Eighty years of Sommerfeld’s radiation condition. {\em Historia Mathematica} {\bf{19}}, 385--401.
\bibitem[Schubauer \& Skramstad(1948)]{SS1948}{\sc Schubauer, G.~B.~, \& Skramstad, H.~K.~}1948 Laminar-boundary-layer oscillations and transition on a flat plate. NACA-TR-909.
\bibitem[Sen \& Venkateswarlu (1983)]{sen}{\sc Sen, P.~K.~ \& Venkateswarlu, D.~} 1983 On the stability of plane Poiseuille flow to finite-amplitude disturbances, considering higher-order Landau coefficients. {\em J.~Fluid Mech.}{\bf{133}}, 179--206. 
\bibitem[Suslov \& Paolucci (1999)]{sus2}
{\sc Suslov, S.~A. \& Paolucci, S.~} 1999 Nonlinear stability of mixed convection flow under non-Boussinesq conditions. Part 1. Analysis and bifurcations. {\em J.~Fluid Mech.\/} {\bf 398}, 61--85.
\bibitem[Suslov (2006)]{sus1}{\sc Suslov, S.~A.~} 2006 Numerical aspects of searching convective / absolute instability transition. {\em J.~Comput. Phys.\/} {\bf{212}}, 188--217.
\bibitem[Trefethen et al (1993)]{tref}{\sc Trefethen, L.~N.~, Trefethen, A.~E.~, Reddy, S.~C.~ \& Driscoll, T.~A.~} 1993 Hydrodynamic stability without eigenvalues. {\em Science\/} {\bf{261}}, 578--584.
\bibitem[Tumin (1996)]{tum}{\sc Tumin, A.~} 1996 Receptivity of pipe Poiseuille flow. {\em J.~Fluid Mech.} {\bf{315}}, 119--137.
\bibitem[Zang \& Krist (1989)]{zang} {\sc Zang, T.~A.~ \& Krist, S.~E.~} 1989 Numerical experiments on 
stability and transition in plane channel flow.
\bibitem[Zhou(1982)]{zhou}{\sc Zhou, H.~} 1982 On the nonlinear theory of stability of plane Poiseuilled flow in the subcritical range. {\em Proc.~R.~Soc.~Lond.~A \/}{\bf{381}}, 407--418.\\
\end{thebibliography}
\end{document}